\documentclass[rmp,aps,reprint,superscriptaddress,nobalancelastpage]{revtex4-1}
\pdfoutput=1

\usepackage{color,amsmath,amssymb,amsthm,amsxtra,amsfonts,dsfont,graphicx,bm,enumerate,csquotes,tikz}

\def\Id{{\openone}}
\newcommand{\be}{\begin{equation}}
\newcommand{\ee}{\end{equation}}
\newcommand{\bea}{\begin{eqnarray}}
\newcommand{\eea}{\end{eqnarray}}

\newtheorem{thm}{Theorem}[section]
\newtheorem{prop}[thm]{Proposition}
\newtheorem{cor}[thm]{Corollary}

\newtheorem{defn}[thm]{Definition}

\newcommand{\ket}[1]{\vert#1\rangle}
\newcommand{\bra}[1]{\langle#1\vert}
\newcommand{\tr}{\mathrm{tr}}

\newcommand{\SO}{\mathrm{SO}}
\newcommand{\SU}{\mathrm{SU}}

\newcommand{\mat}[2]{\left(\begin{array}{#1}#2\end{array}\right)}

\protected\def\ignorethis#1\endignorethis{}
\let\endignorethis\relax
\def\TOCstop{\addtocontents{toc}{\ignorethis}}
\def\TOCstart{\addtocontents{toc}{\endignorethis}}

\begin{document}

\title{Matrix Product States and Projected Entangled Pair States:
Concepts, Symmetries, and Theorems}

\author{J.~Ignacio \surname{Cirac}}
\affiliation{\mbox{Max-Planck-Institut f{\"{u}}r Quantenoptik,
Hans-Kopfermann-Str.\ 1, 85748 Garching, Germany}}
\affiliation{\mbox{Munich Center for Quantum Science and Technology,
Schellingstra\ss{}e~4, 80799 M\"unchen, Germany}}

\author{David Perez-Garcia}
\affiliation{\mbox{Departamento de An\'{a}lisis Matem\'{a}tico, Universidad
Complutense de Madrid,
Plaza de Ciencias 3, 28040 Madrid, Spain}}
\affiliation{\mbox{ICMAT, Nicolas Cabrera, Campus de Cantoblanco, 28049
Madrid, Spain}}

\author{Norbert Schuch}
\affiliation{\mbox{Max-Planck-Institut f{\"{u}}r Quantenoptik,
Hans-Kopfermann-Str.\ 1, 85748 Garching, Germany}}
\affiliation{\mbox{Munich Center for Quantum Science and Technology,
Schellingstra\ss{}e~4, 80799 M\"unchen, Germany}}
\affiliation{\mbox{University of Vienna, Faculty of Physics,  Boltzmanngasse
5, 1090 Wien, Austria}}
\affiliation{\mbox{University of Vienna, Faculty of Mathematics, Oskar-Morgenstern-Platz 1, 1090 Wien, Austria}}

\author{Frank Verstraete}
\affiliation{\mbox{Department of Physics and Astronomy,
Ghent University, Krijgslaan 281, S9, 9000 Gent, Belgium}}

\begin{abstract}
The theory of entanglement provides a fundamentally new language for describing interactions and correlations in many body systems. Its vocabulary consists of qubits and entangled pairs, and the syntax is provided by tensor networks. We review how matrix product states and projected entangled pair states describe many-body wavefunctions in terms of local tensors. These tensors express how the entanglement is routed, act as a novel type of non-local order parameter, and we describe how their symmetries are reflections of the global entanglement patterns in the full system. We will discuss how tensor networks enable the construction of real-space renormalization group flows and fixed points, and examine the entanglement structure of states exhibiting topological quantum order. Finally, we provide a summary of the mathematical results of matrix product states and projected entangled pair states, highlighting the  fundamental theorem of matrix product vectors and its applications.
\end{abstract}

\maketitle

\tableofcontents

\section{Introduction}
\subsection{Setting}

The many body problem has without a doubt been the central problem in physics during the last 150 years. Starting with the discovery of statistical physics, it was realized that systems with symmetries and many constituents exhibit phase transitions, and that those phase transitions are mathematically described by non-analyticities in thermodynamic quantities when taking the limit of the system size to infinity. Quantum mechanics added a new level of complexity to the many body problem due to the non-commutativity of the different terms in the Hamiltonian, but since the discovery of path integrals it has been realized that the equilibrium quantum many body problem in $d$ dimensions can be very similar to the classical many body problem in $d+1$ dimensions. The (quantum) many body problem has been the main driving force in theoretical physics during the last century, and led to a comprehensive framework for describing phase transitions in terms of (effective) field theories and the renormalization group. The central challenge of the many body problem is to be able to predict the phase diagram for physical classes of microscopic Hamiltonians and to predict the associated relevant thermodynamic quantities, order parameters and excitation spectra. A further challenge is to predict the associated non-equilibrium behaviour in terms of quantities such as the structure factors, transport coefficients and thermalization rates.

The main difficulty in many body physics stems from the tensor product structure of the underlying phase space: the number of degrees of freedom scales exponentially in the number of constituents and/or size of the system. A central goal in theoretical physics is to find effective compressed representations of the relevant partition functions or wavefunctions in such a way that all thermodynamic quantities such as energy, magnetization, entropy, etc.\ can efficiently be extracted from that description. A particularly powerful method to achieve this has since long been perturbative quantum field theory: the many body problem is readily solvable for interaction-free systems in terms of Gaussians, and information about the interactive system can then be obtained by perturbing around the best free approximation of the system. This approach works very well for weakly interacting systems, but can obviously break down when the system undergoes a phase transition driven by the interactions. The method of choice for describing phase transitions is the renormalization group introduced by Wilson~\cite{wilson:RG}: here one makes an inspired guess of an effective field theory describing the system of interest, and then performs a renormalization group flow into the space of actions or Hamiltonians by integrating out the high energy degrees of freedom. In the case of gapped systems, such a procedure leads to a fixed point structure described by topological quantum field theory. The full power of this method is revealed when applied to gapless critical systems, where it is able to predict universal information such as the possible critical exponents at  phase transitions. However, the renormalization group  is not well suited for predicting the quantitative information needed for simulating a given microscopic Hamiltonian, and has severe limitations in the strong coupling regime where it is not clear how to integrate out the high energy degrees of freedom without getting a proliferation of unwanted terms. To address those shortcomings, a wide variety of exact and computational methods have been devised.

In the case of 2-dimensional classical spin systems and 1-dimensional quantum spin systems and field theories, major insights into the interacting many body problem have been obtained by the discovery of integrable systems. Integrable systems have an extensive amount of quasi-local conservation laws, and the Bethe ansatz exploits this to construct classes of wavefunctions which exactly diagonalize the corresponding Hamiltonians or transfer matrices. The solution of those integrable systems was of crucial importance. On the one hand, it showcased the inadequacy of Landau's theory of phase transitions for interacting systems and motivated the development of the renormalization group. On the other hand, it showed that the collective behaviour of many bodies such as spinons exhibits intriguing emergent phenomena of a completely different nature as compared to the underlying microscopic degrees of freedom. When perturbing Bethe ansatz solutions and moving to higher dimensions, it is a priori not clear how much of the underlying structure survives. There are very strong similarities between the Bethe ansatz and tensor networks, and tensor networks can in essence be interpreted as a systematic way of extending that framework to generic non-integrable systems.

Computational methods have also played a crucial role in unravelling fascinating aspects of the many body problem. Results of exact diagonalization assisted by finite size scaling results originating from conformal field theory have allowed simulations of a wide variety of spin systems. However, the infamous exponential wall is prohibitive in scaling up those calculations to reasonably sized systems for all but the simplest systems, especially in higher dimensions. A scalable computational method for classical equilibrium problems is given by Monte Carlo sampling: experience has taught that typical relevant Gibbs states have very special properties which allow to set up rapidly converging Markov chains to simulate a variety of local thermodynamic quantities of those systems. This is also  possible for quantum systems provided that the associated path integral does not have the so-called sign problem. Unfortunately, this sign problem shows up in many systems of interest, especially in the context of frustrated magnets and systems with fermions. A powerful and scalable solution to overcome this problem is to assort to the variational method: here the goal is to defined a low-dimensional manifold in the exponentially large Hilbert space such that the relevant states of the system of interest are well approximated by states in that manifold.

The most well known variational class of wavefunctions is given by the class of Slater determinants, and the corresponding variational method is called Hartree-Fock. This method works excessively well for weakly interacting systems, and perturbation theory around the extrema can be done in terms of Feynman diagrams or by coupled cluster theory. Dynamical information can also be obtained by invoking the time dependent variational principle, which can be understood as a least squares projection of the full Hamiltonian evolution on the variational manifold of Slater determinants. Alternatively, the Hartree Fock method can be rephrased as a mean field theory, and dynamical mean field theory (DMFT) extends it by modeling the interaction of a cluster with the rest of the system  as a set of self-consistent equations of the cluster and a free bath. Although this approach works well in 3 dimensions, it is not clear how generally applicable it is to lower dimensional systems. One of the main difficulties for variational methods based on free systems is the fact that the natural basis for free systems is the momentum basis: plane waves diagonalize free Hamiltonians, but the natural basis for systems with strong interactions is the position basis and a phase transition separates both regimes.

This brings us to the concept of tensor networks: tensor networks are a variational class of wavefunctions which allows to model ground states of strongly interacting systems in position space in a systematic way. Just as in the case of (post) Hartree Fock methods, the starting point is a low-dimensional variational class in the exponentially large Hilbert space. This manifold seems to capture a very rich variety of quantum many body states which are ground states of local quantum Hamiltonians, both for the case of spins, bosons and fermions. The defining character of states which can be represented as tensor networks is the fact that they exhibit an area law for the entanglement entropy. Time dependent information can be obtained by applying the time dependent variational principle on the manifold of tensor networks, and spectral information is obtained by projecting the full many body Hamiltonian on tangent spaces of the manifold. The tensor network description can be understood as a compression of the Euclidean path integral as used in quantum Monte Carlo, but then without a sign problem. Both the coordinate and algebraic Bethe ansatz can be reformulated in terms of tensor networks, and tensor networks allow for a systematic exploration of those methods beyond the integrable regime and 1+1 dimensions. Critical properties can be extracted in terms of finite entanglement scaling arguments, and tensor networks allow for a natural formulation of a real-space renormalization group procedure as originally envisioned by Kadanoff. It also turns out that tensor networks provide representations for ground states of a wide class of Hamiltonians exhibiting topological order, hence being many body realizations of Topological Quantum Field Theories (TQFTs), and provide a natural language for describing the corresponding elementary excitations (anyons) and braiding properties by providing explicit representations of associated tensor fusion algebras.

Tensor networks can hence be understood as a symbiosis of a wide variety of theoretical and computational many body techniques. From our point of view, the most interesting aspect of it is that it imposes a new way of looking at quantum many body systems: tensor networks elucidate the need of describing interacting quantum many body systems in terms of the associated entanglement degrees of freedom, and the essence of classifying phases of matter and understanding their essential differences is encoded in the different symmetries of the tensors which realize that entanglement structure. In many ways, tensor networks provide a constructive implementation of the following vision of \citet{Feynman:variational}:

\begin{displayquote}
"Now in field theory, what's going on over here and what's going on over there and all over space is more or less the same. What do we have to keep track in our functional of all things going on over there while we are looking at the things that are going on over here? $\ldots$ It’s really quite insane actually: we are trying to find the energy by taking the expectation of an operator which is located here and we present ourselves with a functional which is dependent on everything all over the map. That’s something wrong. Maybe there is some way to surround the object, or the region where we want to calculate things, by a surface and describe what things are coming in across the surface. It tells us everything that’s going on outside  $\ldots$   I think it should be possible some day to describe field theory in some other way than with wave functions and amplitudes. It might be something like the density matrices where you concentrate on quantities in a given locality and in order to start to talk about it you don't immediately have to talk about what's going on everywhere else."
\end{displayquote}

Tensor networks precisely associate a tensor product structure to interfaces between different regions in space, and the fact that such an interface is always of a dimension smaller than the original space is a manifestation of the famous area law for the entanglement entropy. In the case of 1-dimensional quantum spin chains and quantum field theories, this \emph{virtual} Hilbert space is 0-dimensional, and the different symmetry protected phases of matter (SPT) can be understood in terms of inequivalent ways in which the symmetries act on that Hilbert space. For 2-dimensional systems, the interface is 1-dimensional and provides an explicit local representation for both the entanglement Hamiltonian and the edge modes as appearing in topological phases of matter.

The central goal of this review is to explain how tensor networks describe many body systems from this entanglement point of view, and why it is reasonable to do so. A recurring theme will be that the manifold of tensor network states parameterizes a wide class of ground states of strongly interacting systems, and that all the relevant global information of the wavefunction is encoded in a single local tensor which connects the physical degrees of freedom to the virtual ones (that is the entanglement degrees of freedom). This review does not touch upon variational algorithms for optimizing tensor networks, as those topics have been covered, e.g., by \citet{murg:peps-review,schollwoeck:review-annphys,orus:tn-review,bridgeman:handwaving,haegeman:medley}. For further reading on the more traditional approaches to the quantum many-body problem, as discussed in the previous paragraphs, we refer to the books by \citet{fradkin2013field,wen:book,shavitt_bartlett_2009,becca_sorella_2017,girvin2019modern,chaikin1995principles,anderson2018basic,avella2011strongly}.

\subsection{History}

Let us start with a review of the historic development of the field of tensor networks; this will be complemented by an outlook on ongoing developments and newly evolving directions in Section~\ref{sec:5-outlook}.

\citet{nishino:history} has traced back the history of tensor networks to the works of \citet{kramers1941statistics}.  These authors were studying the 2D classical Ising model, and introduced the concept of transfer matrices (which are nothing but matrix product operators in the language of this review)  and a variational method for finding the leading eigenvector of it by optimizing over a class of wavefunctions which can be interpreted as precursors of matrix product states (MPS). Much later, \citet{baxter1968dimers,baxter1981corner,baxter:book} introduced the formalism of Corner Transfer Matrices, and realized that the concept of matrix product states allowed to make perturbative calculations of thermodynamic quantities of classical spin systems to very high order; to prove his point, he calculated the hard square entropy constant to 42 digits of precision. \citet{accardi1981topics} introduced matrix product states in the realm of quantum mechanics by describing the wavefunctions associated to quantum Markov chains. The most famous matrix product state was introduced by \citet*[][the \emph{AKLT state}]{affleck:aklt-prl}, in an effort to provide evidence for the Haldane conjecture concerning half-integer  vs. integer spin Heisenberg models. They also wrote down a 2-dimensional analogue of the AKLT state \cite{affleck:aklt-cmp}, and provided evidence that it was the ground state of a gapped parent Hamiltonian. Fannes, Nachtergaele and Werner realized that the 1D AKLT state was part of a much larger class of many body states, and they introduced the class of finitely correlated states (FCS)  which corresponds to injective matrix product states. In a series of groundbreaking papers, they proved that all FCS are unique ground states of local gapped parent Hamiltonians, and derived a wealth of interesting properties by exploiting the connection of MPS to quantum Markov chains \cite{fannes:FCS,fannes1989exact,fannes1994finitely,fannes1991valence,fannes1992abundance,fannes1996quantum}.

Independent of this work in mathematical physics, \citet{white:DMRG,white1993density} discovered a powerful algorithm for simulating quantum spin chains, which he called the density matrix renormalization group (DMRG). DMRG revolutionized the way quantum spin chains can be simulated, and provided extremely accurate results for associated ground and excited state energies and order parameters. \citet{nishino1996corner}  soon discovered intriguing parallels between DMRG and the corner transfer matrix method of \citet{baxter1981corner}. Although it was certainly not envisioned and formulated like that, it turns out that DMRG is a variational algorithm in the set of matrix product states \cite{ostlund-rommer,dukelsky1998equivalence,verstraete:dmrg-mps}. The reason for the success of DMRG was  only understood much later, when it became clear that ground states of local gapped Hamiltonians exhibit an area law for the entanglement entropy \cite{hastings:arealaw}, and that all states exhibiting such an area law can faithfully and efficiently be represented as matrix product states \cite{verstraete:faithfully}.  The family of matrix product states was rediscovered multiple times in the community of quantum information theory. First, Vidal devised an efficient algorithm for simulating a quantum computation which produces at most a constant amount of entanglement \cite{vidal:simulation-of-comput}; it turns out that the same algorithm can be reinterpreted as a time-dependent version of DMRG, thereby opening up a whole new set of applications for DMRG \cite{vidal:iTEBD,daley2004time,white2004real,verstraete:finite-t-mps}.

From the point of view of entanglement theory, matrix product states were rediscovered in the context of quantum repeaters, where it was understood that degeneracies in the entanglement spectrum such as occurring in the AKLT model lead to novel length scales in quantum spin systems as quantified by the localizable entanglement \cite{verstraete2004entanglement,verstraete:div-ent-length}. A fundamental structure theorem for matrix product states \cite{perez-garcia:stringorder-1d,cirac:mpdo-rgfp,molnar:normal-peps-fundamentalthm} has clarified such degeneracies as being the consequence of the presence of projective representations in the way the entanglement degrees of freedom transform under physical symmetries, and this has led to the classification of all possible symmetry protected topological phases (SPT) for 1D quantum spin systems \cite{pollmann:symprot-1d,chen:1d-phases-rg,schuch:mps-phases}.

Soon after the study of localizable entanglement in matrix product states
in 2003, a two-dimensional version of MPS was introduced, and  it was
realized that the entanglement degrees of freedom can play a fundamental
role by demonstrating that measurement based quantum computation
\cite{raussendorf:cluster-short} proceeds by effectively implementing a standard quantum circuit on those virtual degrees of freedom \cite{verstraete:mbc-peps}. Those states were subsequently called projected entangled pair states (PEPS), and it was quickly understood that the corresponding  variational class provides the natural generalization of MPS to 2 dimensions in the sense that they parameterize states exhibiting an area law and that there is a systematic way of increasing the bond dimension, i.e., the number of variational parameters \cite{verstraete:2D-dmrg}. Subclasses of PEPS states were considered before: the 2D AKLT state was studied by \citet{affleck:aklt-cmp}; 
Richter and Werner \cite{richter:phd-contour-correlated-states} introduced and studied a 2D generalization of FCS based on isometric tensors;
\citet*{niggemann1997quantum} studied a 2D PEPS where the tensors satisfied the Yang-Baxter equation; 
\citet{sierra1998density}, and Nishino and collaborators \cite{nishino:tps,Nishino2:tps} introduced a generalization of MPS to 2D where the tensors could be interpreted as Boltzmann weights of a vertex model.  

Just as in the 1D case, entanglement theory was the key in  formulating this ansatz in full generality. This led to the introduction of variational matrix product state algorithms for optimizing PEPS \cite{verstraete:2D-dmrg} and infinite versions of it \cite{jordan:iPEPS}.  It was found that PEPS form a very rich class of wavefunctions, and a plethora of interesting quantum spin liquid states were written in terms of PEPS tensors: the resonating valence bond states (RVB) of Anderson, the toric code state of Kitaev, and any ground state of a local frustration-free commuting quantum Hamiltonian \cite{verstraete:comp-power-of-peps} such as any ground state of stabilizer Hamiltonians \cite{verstraete:mbc-peps} or string nets \cite{buerschaper:stringnet-peps}. In \cite{gu:TERG,gu2009tensor}, it was realized that the local symmetries of the tensors are of primordial importance for describing long range topological order, and in \cite{schuch:mps-phases} this was formalized by the crucial concept of G-injectivity and later by the more general concept of matrix product operator (MPO)-injectivity \cite{sahinoglu:mpo-injectivity,bultinck:mpo-anyons}. This opened the way of simulating systems exhibiting topological quantum order and the associated anyons in terms of tensor networks. Similarly, tensor networks and the associated local symmetries turned out to provide a very natural language for describing SPT phases in 2D \cite{chen:spt-order-and-cohomology,buerschaper:twisted-injectivity,williamson:mpo-spt}.

In a different development, \citet{vidal2007entanglement,vidal:mera} discovered the multiscale entanglement renormalization ansatz (MERA). This generalizes tree tensor networks (TTNs), yet another type of tensor network that naturally arises in the context of real space renormalization \cite{fannes1992ground,shi2006classical,murg2010simtinulag,silvi2010homogeneous}. Unlike MPS, MERA and TTNs are meant to describe scale invariant wavefunctions, and capture the scale invariance exhibited in conformally invariant theories by a real space construction of scale invariant tensors. The full richness of MERA is only starting to be explored, but extremely intriguing connections with e.g. AdS/CFT and operator product expansions \cite{pfeifer:conformal-MERA, evenbly2016local} have been uncovered. We will only discuss a few limited aspects of MERA here in the context of the holographic principle and renormalization, and refer to \citet{evenbly2014algorithms} for a full review.
Finally, let us remark that ideas from renormalization have also led to a
range of renormalization-based algorithms for tensor network contraction,
such as the Tensor Renormalization Group (TRG), the Tensor Entanglement
Renormalization Group (TERG), or Tensor Network Renormalization (TNR)
\cite{xie2012coarse,xie2009second,zhao2010renormalization,jiang2008accurate,gu:TERG,PhysRevB.95.045117,Levin_2007,evenbly:tensor-renormalization}.

\subsection{Outline}

There already exists an extensive literature on the application of the different kind of tensor networks (TN) to quantum many-body systems. While there are many reviews on that topic \cite{hallberg2006new,schollwoeck:rmp,schollwoeck:review-annphys,bridgeman:handwaving,orus:tn-review,murg:peps-review,cirac2009renormalization,biamonte2019lectures} the vast majority focus on the practical aspects of tensor networks; in particular, on how to use them in numerical computations in order to approximate ground, thermal equilibrium or dynamical states corresponding to Hamiltonians defined on lattices. However, as it has been emphasized above, tensor networks have also been key to the description or even discovery of a wide variety of physical phenomena, as well as to construct simple examples displaying intriguing properties. This has been achieved through the development of a theory of tensor networks.  The present paper aims at reviewing such a theory, including both core  results and their applications.

We will concentrate on translational invariant systems in 1- and 2-dimensional lattices, since most of the analytical results have been obtained for such systems. We notice, however, that many of the results covered in this review extend naturally to higher spatial dimensions, other lattice geometries, and also to the non-translational invariant case. This restriction, in the context of TN, implies that a single tensor, $A$, encapsulates the physical properties of the many body system. As we will see, quantum states as well as operators (eg, characterizing mixed states, Hamiltonians, or dynamics) are constructed in terms of such tensors. For states (operators), the restriction to translationally invariant systems also implies that we will focus this review on MPS (MPOs) and PEPS (PEPOs).  Basic questions we will address are: Is this construction unique -- that is, can two tensors give rise to the same state or operator? If they do, what is the relation between those tensors? How are the physical properties of the states encoded in the tensor? For instance, how do the local symmetries of the tensors reflect local and/or global symmetries, or topological order? Or vice-versa, how are the symmetries in the tensor reflected in the physical properties of the many-body state, or in the dynamics it describes? There are many other questions that have been resolved in the last years about tensor networks, and it would be impossible to cover them in detail in this review. We will nevertheless go over most of them, and give the original references where they can be found. Apart from that, the reader may also want to consult \citet{zeng2019quantum} which complements in many aspects this review.

This review is organized in four sections and an appendix. Section II motivates the use of TN to describe quantum many-body systems, introduces different TN, and analyzes some of the most relevant properties. The basic structure of TN stems from the entanglement structure of the ground states of many-body Hamiltonians fulfilling an area law, which basically dictates that they exhibit very little entanglement in comparison to typical states. Tensor networks provide us with efficient ways of describing systems with small amounts of entanglement, and they are thus ideally suited for parameterizing states satisfying an area law. We will introduce the basic notions of MPS for 1D systems, and their generalization to higher dimensions, PEPS. We also consider the fermionic versions of those TN states, where the physical systems are fermionic modes. While most of the review concerns pure states, we include some analysis for mixed states and evolution operators, and for that purpose we also introduce matrix product operators (MPOs) and 
projected entangled pair operators (PEPOs). Even though we focus our attention to translational invariant systems, we will briefly mention some connections between MPS and MERA, as they both can be viewed as being created by special quantum circuits. We also argue that  MPS and PEPS not only approximate ground states of local Hamiltonians, but for any of them one can find a special (set of) Hamiltonian(s), the so-called parent Hamiltonian, which is frustration-free and for which they are the exact ground states. In particular, we list the conditions under which the Hamiltonian is degenerate, and also discuss how to describe low-energy excitations. Next we discuss an intriguing property of PEPS, namely that one can explicitly build a bulk-boundary correspondence with them. That is, for any region of space it is always possible to define a state which encodes all the physical properties of the first, but \emph{lives} in a smaller spatial dimension. This is a version of the  holographic principle and enables the use of dimensional reduction, meaning that one can fully characterize the properties of PEPS by a theory that is defined in the boundary. We finish this section by introducing a very powerful technique in the TN context, namely renormalization. The basic idea is to block tensors into others that can be assigned to blocks of spins, in much the same way as real space renormalization is used in statistical physics. The fixed point of such a procedure gives rise to very special TN, that can be viewed as the ones that appear if one looks at big scales. They have a simple form, so that one can very easily deal with them and apply them, for instance, to the classification of phases of gapped Hamiltonians. This procedure can be applied to pure or mixed states, as well as unitary operators. 

Section III analyzes how the symmetries of the tensor generating a MPS or PEPS can be associated to the symmetries of the states they generate, or to their topological order. This statement leads to one of the most celebrated successes of tensor networks, namely the classification of phases by relating those to the representations of the symmetries of the tensors generating them. In the case of global symmetries, this leads to symmetry protected phases (SPT), whereas topological phases are characterized by purely virtual symmetries. The combination of those results can also be used to characterize symmetry enriched phases (SET) for TN. Attending to the (global) symmetries of the states with a certain symmetry group, the generating tensor also possess that symmetry, with the same symmetry group but with a representation that is possibly projective. This is why the classification of SPT phases are intimately related to the corresponding cohomology classes. Topological order, however, is related to purely \emph{virtual} symmetries of the tensor, and we will discuss how those virtual symmetries give rise to notions such as topological entanglement entropy and anyons. We also consider local gauge symmetries, and ways of gauging a global into a local symmetry within the language of TNs. 

Section IV is more mathematical and contains a review of the basic theorems of MPS and PEPS. Of particular importance is the so-called \emph{Fundamental Theorem}, which lists the conditions under which two tensors generate the same state. This theorem is widely used in many of the analytical results obtained  for TN, such as in the characterization of the fixed points of the renormalization procedure of Section II, and  in the classification of symmetries and phases of Section III. It implies that the same states can be generated by many tensors, so it is very useful to find a canonical form, namely a specific property of the tensor we can demand such that it is basically uniquely associated to the state. While this is possible for MPS, and a full theory for such a canonical form and fundamental theorem exists, the situation for PEPS is not yet complete, and we discuss the state of the art. 

Finally, we collect a number of prototypical examples of MPS and PEPS appearing in the context of quantum information and/or condensed matter theory in the appendix.

\section{Many-body quantum systems: entanglement and tensor networks}

\subsection{Entanglement structure in quantum Hamiltonians}

A central feature of many-body quantum systems is the fact that the dimension of the associated Hilbert space scales exponentially large in the number of modes or particles in the system. The natural way of describing materials, atomic gases or quantum field theories exhibiting strong quantum correlations is to discretize the continuous Hilbert space by defining a lattice and an associated tensor product structure for the modes which represent localized orbitals such as Wannier modes.   Such systems can therefore be described in terms of an effective Hamiltonian acting on a tensor product of these local modes. In the case of bosons, one can typically restrict the local occupation number to be bounded (let's say $d$-dimensional), such that we get a Hilbert space of the form $\otimes_{k=1}^N C^d$. In the case of fermions, the tensor product has to be altered to a graded tensor product.

In this review, we will mainly consider local translationally invariant quantum spin Hamiltonians defined on a lattice with the geometry of a ring or torus  of the form
 \[
H=\sum_{i=1}^N h_{i,n}\ ,
 \]
where $h_{i,n}$ is a \emph{local} observable centered at site $i$, and acting nontrivially only on the $n-1$ closest sites of $i$. As an example, a nearest neighbor Hamiltonian such as the Heisenberg model has $n=2$. As $n$ is finite, it is always possible to block several sites together such that $h_{i,n}$ is acting only on next nearest neighbors according to the underlying lattice. We will be mostly interested in the ground state   and the lowest energy excitations of such a Hamiltonian in the thermodynamic limit  $N\to \infty$. 

The  gap, $\Delta$, plays an important role in such spin systems. It measures the energy difference of the first excited state and the ground state. If it vanishes in the thermodynamic limit $N\rightarrow\infty$ we say that the Hamiltonian is gapless and otherwise gapped. The first case occurs for critical systems, wheres the latter implies the existence of a finite correlation length.

Just as in quantum field theories, the central object of interest in strongly correlated quantum spin systems is the ground state or vacuum, as the quantum features are most pronounced at low temperatures. The vacuum quantum fluctuations hold the key in unraveling the low temperature properties of the material of interest, and the structure of the ground state wavefunction dictates the features of the elementary excitations or particles which can be observed in experiments. Determining the smallest eigenvector of an exponentially large matrix is in principle an intractable problem. Even a relatively small system, such as a 2D Hubbard model with $12\times 12$ sites, has a Hilbert space of dimension $2^{288}\approx 5\times 10^{86}$, which is much larger than the number of baryons in the universe, and hence writing down the ground state wavefunction as a vector is an impossible feat.  The key which allows us to circumvent this impasse is to realize that the matrices corresponding to Hamiltonians of quantum spin systems are very sparse due to the fact that they exhibit a tensor product structure and are defined as a sum of local terms with respect to this tensor product. This will force the ground state to have a very special structure, and tensor networks are precisely constructed to take advantage of that structure. Equally importantly, the locality of the Hamiltonian forces the other eigenvectors with low energy to be simple local perturbations of the ground state \cite{haegeman:mps-ansatz-excitations}, and this feature is responsible for the existence of localized elementary excitations which we observe as particles, and hence for the fact that the ground state is such a relevant object even if the system under consideration is not at zero temperature.  This has to be contrasted to a generic eigenvalue problem where knowledge of the extremal eigenvector does not give any information about the other eigenvectors except for the fact that they are orthogonal to it. Without locality, physics would be wild.

The locality and tensor product properties of the Hamiltonians from which we want to determine the extremal eigenvectors are clearly the keys to unravelling the structure of the corresponding wavefunctions. This tensor product structure and locality also play the central role in the field of quantum information \cite{nielsen-chuang} and entanglement theory \cite{horodecki:rmp-entanglement}, whose original aim was to exploit quantum correlations to perform novel information theoretic tasks. The study of entanglement theory introduced a new way of quantifying quantum correlations in terms of elementary units of entanglement (ebits), and of describing local operations which transform states into each other. The key insight in entanglement theory has been the fact that any pure bipartite states with an equal amount of entanglement (as measured by the entanglement entropy) can be converted into each other by local quantum operations and classical communication \cite{bennett1996concentrating}. These fundamental facts of the theory of entanglement were the original inspiration for defining tensor networks: ground states of local Hamiltonians turn out to exhibit an area law for the entanglement entropy, just as entangled pairs of particles distributed among nearest-neighbours on a lattice have. There should therefore exist local operations which transform both sets of states into each other. This construction precisely gives rise to the classes of matrix product states (MPS) and projected entangled pair states (PEPS), which are the main characters of this review.

\subsubsection{Local Reduced Density Matrices}

The energy of a wavefunction with respect to a local Hamiltonian is completely determined by its marginal or local reduced density matrices $\rho_{i,n}$ defined as the density matrix obtained by tracing out all degrees of freedom outside of the region $n$ around site $i$:  $E=\sum_i\mathrm{ Tr}\big[h_{i,n}\rho_{i,n}\big]$. In the case of a translationally invariant Hamiltonian and a unique ground state, the ground state inherits all symmetries of the Hamiltonian, including the translational invariance, and hence we can drop the dependence on $i$ and the ground state energy 
${\rm Tr}\left[h_n\rho_n\right]$
is a linear functional in the reduced density matrix $\rho_n$.
The question of finding ground states is hence equivalent to finding a many-body state whose marginal is extremal with respect to $h_n$. The set of all possible marginals of translationally invariant quantum many-body states is  convex. Any state whose marginal is an extreme point in this convex  set  must hence be the ground state of a local Hamiltonian defined by the tangent plane on that convex set.  The ground state problem is therefore equivalent to characterizing the set of all possible extremal points of local reduced density matrices. The problem would hence easily be solved if such a characterization were possible, but  this problem is known as the $N$-representability problem \cite{coleman1972necessary} and is well known to be intractable for generic systems \cite{liu:n-rep-qma}.

The important message however is that ground states are very special: they have extremal local reduced density matrices, and all the global features such as correlation length, possible topological order, and types of elementary excitations, follow from this local extremality condition. In other words: these global features emerge from the requirement that the local reduced density matrix is an extreme point in the set of all possible reduced density matrices compatible with the symmetries of the system. It will turn out that these extremal points can only correspond to states with very little entanglement, and all of them satisfy an area law for the entanglement entropy \cite{verstraete:faithfully,zauner2016symmetry}.

It is instructive to consider the example of the Heisenberg spin $\tfrac12$ antiferromagnetic Hamiltonian $\sum_{\langle i,j\rangle}\vec{S}_i\vec{S}_j$ where the sum is restricted to nearest neighbors, and the $\vec{S}=(S_x,S_y,S_z)$ are the standard spin operators.
If we only consider 2 sites, then the ground state is obviously equal to the spin singlet, which is maximally entangled, and with associated energy $-1$. The case of a chain of $N$ sites is much  more complicated: due to the non-commutativity, it is impossible to find a state whose ground state energy is equal to $-1$ per interaction term. This non-commutativity leads to ``frustration'': the closer the reduced density matrix of e.g.\ sites 1 and 2 is to a singlet, the further it will have to  be from a singlet for the reduced density matrices of sites 2 and 3.  This effect can also be understood in terms of the monogamy property of entanglement \cite{terhal2004entanglement}: a spin $\tfrac12$ has only the capacity of $1$ ebit of entanglement (an ebit being defined as the amount of entanglement in a maximally entangled state of two spin $\tfrac12$ systems~\cite{nielsen-chuang}), and if it has to share this 1 ebit with its neighbors, the corresponding reduced density matrices will have at most $\tfrac12$ ebit of entanglement. The more neighbors a spin has, the less entanglement it can share with each individual one. This can be formalized in the  quantum de Finetti theorem and is the reason that mean field theory becomes exact in high dimensionsal lattices \cite{raggio1989quantum,brandao:mean-field-approx}. This is also the reason that one and 2D systems exhibit some of the most interesting quantum effects: in general, the marginals of quantum many-body states in 3D lattices are already well approximated by the ones obtained by product or mean field solutions, while this is not the case for low dimensional systems.

The physics of ground states is  completely determined by the competition between translational invariance and extremal local reduced density matrices (for the case of the Heisenberg model, the density matrices will be as close as possible to the singlet).  Monogamy of entanglement is precisely the property which gives rise to interesting physics: in the case of classical statistical mechanics, the competition of energy versus entropy gives rise to cooperative phenomena and phase transitions. In the quantum case, the non-commutativity of the different terms in the Hamiltonian leads to monogamy, which plays a similar role and makes such phase transitions possible at zero temperature.

The key to uncover the structure of ground states of local Hamiltonians is to understand how the entanglement is shared between the different degrees of freedom. Intuitively, for a given spin 
it is of no use to have strong correlations with far away spins, as this will only bring marginals further away from the extremal points. The strongest (quantum) correlations it needs to have are with those spins with which the Hamiltonian forces it to interact, namely the nearest neighbors. We can hence imagine that the entanglement between a bipartition of a big system in two regions is proportional to the surface between them, and this area law for entanglement is exactly what is going on in ground states.

In summary: ground states of local Hamiltonians of quantum spin systems are in one to one correspondence with states whose reduced density matrices are extremal points within the set of all possible reduced density matrices with a given translational symmetry. This property forces the entanglement to be localized, giving rise to an area law.

\subsubsection{Area laws for the entanglement entropy}\label{sec2arealaws}

Let us consider a quantum spin system with a local quantum Hamiltonian and ground state $|\psi\rangle$, and a bipartition of the quantum spin system  into two connected regions, $A$ and $B$, such that  $\rho_A$ and $\rho_B$ are the reduced density matrices of the ground state in these regions. The entanglement entropy \cite{bennett1996concentrating}
\be
S(\rho_A)=-{\rm Tr}[\rho_A\log(\rho_A)]=S(\rho_B)
\ee
quantifies the amount of quantum correlations between the two regions, and as argued in the last section, this quantity is expected to be proportional to the surface of the boundary between the two regions, $\partial A$, and hence called the \emph{area law} for the entanglement entropy \cite{eisert:arealaw-review}. This area law should be contrasted to the volume law exhibited by random states in the Hilbert space: quantum states exhibiting an area law for the entanglement entropy are very special; such states are hence very atypical, and it will  be possible to represent them using tensor networks.

The origins of the area law can be traced back to studies of the entanglement entropy in free quantum field theories \cite{holzhey1994geometric}, where the ensuing area laws were related to the Bekenstein-Hawking black hole entropy \cite{bekenstein1973black}. Area laws can rigorously be demonstrated for free bosonic \cite{plenio2005entropy} and fermionic systems \cite{wolf:fermion-arealaw,gioev2006entanglement}, modulo some logarithmic corrections in the presence of Fermi surfaces. They can also be proven in case that  correlations (defined in terms of the mutual information) between two arbitrary regions decay sufficiently fast with the distance independently of their sizes \cite{wolf:mutual-info-arealaw}. 

For interacting quantum systems at finite temperature $T$ and described by Gibbs states $\rho\propto \exp(-H/T)$, an area law for the mutual information
\[
I(A:B)=S(\rho_{A})+S(\rho_{B})-S(\rho)\le c\frac{|\partial A|}{T}
\]
has been proven by \citet{wolf:mutual-info-arealaw} for any local Hamiltonian in any dimension as long as all terms in the Hamiltonian are bounded from above. Here $|\partial A|$ denotes the number of spins in the boundary $\partial A$ between region $A$ and $B$. Recently, \citet{kuwahara:thermal-area-law} have improved the temperature-dependence of this bound to diverge as $T^{2/3}$.

It is much harder to prove the area law for ground states of interacting quantum spin systems, although there is plenty of evidence supporting that claim. In the case of gapped quantum spin chains in one dimension, a remarkable theorem has been formulated by 
\citet{hastings:arealaw}
proving the area law which was later strengthened by 
\citet*{arad:1d-arealaw}:
given a local Hamiltonian of a quantum spin chain of $N$ $d$-dimensional spins whose gap is  given by $\Delta$, then the entanglement entropy in the ground state is bounded above by $\mathcal{O}\left((\log d)^3/\Delta\right)$ for any bipartite cut into two connected regions (see \citet{kuwahara:1d-longrange-arealaw} for a generalization to long-range interactions). Note that this means that the entanglement entropy saturates in the thermodynamic limit for the case of a gapped system.  In the case of a critical quantum spin chain where the gap vanishes as $\mathcal{O}(1/N)$ or faster, this bound yields a volume law, though it seems that nature is much more economical and for all critical spin chains described by a conformal field or a Luttinger liquid theory, the actual entanglement entropy is exponentially smaller and scales as $\mathcal{O}(\log(L))$ for a region $A$ of length $L\le N/2$. When the gap is allowed to vanish must faster as a function of the system size, examples were constructed that saturate the volume law, and hence give rise to novel phase transitions from bounded to extensive entanglement \cite{movassagh2016supercritical,zhang2017novel}. 

Much more precise information about the nature of the entanglement in a system can be obtained by looking at the entanglement spectrum \cite{li:es-qhe-sphere}, defined as the logarithm of the set of eigenvalues of the reduced density matrix $\lambda_i(\rho_A)$. The Schmidt coefficients are the square roots of these eigenvalues, and the convention is to order them in decreasing order. In the case of gapped integrable spin chains in the thermodynamic limit, these Schmidt coefficients decay as $\exp(-\alpha n)$ with $n\in\mathbb{N}$, $\alpha$ a constant, and a degeneracy $a(n)$ equal to the number of ways to partition $n$ in sums of unequal integers \cite{peschel1999density}. Asymptotically, we have $a(n)=\mathcal{O}\left(\exp(\pi\sqrt{n/3})/n^{3/4}\right)$. This result can be obtained by calculating the eigenvalues of the corner transfer matrix, which is a discrete version of the boost operator $H_\mathrm{Mod}$ as used in quantum field theory to calculate the entanglement entropy. For critical systems described by a CFT, the largest Schmidt coefficient seems to encode the information about the full entanglement entropy \cite{orus2006half}.

Alternatively, Renyi entropies $S_\alpha(\rho)=\tfrac{1}{1-\alpha}\log\,
\mathrm{Tr}\left(\rho^\alpha\right)$ with $\alpha\ge 0$ can be used to characterize the decay of the Schmidt coefficients. These Renyi entropies are monotonically decreasing as a function of $\alpha$; $S_0(\rho)$ measures the rank of $\rho$, and the ones with $0\leq \alpha<1$ will be of particular importance for the description of matrix product states.
Improving results and techniques from \cite{hastings:arealaw} and \cite{landau:1d-efficient-algo}, Huang proved in \cite{huang:area-law} that the ground state $\alpha$-Renyi entanglement entropy in gapped 1D systems  is upper bounded by $\tilde{\mathcal{O}}(\alpha^{-3}/\Delta)$, where $\tilde{\mathcal{O}}$ stands for $\mathcal{O}$ up to logarithmically smaller factors. More specifically, it was demonstrated that the residual probability $\epsilon(D)$, defined as the sum of all eigenvalues of the reduced density matrix smaller than the $D'$th largest one, scales as $\exp\left(-c.\Delta^{1/3}\left[\log(D)\right]^{4/3}\right)$ for a general spin chain with gap $\Delta$ \cite{arad:1d-arealaw}. For integrable systems and systems in the scaling regime of a conformal field theory, a faster decay in the form of  $\epsilon(D)\simeq\exp(-c.\left[log(D)\right]^2)$ is obtained \cite{Calabrese_Lefevre,verstraete:faithfully}. 

For higher dimensional quantum spin systems, no general proofs of an area law for ground states exist. It is believed that: (i) Gapped systems always exhibit an area law for the entanglement entropy; (ii) Critical systems without a Fermi surface also satisfy an area law, but get additive logarithmic corrections; (iii) Critical systems with a Fermi surface exhibit an entanglement entropy scaling as $|\partial A|\log |\partial A|$, which is marginally larger than the area law scaling.

In two dimensions, additive corrections also pop up for systems exhibiting topological quantum order. For a region with a perimeter $L$ and ignoring corner effects, the entanglement entropy scales like $cL-\log(\mathcal{D})$ with $\mathcal{D}$ the total quantum dimension of the underlying anyonic theory. As this quantum dimension is always larger than $1$, topologically ordered systems have less entanglement than the ones in a trivial phase. 
This indicates that a topologically ordered system exhibits a certain symmetry which reduces  the support of the local reduced density matrix; it will turn out that such symmetries are naturally described by matrix product operators.

\subsection{Tensor networks}\label{sec2TN}

We will now define different types of states and operators that can be expressed as tensor networks, and analyze their basic properties. Although most of this review will focus on translational invariant states of spin lattices and thus MPS and PEPS will be the main actors, we will also introduce their extension to fermionic systems and make connections to other sets of states like Tree Tensor Networks (TTN) and MERA. 

\begin{figure}
\includegraphics[width=7cm]{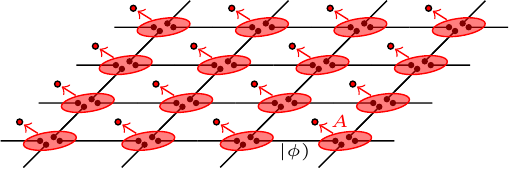}
\caption{Construction of projected entangled pair states on a 2D lattice.}
\label{sec2fig1}
\end{figure}

The discussion on local reduced density matrices made clear that ground states of local quantum Hamiltonians are completely moulded by their desire to have extremal local correlations, on the one hand, and to preserve lattice symmetries, on the other. The discussion on area laws for the entanglement entropy made clear that the entanglement between two neighboring regions is mainly concentrated on the interface between the two regions. The same entanglement pattern can be obtained by distributing maximally entangled pairs of $D$-dimensional spins between all nearest neighbors, and then doing a local projection (or a general linear map)  on all these local spins to obtain one $d$-dimensional spin. This projection involves a linear map from a Hilbert space of $N_i$ $D$-dimensional spins to a $d$-dimensional spin, with $N_i$ the coordination number of the lattice at site $i$, and any such linear map can be represented by a tensor $A^i_{\alpha_1\alpha_2\cdots \alpha_N}$. Translational invariance is obtained if the lattice has periodic boundary conditions and the same projection is chosen on all lattice sites. This construction,  illustrated in Figure~\ref{sec2fig1}, defines the class of projected entangled pair states (PEPS) with bond dimension $D$, and the different states in that family can be obtained by choosing different projections $A$. Below we will give a more precise description of PEPS, and connect them to tensor networks.

This PEPS construction yields quantum many-body states which have strong local correlations, exhibit the translational symmetry of the underlying lattice, and obey an area law for the entanglement entropy with respect to any bipartition. Furthermore, extra symmetries such as global $\mathrm U(1)$ or $\mathrm{SU}(2)$ symmetries can easily be incorporated by defining tensors which transform according to some representation of the corresponding group \cite{singh:tns-decomp-symmetry,perez-garcia:inj-peps-syms}. All those properties are highly nontrivial, and it is especially hard to write down entangled translationally invariant wavefunctions without using the projected entangled pair construction.

Conceptually, PEPS present a way of parameterizing interesting many-body wavefunctions on any lattice with a constant coordination number using a number of parameters which is independent of the system size (at least for translationally invariant states). They hence provide a way of writing down a nontrivial wavefunction in an exponentially large Hilbert space in a compressed form. Of course, this has to come as no surprise. Both product states and Slater determinants provide ways of writing down wavefunctions in an exponentially large Hilbert space using a few parameters. The main difference, however, is the fact that the PEPS construction can represent a wide variety of ground states of strongly interacting systems. Being able to represent such wavefunctions  has the potential of cracking open some of the hardest problems in many-body physics. 

In the case of 1D spin chains, this PEPS construction defines the class of matrix product states (MPS). The area law for entanglement allows to demonstrate that any ground state of a gapped quantum spin chain can be represented efficiently using such an MPS \cite{hastings:arealaw,arad:1d-arealaw}; and vice versa, that any MPS is the ground state of a local gapped Hamiltonian \cite{fannes:FCS, perez-garcia:mps-reps}. Similarly, a wide class of correlated many-body systems in higher dimensions can be represented using PEPS \cite{verstraete:comp-power-of-peps,buerschaper:stringnet-peps}, and the main topic of this review is to report on the mathematical properties of the manifolds of MPS and PEPS and the relevance for the physical properties and classification of strongly correlated systems. In a nutshell, the manifold of MPS and PEPS form very rich classes of many-body systems, and provide a unique window into the physics of strongly correlated quantum many-body systems, both from the theoretical and computational point of view.

\subsubsection{MPS and PEPS} \label{MPSandPEPS}

We will now introduce PEPS for an arbitrary lattice and then particularize the definition to one spatial dimension to obtain MPS. Let us consider a lattice with $N$ vertices, $V=\{1,2,\ldots,N\}$, and a set of edges, $E$, connecting them. We consider a spin at each vertex, with a corresponding Hilbert space $\mathbb{C}^{d_i}$ of dimension $d_i$. Our goal is to construct states of these spins, i.e. $|\psi\rangle\in \otimes_{i=1}^N \mathbb{C}^{d_i}$. 

The elements $e\in E$ are pairs of vertices; for instance, $e=(1,2)$ represents the edge connecting vertices 1 and 2. We will further denote by $S_i\subset V$ the set of vertices that are connected to the vertex $i$, i.e. $S_i=\{j\in V \text{, s.t. } (i,j)\in E\}$ with $z_i=|S_i|$ the coordination number. To construct $\ket\psi$, we first assign to each vertex $i$, several auxiliary spins (one for each edge connecting that vertex to another one) that are in a maximally entangled states with their neighbors. More explicitly, for each $i\in V$ and $j\in S_i$ we denote by $a_{i,j}$ the ancilla, which has an associated Hilbert space $\mathbb{C}^{D_{i,j}}$ with dimension $D_{i,j}=D_{j,i}\in \mathbb{N}$. 
To distinguish states of the auxiliary spin from the physical ones, we use the notation $\vert\cdot)$, as opposed to $\vert\cdot\rangle$.
The ancillas $a_{i,j}$ and $a_{j,i}$ form a maximally entangled state
\begin{equation}
    \label{eq:ancilla-state-bipartite}
|\phi)_{i,j} = \sum_{n=1}^{D_{i,j}} \vert n)_{a_{i,j}}
\otimes \vert n)_{a_{j,i}}\ ,
\end{equation}
where the $\{|n)\}$ form an orthonormal basis; note that this fixes a preferred basis, making the objects in the construction basis-dependent. Thus, the state of the ancillas is
\begin{equation}
    \label{eq:ancilla-state-Phi}
 |\Phi) = \bigotimes_{e\in E} |\phi)_e\ .
\end{equation}
Next, to each vertex $i$, we assign a linear map 
$$
A[i]: \bigotimes_{j\in S_i} \mathbb{C}^{D_{i,j}}\to \mathbb{C}^{d_{i}}\ .
$$
We define the PEPS as
 \be
 |\psi\rangle = \bigotimes_{i\in V} A[i]\: |\Phi)
 \ee
That is, the state is obtained by a linear map of the entangled pairs of ancillae into the physical spins at each vertex, cf.\ Fig.~\ref{sec2fig1}. The final state will in general be entangled, since the entanglement in the ancillae is transferred to the spins through the mapping. This entanglement can lead to long-range correlations, even though the ancillae are only entangled locally. This is a simple consequence of entanglement swapping, which allows to entangle remote particles by a sequence of projections on entangled pairs. Note that the whole state is completely determined by the maps $A[i]$: since each of them is characterized by $p_i=d_i \prod_{j\in S_i} D_{i,j}$ parameters, we just need $\sum_{i\in V} p_i$ parameters to specify the state.

The map $A[i]$ is characterized by the coefficients in a basis:
 \be\label{sec2coefA}
A^s_{\alpha_1,\ldots,\alpha_{z_i}}=
\bra{s}A[i]\vert\alpha_1,\ldots,\alpha_{z_i})
\ee
and thus, by a tensor (whose entries depend on the basis choice). We will indistinguishably call $A$ map or tensor in the following.  A concept that will play a chief role in this review is injectivity and its generalizations: we say that the tensor $A[i]$ is injective if the corresponding map is injective; that is, if there exists another map, $A[i]^{-1}:\mathbb{C}^{d_{i}} \to \bigotimes_{j\in S_i} \mathbb{C}^{D_{i,j}}$ such that $A[i]^{-1}A[i] = \openone$.

There are other equivalent ways of defining PEPS that will be used later on. One particularly interesting one consists of associating to each vertex $i\in V$ a fiducial state, $|\phi_i\rangle$, of the spin and the virtual system (i.e., $\ket{\phi_i}\in \mathbb{C}^{d_{i}} \otimes_{j\in S_i} \mathbb{C}^{D_{i,j}} $), and define the PEPS to be
 \be
 |\psi\rangle = \langle \Phi| \Big[\textstyle\bigotimes\limits_{i\in V} |\phi_i\rangle\Big]\ .
 \ee
This state coincides with the one above if we write
 \be
 |\phi_i\rangle = \sum_{s,\alpha_1,\ldots,\alpha_{z_i}} A^s_{\alpha_1,\ldots,\alpha_{z_i}} |s\rangle\otimes |\alpha_1,\ldots,\alpha_{z_i}\rangle
 \ee
and choose as $A^s_{\alpha_1,\ldots,\alpha_{z_i}}$ the elements of the map $A[i]$ in the physical ($|s\rangle$) and virtual ($|\alpha_{j}\rangle$) basis. In this case, the fiducial states $|\phi_i\rangle$ completely determine the many-body state. 
Finally, yet another equivalent definition is obtained by replacing the state $|\phi)$ of the ancillas in Eqs.~(\ref{eq:ancilla-state-bipartite},\ref{eq:ancilla-state-Phi}) by some tri- or multipartite local states -- such an ansatz is yet again equivalent to the original construction, but can be advantageous e.g.\ in the numerical simulation of frustrated spin systems~\cite{xie:pess,schuch:rvb-kagome}.

Although the above construction applies to any lattice, we will exclusively consider regular lattices, with the same coordination number and the same physical dimension at each vertex ($z_i=z$ and $d_i=d$). We will call $d$ the physical dimension. We will be particularly interested in square lattices in 2 dimensions, or in 1D lattices, where we recover MPS. In the first case, we will use the convention in the tensors (\ref{sec2coefA}) that $\alpha_1,\ldots,\alpha_4$ are taken clockwise (top, right, down, left). In the latter, it is useful to define matrices $A^{s_i}[i] \in \mathbb{C}^{D_{i-1,i}\otimes D_{i,i+1}}$ with elements  $A^{s_i}_{\alpha\beta}[i]$, and the above expression is equivalent to
\be \label{eq:overlap-MPS-product}\langle s_1,s_2,\ldots |\psi\rangle={\rm Tr}\left[A^{s_1}[1]A^{s_2}[2]\cdots A^{s_N}[N]\right]
\ .
\ee
Every probability amplitude is given by the trace of a product of $N$ matrices, hence the name Matrix Product State (MPS). 

In regular lattices, translationally invariant (also named {\it uniform}) states are obtained by choosing the same map at every site, $A[i]=A$ and thus $D_{i,j}=D$, the bond dimension. By construction, it is clear that the PEPS is invariant under translations. For any lattice size, the state is completely determined by a single map, $A$ or, equivalently, a single tensor. We will say that the tensor $A$ generates the state $|\psi(A)\rangle$. Thus, we can associate to any tensor $A$ a set of states $|\psi(A)_N\rangle$ corresponding to each lattice size. This map from a tensor to a set of states is not one-to-one, which will be the basis for many of the features of MPS and PEPS descriptions. Apart from that, note that all the physical properties (like criticality, symmetries, topological order, etc.) of the states are completely determined by $A$, and thus are somehow encoded in that tensor. A main goal of the theory of tensor networks is to obtain such properties directly from the tensor.

\begin{figure}
(a)\, \includegraphics[width=4.5cm]{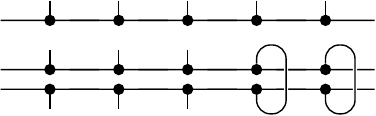}

\vspace{1cm}

(b) \, \includegraphics[width=4.5cm]{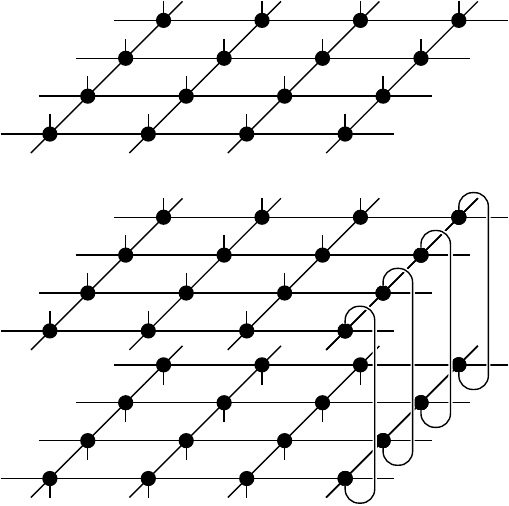}
\caption{Tensor network description of an MPS (a), a PEPS on a square lattice (b), and their corresponding marginals.}\label{sec2fig4}
\end{figure}

Instead of working with notation (\ref{sec2coefA}) and corresponding proliferation of indices, it turns out to be much more useful to work with a graphical tensor notation, and to represent MPS and PEPS as a tensor network. A tensor network consists of vertices and edges that have the same geometry as the lattice. Every vertex represents a tensor with a number of legs equal to the number of edges.  An edge with an open end represents an open index, while an edge which is sandwiched between two vertices is to be contracted and hence summed over.  For example, the tensor $\psi_{ijk}=\sum_{\alpha\beta\gamma}A^{i}_{\alpha\beta}A^{j}_{\beta\gamma}A^{k}_{\gamma\alpha}$ is represented by three vertices, three open lines, and three closed ones as
$$\label{sec2fig3}
 {\includegraphics[width=3.5cm]{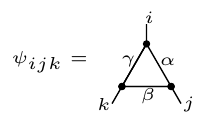}} 
$$
With this tensor network notation, we can readily represent any MPS and PEPS,  shown for a spin chain and a square lattice in Figure~\ref{sec2fig4}. The marginal or reduced density matrix of an MPS or PEPS can be obtained by summing over or contracting the physical indices. Similarly, we can represented local expectation values in the form of a tensor network contraction.

An important practical consideration is the question of the computational complexity of contracting such tensor networks. 
Generally, contracting a generic PEPS network is as hard as calculating the partition function of a spin glass and can hence be \textsf{\#P}-hard in the number of tensors \cite{verstraete:comp-power-of-peps,schuch:cplx-of-PEPS,haferkamp:cplx-of-avg-contraction}. In practice however, PEPS tensor networks have a high degree of homogeneity (e.g. translational invariance), and powerful algorithms are being developed to contract them.  Contracting tensor networks made of  matrix product states is much cheaper, as the cost of calculating any expectation value scales linearly in the number of sites and as a cube in the bond dimension ($\sim ND^3$): one can contract the tensor network starting from one end and progress to the other end while contracting all tensors along the way. This difference in complexity of contracting 1D versus higher dimensional tensor networks is responsible for the big discrepancy that currently exists in the accuracy for simulating 1D spin chains using DMRG versus 2D systems using PEPS.  It is however a very active area of research how to speed up these higher dimensional tensor contractions. For the state of the art algorithms, we refer to \onlinecite{corboz2016variational,vanderstraeten2016gradient,liao2019differentiable}.

Another important consideration is the fact that MPS and PEPS representations are not unique: as we mentioned before, two tensors may generate the same set of states. This fact will play a central role in understanding how symmetries are represented in tensor networks, in the classification of different phases of matter, and in the process of devising efficient numerical methods for dealing with tensor networks.  Let us consider the case of MPS, and define a translationally invariant MPS $|\psi(A)\rangle$ with periodic boundary conditions generated by the tensor $A$; due to the cyclic nature of the trace, it is clear that $|\psi(A)\rangle=|\psi(B)\rangle$ if $B^i$ is related to $A^i$ by a ``gauge transform'' $X$: 
\be
 \label{PullCondMPS}
 B^i=XA^iX^{-1}
 \ee
where $X$ is any invertible $D\times D$ matrix. This is a specific instance of the fundamental theorem of MPS \cite{perez-garcia:mps-reps, perez-garcia:stringorder-1d}, which will be reviewed in Section \ref{sec:4}, and that basically states that this is the only possibility as long as the tensors are expressed in some canonical form. 

Many familiar states in the context of quantum information and condensed matter theory have simple descriptions in terms of MPS and PEPS. One can also construct PEPS that are closely connected to classical Gibbs distributions: that is, for any classical spin system with short-range interactions one can build a quantum state such that the expectation values of the operators diagonal in the computational basis coincide with these of the classical distribution \cite{verstraete:comp-power-of-peps}.

\subsubsection{MPO and PEPO \label{sec:2:MPU}}

\begin{figure}[t]
(a) \, \includegraphics[width=5cm]{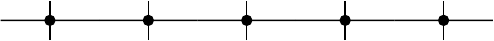}
\vspace{1cm}
(b)\, \includegraphics[width=5cm]{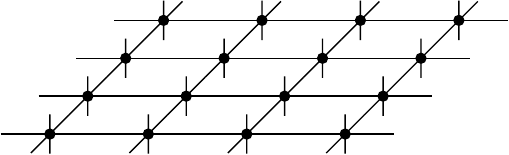}
\caption{Definition of (a) Matrix Product Operators (MPO)  and (b)  Projected Entangled Pair Operators (PEPO)}\label{sec2fig5}
\end{figure}

An important generalization of the class of MPS and PEPS are matrix product operators (MPO) and  projected entangled pair operators (PEPO). They are readily defined by the tensor network depicted in Figure~\ref{sec2fig5}. When the operators that they represent are translationally invariant, they are fully characterized by a single tensor, just as PEPS, but now with two physical indices: one corresponding to the bra and the other to the ket of the local action of the operator. Analogously to their pure state counterparts, they allow us to encode relevant many-body operators in a very economical way. In particular, matrix product operators  \cite{verstraete:finite-t-mps,zwolak:mixedstate-timeevol-mpdo} describe mixed states (like those corresponding to thermal equilibrium, or open quantum systems), Hamiltonians \cite{pirvu2010matrix,mcculloch2007density,crosswhite2008finite}, or unitary evolution \cite{cirac2017matrix,sahinoglu:mpu}.

MPO and PEPO relate to other operators appearing in the context of statistical physics. First and foremost, they appear as transfer matrices in 2- and 3D  classical statistical mechanical models, where the free energy can be inferred from its leading eigenvalue. The exact diagonalization of such transfer matrices in the former case is the main aim of the field of integrable models, and beautiful algebraic structures in integrable systems have been uncovered by invoking the Bethe ansatz and the associated Yang-Baxter relations. Similarly, MPOs are obtained as the transfer matrix in the path integral formulation of 1D quantum spin systems. They also appear in the description of cellular automata and as transfer matrices in non-equilibrium statistical physics, in the realm of percolation theory and the asymmetric exclusion process. We refer to the review \cite{haegeman:medley} for a detailed exposition of these connections. 

\begin{figure}[b]
\includegraphics[width=3.5cm]{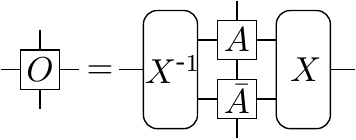}
\caption{Sufficient condition for a translationally invariant MPO $\rho$ to be positive.}\label{sec2fig6}
\end{figure}

MPOs are widely used in different scenarios in the field of tensor networks. Although all these different roles can be extended to higher dimensions using PEPOs, let us discuss them here in the context of MPOs.

Let us first focus on MPOs as density matrices, it is  mixed state analogues of a pure MPS. In this case they are called Matrix Product Density Operators (MPDO). From the computational point of view, they arise in simulations at finite temperature or in the presence of dissipation. A sufficient local condition for a MPO represented by the 4-leg tensor $O$ to be a global positive operator (in the semidefinite sense, hence representing a density matrix) is the existence of a 4-leg tensor $A$ (the "purification") and a 3-leg tensor $X$ (the "gauge transform") such that the property depicted in Figure \ref{sec2fig6} is satisfied. Note that this is only a sufficient condition for positivity. Indeed, it has been shown that there can be an arbitrary tradeoff in the bond dimension of the purification~\cite{delascuevas:mpdo-purification-tradeoff}, and that there exist translationally invariant MPDOs which do not possess MPO purifications as in Fig.~\ref{sec2fig6} valid for all system sizes~\cite{delascuevas:mpdo-purification-tinv}.

\begin{figure}[t]
(a) \includegraphics[width=2.5cm]{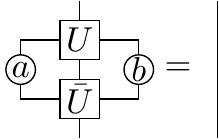}\, ,\quad
\includegraphics[width=5cm]{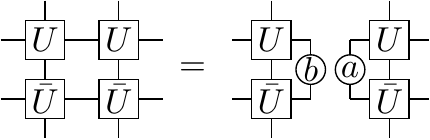}

\vspace{0.5cm}

(b) \, \includegraphics[width=5cm]{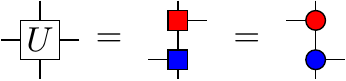}

\vspace{0.3cm}

\includegraphics[width=8cm]{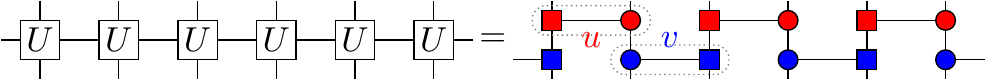}
\caption{(a) Tensors generating an MPU after blocking. (b) the MPU can be described as a quantum circuit, with alternating layers of unitary operators $u$ and $v$ acting on nearest neighbors withe even-odd indices or odd-even indices, respectively.}\label{sec2fig7b}
\end{figure}
\begin{figure}[b]
(a)\, \includegraphics[width=5cm]{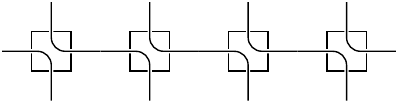}

\vspace{0.5cm}

(b) \, \includegraphics[width=5cm]{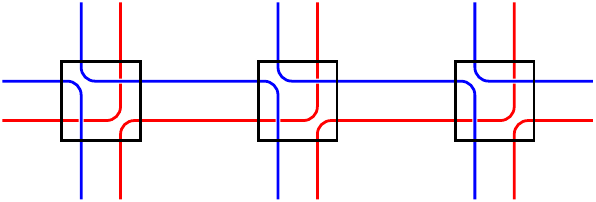}
\caption{(a) MPU representation of the left-moving shift operator; (b) MPU made of two shift operators, one right- and another left- moving.}\label{sec2fig7c}
\end{figure}

Second, MPO can also describe the dynamics of a quantum many-body systems. In that case they are called Matrix Product Unitaries (MPU) \cite{cirac2017matrix}, as they generate a unitary operator $U$ fulfilling $UU^\dagger=\Id$. As in the case of MPDOs, this extra condition imposes a restriction on the tensor generating $U$. For this case, it is possible to fully characterize them. In fact, by blocking at most $D^4$ spins, the resulting tensors have a very simple structure (Fig.~\ref{sec2fig7b}). The MPU can thus be viewed as a quantum circuit with two layers of unitary operators acting on nearest neighbors. In fact, MPUs can be shown to be equivalent to 1D quantum cellular automata; that is, unitary operators that transform local operators into local operators, where by local we mean acting non-trivially in a finite region only. That is, any MPU possesses that property and any quantum cellular automaton can be written as an MPU with finite bond dimension. Furthermore, an evolution operator generated by a local Hamiltonian in finite time can be approximated by an MPU, since the Lieb-Robinson bound for the propagation of correlations ensures that it behaves as a quantum cellular automaton, and thus as an MPU, up to some small corrections.
There are also some MPUs that cannot be approximated by a local time-evolution operator. A particular example is the shift operator sketched in Fig.~\ref{sec2fig7c}(a) (see Appendix~\ref{sec:app-examples:mpo-mpu}), which in each application translates the state by one site to the left. The fact that this operator cannot be obtained (or even approximated) by the evolution of a local Hamiltonian is a direct consequence of the index theorem, originally proven for 1D quantum cellular automata \cite{gross:index-theorem}, which states that MPUs can be classified in terms of an index, where the equivalence relation is that the tensors generating the MPU can be continuously transformed into one another. The index measures how quantum information is transported to the right (positive index) or to the left (negative index), and can only take discrete values. The dynamics generated by local Hamiltonians has zero index, whereas the one of the shift operator is $\pm 1$. In Fig. \ref{sec2fig7c}(b) we give an example of an MPU where the local Hilbert space has dimension 4 (and thus, it acts on pairs of qubits), with zero index as it moves the same information to the left as to the right.

\begin{figure}
\includegraphics[width=4.5cm]{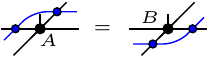}
\caption{A sufficient condition for two PEPS to be equal to each other is the existence of a MPO satisfying the Pulling Through equation }\label{sec2fig7}
\end{figure}

Third, MPOs play a fundamental role in describing symmetries of PEPS \cite{chen:2d-spt-phases-peps-ghz,sahinoglu:mpo-injectivity,bultinck:mpo-anyons}.  In particular, the generalization of Eq. \eqref{PullCondMPS} to PEPS is the pulling through equation depicted in Figure \ref{sec2fig7}. It gives a sufficient condition for two tensors to  generate the same PEPS. In the case of systems exhibiting topological quantum order, similar pulling through equations characterize the symmetries of the underlying tensors; these symmetries form an algebra and provide an explicit representation of  tensor categories describing the topological phase and its emerging anyons (see Section \ref{sec:3:2D}).

Finally, MPOs are also key in the bulk-boundary correspondence \cite{cirac:peps-boundaries}, where the physical properties of PEPS in 2 dimensions can be mapped into these of a theory defined at the boundary by an MPO. As we will show, the classification of the renormalization fixed points of MPO will thus allow us to characterize the topological order of 2D systems \cite{cirac:mpdo-rgfp}.

\subsubsection{Correlations, Entanglement, and the Transfer Matrix\label{sec2TN3}\label{sec:2:CorrsTOp}}

\begin{figure}
(a) \raisebox{-.5cm}{\includegraphics[width=6cm]{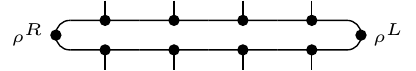}}

\vspace{0.5cm}

(b) \raisebox{-.3cm}{\includegraphics[width=6cm]{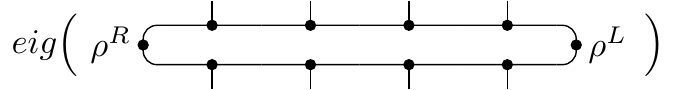}} $= $

$=$ \raisebox{-.5cm}{\includegraphics[width=6cm]{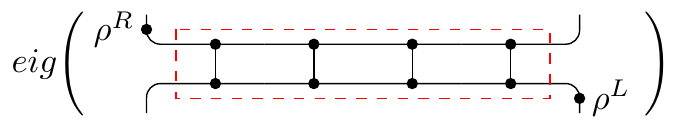}} $\simeq$

$\simeq$ \raisebox{-.5cm}{\includegraphics[width=4.5cm]{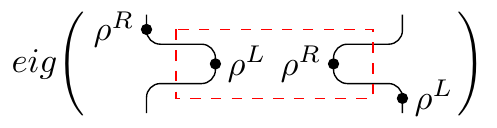}}
\caption{(a) Tensor Network description of the reduced density matrix of $n$ spins in an infinite translationally invariant MPS. $\ket{\rho^{R,L}}$ correspond to the right and left fixed points, respectively, of the transfer matrix (see main text). (b) Argument showing that the eigenvalues of the reduced density matrix of $n$ sites in a MPS coincide with those of $\left(\rho^L\rho^R\right)^{\otimes 2}$ in the limit of large $n$.}\label{sec2fig8}
\end{figure}

\paragraph{Matrix Product States}
All matrix product states satisfy an area law for the entanglement entropy. This follows directly from the projected entangled pair construction, where the MPS was obtained by applying a local map on a tensor product of $D$-dimensional maximally entangled states (see Fig.\ref{sec2fig1}). Let us take a region of contiguous spins in the chain. Before the map, it is clear that only the two pairs that are at the boundary contribute to the entanglement entropy. In fact, the rank of the reduced state in that region is equal to $D^2$. Since the map does not change the rank of the reduced state, it follows that the entropy is at most $2\log(D)$. For a generic infinite translationally invariant MPS, it is possible to calculate all eigenvalues of this reduced density matrix exactly. This density matrix can be represented by the tensor network depicted in Figure \ref{sec2fig8}. Before we show how to determine this spectrum, some basic algebraic properties of MPS have to be introduced.

A central object for a MPS is its corresponding transfer matrix $E$, defined as $E_{\alpha\alpha',\beta\beta'}=\sum_i A^i_{\alpha\beta}\bar{A}^i_{\alpha'\beta'}$. It is called the transfer matrix as it plays a role similar to the transfer matrix in classical 1D statistical mechanics models. The eigenvalues and eigenvectors of this transfer matrix are of importance, as we will see in several places throughout this review. The eigenvalues can be complex, as $E$ is not necessarily hermitian, and it may have some Jordan blocks. 
For a generic MPS, however, the largest eigenvalue in magnitude is unique and there are no Jordan blocks associated to it. We will assume this to be the case for the time being. We find it convenient to write the corresponding right and left eigenvectors $|\rho^{R,L}\rangle$ as operators $\rho^{R,L}$, so that $(\rho^{R,L})_{\alpha,\alpha'}=\langle \alpha,\alpha'|\rho^{R,L}\rangle$. The quantum Perron Frobenius theorem \cite{albeverio1978frobenius,wolf:channel-notes} then guarantees that this largest eigenvalue is positive, as the eigenvalue equation can be written in the form of a completely positive map (the quantum version of a stochastic matrix): $\sum_i A^i\rho^R A^{i\dagger}=\lambda_1\rho^R$ and   $\sum_i A^{i\dagger}\rho^L A^{i}=\lambda_1\rho^L$.  Note that the left eigenvector and right eigenvector do not have to be equal to each other, but Perron Frobenius theory guarantees that both $\rho^L$ and $\rho^R$ can be chosen to be positive semidefinite. As we will see in Section \ref{sec_4}, the MPS fulfilling that the largest eigenvalue (in magnitude) of the transfer matrix is unique and both $\rho^L$ and $\rho^R$ are positive definite is called \emph{normal}. It is not difficult to show that by blocking a finite number of sites, any normal MPS becomes injective (we refer the reader to Section \ref{sec_4} for a careful discussion). 

In the case of an injective MPS with periodic boundary conditions, the Euclidean norm of the MPS is given by ${\rm Tr}\,E^N$ and hence scales as $\lambda_1^N$ with $N$ the number of sites. In the limit of large $N$, this norm should be equal to $1$, and we  rescale the tensors $A^i\rightarrow A^i/\sqrt{\lambda_1}$ to achieve this. We henceforth assume that $\lambda_1=1$. The second largest eigenvalue $\lambda_2$ of the transfer matrix defines the correlation length of the state: $\xi=-1/\log(|\lambda_2|)$. For a general connected correlation function 
$C(X,Y):=\langle XY\rangle-\langle X\rangle \langle Y\rangle$ 
of two operators $X$ and $Y$
with a distance $n$ between them, 
the expectation value will be of the form $\sum_{i\geq 2}^{D^2} c_{XY}(i)\lambda_i^n$ and is hence a sum of $D^2-1$ pure exponentials, with $D$ the bond dimension. MPS can henceforth not reproduce algebraic correlations at long distances, as a sum of exponentials cannot reproduce the tail of an algebraic function. For a finite system with $N$ sites, however, it is  enough to choose $D$ as a polynomial in $N$ to reproduce all correlation functions faithfully \cite{verstraete:faithfully}. Similarly, Ornstein-Zernike type corrections of the form $\exp(-n/\xi)/\sqrt{n}$ can be taken into account by taking $D$ large enough \cite{zauner:tm-dispersion,rams2015truncating}. 

Let us now go back to the calculation of the eigenvalues of a reduced density matrix of $n$ sites of an injective MPS, see Figure \ref{sec2fig8}. Using the basic fact in linear algebra that the eigenvalues of the matrix $A\,B$ are equal to the eigenvalues of the matrix $B\,A$, the problem reduces to finding the eigenvalues of the $D^2\times D^2$ matrix $F_{\alpha\beta,\alpha''\beta''}=\sum_{\alpha'\beta'}\rho^L_{\alpha\alpha'}\rho^R_{\beta\beta'}\left(E^n\right)_{\alpha'\alpha'',\beta'\beta''}$. In the limit of large $n$, $E^n$ factorizes in $|\rho^R\rangle\langle\rho^L|$ and we see that the eigenvalues of the reduced density matrix are given by the eigenvalues of $\left(\rho^L.\rho^R\right)^{\otimes 2}$; in other words, the contribution from the left and right side of the block disentangle for distances larger than the correlation length. The eigenvalues of the matrix $\rho^L\rho^R$ are the squares of the Schmidt coefficients of an injective MPS with open boundary conditions. The entanglement spectrum is defined as the logarithm of these eigenvalues. 

\paragraph{PEPS}
PEPS automatically fulfill the area law, as one can argue in the very same way as we did with MPS. In this case, the entanglement entropy of a $L\times L$ block on a square lattice is upper bounded by $4L\log(D)$.  

Calculating correlation functions and the entanglement spectrum of a PEPS is much more involved than the MPS case. This follows from the fact that the contracted tensor network looks very much like the partition function of
a classical statistical mechanical model, but then with 2 layers and complex numbers. The calculation of the leading eigenvector of this transfer matrix correponds to finding fixed points of completely positive maps acting on an infinite spin chain.   The entanglement spectrum is then obtained by the eigenvalues of the corresponding fixed-point density matrix. Note that for topological states of matter, an additive negative correction to this entanglement entropy emerges \cite{kitaev:topological-entropy,levin:topological-entropy}; this is called  topological quantum entanglement entropy, and is a signature of the fact that the PEPS tensors exhibit nontrivial symmetries by which they do not have full support on the physical Hilbert space. This will be discussed in Sec. \ref{sec:3:2D}.

Unlike the 1D case, correlation functions can in principle decay following a power law, for example in the case of the so-called Ising PEPS tuned at criticality \cite{verstraete:comp-power-of-peps}, discussed in the Appendix.

\subsubsection{Extension to fermionic, continuous and infinite tensor networks \label{FandCTN}}

Tensor networks have also been defined for fermionic systems \cite{kraus:fPEPS,corboz:fMERA,barthel2009contraction,bultinck:fermionic-mps-phases}, have been formulated directly in the continuum \cite{verstraete2010continuous}, and nontrivial MPS with infinite bond dimension have been constructed using vertex operators of conformal field theories \cite{cirac2010infinite}.

\paragraph{Fermionic tensor networks}

Defining tensor networks for fermionic systems presents two new difficulties. First of all, the tensor product structure is altered due to the anti-commutation relations of the creation and annihilation operators, and second, a new superselection rule emerges in the form of parity conservation.  The projected entangled pair construction can  however readily be extended to the fermionic case by considering virtual maximally entangled modes of fermions as opposed to maximally entangled $D$-level systems: $|I\rangle=\sum_{\alpha}a_\alpha^\dagger b_\alpha^\dagger|\Omega\rangle$. The parity constraint can be enforced by choosing the projection operator $\hat{A}^i=\sum_{\alpha\beta\gamma\ldots}A^i_{\alpha\beta\gamma\ldots}a_\alpha b_\beta c_\gamma \cdots$ to have a fixed parity. This parity constraint ensures that the locality of the tensor network is conserved, which makes it possible to contract tensor networks built from such tensors efficiently. Alternatively, the construction can be made using Majorana modes, and this will be useful to construct fermionic PEPS with a chiral character.

From the mathematical point of view, working with fermions amounts to
changing the convention of working in vector spaces with a tensor product
structure to working in supervector spaces with a $\mathbb{Z}_2$ graded
tensor product. In essence, the Hilbert space is split into a direct sum
of two vector spaces, $V_0\oplus V_1$, and every vector $|i\rangle$ in the
Hilbert space has to be fully supported in one of these spaces and has
therefore a parity $|i|$ associated to it. Given the graded tensor product
of two vectors $|i\rangle\otimes_g|j\rangle$, swapping the vectors amounts
to the relation $|i\rangle\otimes|j\rangle\rightarrow
(-1)^{|i|.|j|}|j\rangle|i\rangle$. Matrix product states can now be
defined in this supervector space \cite{bultinck:fermionic-mps-phases} in the form of $\hat{A}=\sum_{i\alpha\beta}A^i_{\alpha\beta}|\alpha\rangle\otimes_g\langle i|\otimes_g\langle\beta|$, and using the sign rules of grading when moving vectors around each other such as to contract the virtual indices,  any bosonic tensor network can readily be fermionized.

Interestingly, the notion of injectivity has to be altered in fermionic tensor networks because of the fact that the parity superselection rule cannot be broken. As a consequence, different boundary conditions have to be chosen to construct translationally invariant states with an even or an odd parity. These 2 distinct possibilities relate to the fact that there are 2 distinct types of $\mathbb{Z}_2$ graded tensor algebras, and these are characterized by the absence or presence of Majorana edge modes. The prime example of a fermionic spin chain with Majorana edge modes is the Kitaev wire \cite{kitaev:majorana-chain}. When putting the Kitaev wire on a ring with periodic boundary conditions (and hence translationally invariant), one gets a system with odd parity, and the MPS description is given by
\be |\psi\rangle=\sum_{i_1\cdots i_N}{\rm Tr}\left[YA^{i_1}A^{i_2}\cdots A^{i_N}\right]|i_1\rangle\otimes_g|i_2\rangle\otimes_g\cdots\label{Majchain}\ee
with $A^0=\mat{cc}{1&0\\0&1},A^1=\mat{cc}{0&1\\-1&0}=Y$. Contrary to the
bosonic case, this MPS description is irreducible and hence injective. By
considering more copies of this Kitaev chain, it is possible to study the
entanglement spectrum of all states in the $Z_8$ classification of gapped
fermionic spin chains
\cite{fidkowski:1d-fermions,bultinck:fermionic-mps-phases}. This construction has been extended to fermionic MPU \cite{Piroli:FermionMPU}, which include all quantum celullar automata in one dimension.

\paragraph{Continuous MPS}

Continuous matrix product states \cite{verstraete2010continuous,haegeman2013calculus} can be defined by taking the limit of the lattice spacing going to zero, whilst rescaling the matrix product tensors in an appropriate way. This enables to write down wavefunctions for quantum field theories without a reference to an underlying lattice discretization, and this is very useful for doing numerical simulations of e.g. cold atoms and of quantum field theories.

Let us consider a bosonic  system on a ring of length $L$ and with creation and annihilation operators of type $\alpha$ satisfying $\left[\psi^\alpha_x,\psi^{\beta\dagger}_y\right]=\delta_{\alpha\beta}\delta(x-y)$. The cMPS wavefunction of bond dimension $D$ is defined as
\be |\psi\rangle={\rm Tr}\left[\mathcal{P}\exp\left(\int_{-L/2}^{L/2} \left( Q(x)+\sum_\alpha R^\alpha(x)\hat{\psi}^\dagger_x(\alpha)\right)dx\right)\right]|\Omega\rangle \ee
where $Q(x),R^\alpha(x) \in \mathbb{C}^{D\times D}$, $\mathcal{P}$ denotes a path ordered exponential, and where $[R^\alpha(x),R^\beta(x)]=0$. The path ordered exponential is the continuous limit of an MPS with tensors given by $A^0[i]=\openone+\epsilon Q[i]$ and $A^\alpha[i]=\sqrt{\epsilon}R^\alpha[i]$ with $\epsilon$ the lattice parameter. Fermionic cMPS are defined by replacing the commutation by anti-commutation conditions.

A translationally invariant cMPS is obtained by choosing the $Q,R^\alpha$ independent of $x$, and choosing $L\rightarrow\infty$. An intriguing property exhibited by these cMPS is the fact that the diagonal elements of the one-particle reduced density matrix in momentum space $n(k)$ decay as $\mathcal{O}(1/k^4)$. This implies that they are sufficiently smooth such as to not suffer from UV divergencies, and can therefore be used
as variational wavefunctions without suffering the UV catastrophes experienced by other variational methods \cite{haegeman2010applying}.

\paragraph{Infinite MPS}

Instead of taking the continuum limit in the space direction, it is also possible to consider matrices $A^i$ as operators in a general infinite dimensional Hilbert space as opposed to a finite dimensional one (such as in the case of MPS). A particularly interesting choice in the case of spin-$1/2$ systems is to define those operators in terms of normal ordered vertex operators appearing in conformal field theory \cite{cirac2010infinite,nielsen2012laughlin,nielsen2013local,tu2014lattice},
\[A^{s_i}=\::e^{i\sqrt{4\alpha}.s_i.\phi(z_n)}:\]
where $\alpha>0$ is a free parameter, $s_i=\pm 1/2$, and $\phi(z_n)$ is the field of a free massless boson located on position $z_n=x+i.y$ of the lattice. For a spin chain with $N$ sites with periodic boundary condition, we can choose $z_n=L.\exp(2\pi i.n/N)$, but it is also possible to parameterize 2D wavefunctions in this form. By taking the expectation value of the virtual bosonic fields with respect to the vacuum,  the MPS wavefunction is then equivalent to 
\[\psi_{s_1s_2\cdots s_n}=\delta_{\sum_i s_i}\prod_i\chi_{i,s_i}\prod_{i<j}\left(z_i-z_j\right)^{4\alpha s_i.s_j}.\]
These wavefunctions are effectively lattice versions of the Laughlin state, and a wealth of interesting critical and chiral states such as the ground state of the Haldane-Shastry or Kalmeyer-Laughlin type wavefunctions can be constructed in this form. Additionally, it is possible to define the corresponding parent Hamiltonians, and to calculate exact expressions for the entanglement entropy.

A similar construction allows to study ground states of the fractional quantum Hall effect on a cylinder by using different CFTs, and such a systematic program of variational calculations for quantum Hall systems has been pursued by \citet{zaletel2012exact,estienne2013matrix,zaletel2013topological,zaletel2015infinite}.

\subsubsection{Tensor networks as quantum circuits: tree tensor states and MERA \label{TTNMERA}}

\begin{figure}
(a) \includegraphics[width=2cm]{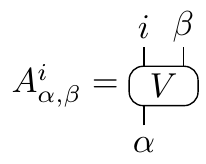} \hspace{1cm}
\includegraphics[width=2cm]{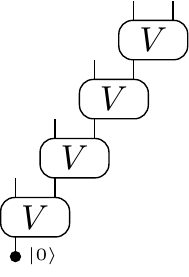}

\vspace{0.5cm}

(b) \, \includegraphics[width=2.5cm]{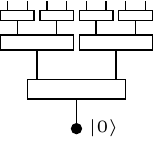}

\caption{(a) Staircase quantum circuit representing a MPS, (b) Tree tensor network}\label{sec2fig9}
\end{figure}

In the previous sections, we have advocated the picture that MPS and PEPS parameterize the quantum correlations in the corresponding many-body wavefunctions, and therefore that the virtual $D$-level systems represent the entanglement degrees of freedom. In the case of an MPS with open boundary conditions, there is however an alternative way of interpreting the wavefunction in terms of a quantum circuit acting on a product state, and assisted by a $D$-level ancillary system (Figure \ref{sec2fig9}(a)). The unitaries (or rather isometries) building up the circuit are obtained by bringing the MPS into canonical form by appropriate gauge transformations \cite{schoen:hen-and-egg}.

Instead of using a circuit in the form of a staircase, one could also envision using a circuit in the form of a tree (Figure \ref{sec2fig9}(b)). Although translational invariance is lost in this construction, it is possible to parameterize quantum states using this construction with an entanglement entropy that scales logarithmically in certain bipartitions \cite{fannes1992ground,shi2006classical,murg2010simtinulag,silvi2010homogeneous}. This makes this ansatz particularly well suited for simulating critical 1D systems. Expectation values can still be calculated efficiently as one can always choose a contraction sequence for which there is no proliferation of indices.

A more powerful and sophisticated ansatz can be made by allowing for loops in the quantum circuit of the tree tensor network, but at the same time ensuring that the circuit is efficiently  contractable. This gives rise to the concept of the multiscale entanglement renormalization ansatz \cite{vidal2007entanglement,vidal:mera}. We refer to \citet{evenbly2014algorithms} for an authoritative review.

\subsection{The ground state manifold of local Hamiltonians }

Matrix Product States and PEPS form a low dimensional manifold in an exponentially large Hilbert space. More precisely, the set of uniform MPS forms a K\"ahler manifold \cite{haegeman2014geometry}. The question addressed in this section is how this manifold is related to ground states of local quantum spin Hamiltonians. It will be shown that the manifold of MPS is in one to one correspondence with ground states of local gapped Hamiltonians. This is a very strong justification for using this manifold for variational calculations. Using concepts of differential geometry, it is then  possible to relate the linear Schrodinger equation on the exponentially large Hilbert space to nonlinear differential equations on the compressed MPS manifold, as implemented in DMRG algorithms. Additionally, the concept of the tangent plane of the MPS manifold leads to a clear characterization of the elementary excitations on top of the ground state vacuum.

\subsubsection{States to Hamiltonians: Parent Hamiltonians}

As argued in section \ref{sec2TN},  MPS and PEPS were explicitly constructed such as to mimic the entanglement structure of ground states of local quantum Hamiltonians: they have strong local correlations, possess all the symmetries of the problem, and obey an area law. Conversely, to every MPS and PEPS, there exists a local frustration-free quantum spin Hamiltonian for which the MPS or PEPS is the ground state, called the parent Hamiltonian. (Here, \emph{frustration-free} refers to the fact that the ground state minimizes the energy of each Hamiltonian term locally.) In the case of injective MPS and PEPS it is furthermore the unique ground state, and in the case of MPS, the parent Hamiltonian can be proven to be gapped (see Section \ref{sec:4}).

The existence of such a parent Hamiltonian is a consequence of the fact that the rank of the reduced density matrix in a contiguous region $B$ scales as $D^{|\partial A|}$, with $|\partial A|$ the length of the boundary of that region, as shown in section \ref{sec2TN3} (in the case of 1D systems, $|\partial A|=2$). The total dimension of the Hilbert space spanned by all spins in the region $B$ scales as $d^V$, with $V$ the volume of that region (in the 1D case, this is the length of the block $L$). On any regular lattice, there exists a size and shape of the block for which $D^A<d^V$, independent of the lattice size,  and this region will be the support of one local term of the parent Hamiltonian defined as the projector orthogonal to the support of the local reduced density matrix in that region. The full parent Hamiltonian is then obtained by summing up all these projectors over all possible regions. Clearly, this Hamiltonian is a sum of positive semidefinite terms, and every term in the Hamiltonian annihilates the tensor network state under consideration. It is therefore frustration-free. We will further elaborate on the properties of this  parent Hamiltonian in section \ref{sec:4}.

Conceptually, it is very interesting that any MPS or PEPS is the ground state of a local frustration-free Hamiltonian. However, these states are also approximate ground states of completely different frustrated Hamiltonians. How would all those different Hamiltonians be related to each other? From the linear algebra point of view, it indeed seems that knowledge about one extremal eigenvector does not provide much information about the other ones, let alone about the spectrum of the full Hamiltonian. However, the locality of the Hamiltonian puts very stringent constraints on the possible elementary excitations or eigenvectors with eigenvalues close to the ground state energy. The excitation spectrum is indeed completely reflected into the correlation functions in the ground state. If the quantum states under consideration would exhibit Lorentz invariance, this would be an obvious statement, but it is not obvious that this argument survives for lattice systems. It has indeed been observed \cite{zauner:tm-dispersion,haegeman:shadows} that the logarithms of the eigenvalues of the transfer matrix reproduce the dispersion relations of the full quantum Hamiltonian. This is a clear indication of the fact that it is enough to understand the ground state to deduce all low energy physics of a full quantum Hamiltonian: all information is encoded in the ground state.

An even stronger argument can be made in the case of topologically ordered systems and associated anyonic excitations. In that case, the full fusion and braiding statistics of the anyons can be deduced from studying the symmetries of the PEPS tensors representing the ground states. No knowledge of the Hamiltonian is needed, as the algebraic features of the excitations follows from the entanglement structure present in the ground state. This will be discussed in Section~\ref{sec:3:2D}.

\subsubsection{Hamiltonians to states}\label{sec:H-to-states}

From the traditional point of view of quantum many-body physics, the central question of interest concerning tensor networks is whether any ground state of a local quantum Hamiltonian can be well approximated with an MPS or a  PEPS.  From a practical point of view, the crucial question is to understand how large the bond dimension $D$ has to be chosen such that an MPS  or PEPS exists which provides a faithful approximation to the true ground state. Given a ground state $|\psi_N\rangle$ of a quantum spin system on $N$ sites, the question is how large the bond dimension $D\equiv D(N,\epsilon)$ has to be chosen as a function of  $N$  and error measure $\epsilon$ such that there exists an MPS/PEPS $|\psi(A)\rangle$ for which the fidelity or overlap with the exact ground state $|\langle \psi(A)|\psi_N\rangle|\geq 1-\epsilon$. For the tensor network description to be useful, the dependency on $N$ and $1/\epsilon$ should  be polynomial (note that simple strategies based on exact diagonalization of small blocks yields an exponential scaling).

We will discuss several approaches towards this problem, as each different approach highlights a complementary aspect of this issue.  In summary, the results for the case of spin chains and MPS are as follows:

\begin{itemize}
\item If the Renyi entropy $S_\alpha$, $\alpha<1$, with respect to any bipartition is bounded above by $c.\log(N)$, then an efficient polynomial approximation exists of the ground state in terms of an MPS. Note that this includes the case of critical systems \cite{verstraete:faithfully,schuch:mps-entropies}.
\item If the quantum spin chain is gapped and has a unique ground state, then an efficient polynomial approximation of the ground state exists in terms of an MPS \cite{hastings:arealaw,arad:1d-arealaw}.
\item if the state to be approximated has exponential decay of correlation functions with respect to every pair of observables (with potentially unbounded support) then an efficient polynomial approximation exists in terms of an MPS \cite{brandao2013area}.
\end{itemize}

The fact that MPS are powerful enough to represent ground states does not mean that an efficient algorithm can be found to find them. Examples of quantum Hamiltonians can indeed be constructed for which the exact ground state is an MPS, but for  which it is NP-hard to find it \cite{schuch:mps-gap-np}; the catch is that such Hamiltonians have a very small gap scaling as $1/\mathrm{poly}(N)$. Provided that the system under consideration has a gap which does not scale with the system size, it was proven by \citet*{landau:1d-efficient-algo} that an efficient algorithm scaling as a polynomial in the system size can be constructed to find an approximate MPS ground state. This result was extended by \citet{arad:rg-algorithms-and-area-laws} to the case in which the Hamiltonian has a small density of low-energy states. However, the question whether such an algorithm is possible for all critical systems with a gap scaling as $\mathcal{O}(1/N)$ is still open.

In the case of 2D PEPS, the strongest approximation result is weaker but still certifies that we can simulate quantum spin systems efficiently:  a Gibbs state $\exp(-\beta H)$ can be approximated by a PEPO with bond dimension $D=(N/\epsilon)^{\mathcal{O}(\beta)}$ \cite{hastings:locally,molnar:thermal-peps}. If however the Hamiltonian consists of terms which all commute, such as in the case of stabilizer Hamiltonians and all string nets, an exact representation of the ground states can be found in terms of a PEPS with a bond dimension which only depends on the size of the support of the local terms in the Hamiltonian terms and not on the system size \cite{verstraete:comp-power-of-peps}.

\paragraph{Area laws and approximability }

Using arguments related to the decay of the Schmidt coefficients in 1D spin chains, it has been proven that an efficient approximation of the ground state for a finite spin chain with $L$ sites can be obtained whenever the Renyi entanglement entropy of half a chain of length $N$ is bounded by $S_\alpha(N)\leq c.\log(N)$ for an $\alpha<1$.

To show this, consider an exact ground  state $|\psi_N\rangle$ defined on a spin chain with $N$ sites and with open boundary conditions, and assume that its Schmidt spectrum or eigenvalues of the reduced density matrix across a cut between sites $i$ and $i+1$ are given by $\left\{\mu_x(i)\right\}$. The main idea lies in the fact that for ground states, these Schmidt coefficients decay fast, and therefore only a small error will be made when cutting the Schmidt decomposition after the $D$ largest Schmidt coefficients. We define the residual probability $\epsilon_i(D)=\sum_{x=D+1}^\infty \mu_x(i)$ and $\epsilon_{\rm total}(D)=\sum_{i=1}^N\epsilon_i(D)$. In \citet{verstraete:faithfully}, it was shown that an MPS with bond dimension $D$ is guaranteed to exist for which the fidelity $|\langle\psi(A)|\psi_N\rangle|\geq 1-\epsilon_\mathrm{total}(D)$. If the Renyi entanglement entropy $S_\alpha$ with $\alpha<1$ for the system of size $N$ maximized over all bipartitions of the spin chain is given by $S_\alpha(N)$, there exists an MPS with bond dimension $D$ for which
\be \frac{\epsilon_{\rm total}(D)}{N}\leq \left(\frac{D}{1-\alpha}\right)^{-\frac{1-\alpha}{\alpha}}\exp\left(\frac{1-\alpha}{\alpha}S_\alpha(N)\right)\ . \ee
This proves that the scaling of $D$ to achieve a fidelity $1-\epsilon$ is polynomial in $1/\epsilon$ and $N$ provided that the Renyi entropy of half a chain satisfies $S_\alpha(N)\leq c\log N$. As shown in Section~\ref{sec2arealaws}, this is indeed the case for all gapped quantum spin systems, and also for all integrable critical spin chains with a logarithmic term. An MPS description therefore provides an exponential compression for a wide variety of ground states of quantum spin systems, including critical ones.

The fact that Renyi entropies with $\alpha<1$ had to be used stems from the fact that these are more susceptible to the tails of the distribution of  Schmidt values. Conversely, it holds that a state with an entanglement entropy satisfying a volume law for the von Neumann entropy cannot be approximated faithfully using MPS, but this problem is undetermined for the $\alpha<1$ Renyi entropies \cite{schuch:mps-entropies}.

\paragraph{No low tensor rank Ansatz}

It has been argued in that the MPS manifold is in one to one correspondence with the set of ground states of local gapped Hamiltonians. One may question whether one can obtain a similar identification with a simpler ansatz, such as linear combinations of polynomially many product states, as it is the case for instance in \cite{beylkin2002numerical,hackbusch2012tensor,kolda2009tensor,de2000multilinear}. Let us prove that this is  not possible, since one can show that {\it any} injective MPS has exponentially small overlap with {\it any} product state. 

Indeed, consider an injective MPS in canonical form so that the right (resp. left) fixed point of the associated transfer operator, seen as a completely positive map,  is the identity matrix (resp. a positive definite full matrix $\Lambda$ with ${\rm Tr}(\Lambda)=1)$. See Section \ref{sec:4} for more details. This implies that the operator norm  $\|\Lambda\|_{\rm op}<1$. Consider $\epsilon=\frac{1}{2}(1-\|\Lambda\|_{\rm op})$. 
Then, block the tensors until the transfer operator $E$ is, in operator norm, $\epsilon$-close to the rank-one projector $|\Id)(\Lambda|$. 

Let us name $A$ to the corresponding (blocked) tensor, as an element of the tensor product of the physical space $\mathbb{C}^d$ with the set of $D\times D$ matrices and consider $\lambda$ to be the maximum of  the operator norm of  $A^{w}=\bra{w}A$ (compare with \eqref{eq:overlap-MPS-product}), where $\|w\|\le 1$ in the Hilbert norm. Clearly $\lambda$ can be rewritten as the maximum of $|(u|A^w|v)|$, where $u,v,w$ are again normalized in Hilbert norm, .

By the Cauchy-Schwarz inequality and the hypothesis on the transfer operator,
$$|(u|A^{w}|v)|\le |(u|\bar{(u|}E|v)\bar{|v)}|\le (u|u)(v|\Lambda|v) +\epsilon $$
$$\le \|\Lambda\|_{\rm op}+\epsilon<1,$$
which implies that $\lambda<1$ (by a  standard compactness argument that holds due the the crucial fact that both the physical and bond dimensions are finite).

Therefore, we can conclude (see \eqref{eq:overlap-MPS-product})
$$|\langle w_1w_2\cdots w_N|\psi(A)\rangle|=|{\rm Tr}{A^{w_1}A^{w_2}\cdots A^{w_N}}|$$
$$\le D\prod_{j=1}^N\max_w\|A^w\|_{\rm op}= D\lambda^N\; .$$

\paragraph{Efficient descriptions in thermal equilibrium}

The ground state of a gapped Hamiltonian can be well approximated by evolving the Euclidean path integral $\exp(-tH)|0\rangle$ for a time $t\simeq 1/\Delta$ with $\Delta$ the gap of the system and $|0\rangle$ a state with non-zero overlap with the ground state. If the Hamiltonian is frustration-free and only consists of commuting terms, this expression is equal to the product of the exponentials of the local terms acting on the state $|0\rangle$, even in the limit of  $t\rightarrow\infty$, and therefore automatically produces a MPS/PEPS with a bond dimension related to the size of the local support of the individual commuting terms. This e.g. implies that the ground states of all  local stabilizer Hamiltonians and string nets have a very simple exact description in terms of PEPS with a bond dimension independent of the system size.

One can extend this result to the non-commutative case by using a Trotter expansion 
of  $\exp(-tH)\approx \left(\prod_k \exp(-\beta h_k/M)\right)^M$ (with $H=\sum h_k$ the local terms in the Hamiltonian), with $M$ sufficiently large. Each term $\exp(-\beta h_k/M)$ can be written as a local tensor network with constant bond dimension. However, when put together naively, this construction yields a tensor network with a bond dimension which scales exponentially in $M$.  Fortunately, since each $\exp(-\beta h_k/M)\approx \openone -\beta h_k/M$ is very close to the identity, it is possible to compress this representation by choosing a suitable subset of terms in the full expansion of $\left(\prod (1-\beta h_k)\right)^M$.  This way, one obtains a PEPO $\sigma_D$ with bond dimension $D$ which approximates the Gibbs state $\rho_\beta=e^{-\beta H}/Z$  up to error $\epsilon:=\|\sigma_D-\rho_\beta\|_1$ in trace norm with a bond dimension $D=\exp(O(\log^2(N/\epsilon))$, as long as $\beta\le O(\log N)$ \cite{molnar:thermal-peps}, independent of the spatial dimension.  In order to obtain a polynomial scaling in $N$ for fixed temperature, one starts instead from the Taylor series $e^{-\beta H}
= \sum (-\beta)^\ell (\sum h_k)^\ell/\ell!$, expands $(\sum h_k)^\ell$, and finds that only clustering terms in this expansion are relevant.  This way, one arrives at a PEPO approximation of the Gibbs state for which $D=(N/\epsilon)^{O(\beta)}$ for a fixed temperature
\cite{hastings:locally,molnar:thermal-peps} (for a practical implementation, see \citet{vanhecke2019symmetric}); 
using a refined approximation of the Taylor series, an improved scaling of $D\sim\exp[O(\sqrt{\beta\log(n/\epsilon)})]$ has recently been shown~\cite{kuwahara:thermal-area-law}.
From here, one can construct a PEPS approximation for ground states by applying $e^{-\beta H}$ to a suitable initial product state at a temperature $\beta=O(\log N)$, which yields sufficient overlap with the ground state as long as the density of states scales at most as $N^cE$ for some constant $c$, that is, polynomial in $N$ \cite{hastings:locally}, which is the case for gapped systems with particle-like excitations. The reverse direction has been analyzed  by \citet{chen2020matrix}, where it is shown that generic MPDOs can be well approximated by Gibbs states of (quasi-)local Hamiltonians.

A different approach can be taken if, instead of assuming a low density of low-energy states, one restricts to Hamiltonians which belong to a phase with a zero-correlation renormalization fixed point, as it is the case for all known non-chiral topological phases in 2D. As derived by \citet{coser:phases-mixed}, following an idea of \citet{osborne:1d-gs-approx}, the quasi-adiabatic theorem \cite{hastings:quasi-adiabatic, bachmann:quasi-adiabatic} gives a finite depth quantum circuit whose gates act on $\log^{2+\delta}N$ neighboring sites ($\delta>0$ but arbitrarily small). When applied on the renormalization fixed point state (which is an exact PEPS with finite bond dimension), the circuit gives an approximation of the ground state of the target Hamiltonian with an error that goes to zero with $N$ faster than any polynomial. This approximation is a PEPS with bond dimension scaling as $e^{\mathcal{O}(\log^{2+\delta}(N/\epsilon))}$. Moreover, it keeps all the virtual symmetries present on the initial renormalization fixed point PEPS and hence the information about the topological phase it belongs to (see Section \ref{sec:3}). A similar argument has been used by \citet{huang:approx-local-observable-trivial-phase} to propose quasi-polynomial algorithms to compute local observables in systems belonging to the trivial phase.

In the case of gapped 1D systems, the strongest results along these lines have been obtained by randomizing the path integral and invoking a Chebyshev-based Approximate Ground Space Projector \cite{arad:1d-arealaw,arad:rg-algorithms-and-area-laws,huang:area-law}. Given an infinite chain of $d$-level systems with $r\le\mathrm{poly(n)}$-fold degenerate ground space and gap $\Delta$ and considering one cut in an infinite spin chain, it yields that  

\begin{equation} \label{eq-AGSP}
\log(\epsilon(D))= -\frac{\Delta^{1/3}\tilde\Omega(\log(D/r)^{4/3})}{(\log d)^{4/3}}
+\frac{(\log d)^{8/3}}{\Delta}
\end{equation}
where $\epsilon(D)$ is defined as above, that is, the sum of the square of all but the first $D$ Schmidt coefficients on that cut, and $g=\tilde\Omega(h)$ exactly if $h=\tilde O(g)\equiv O(g\,\log(g)^k)$ for some $k$ -- that is, the truncation error $\epsilon(D)$ decreases superpolynomially with the number of Schmidt coefficients kept, i.e., the bond dimension. 

\begin{figure}
\includegraphics[width=8cm]{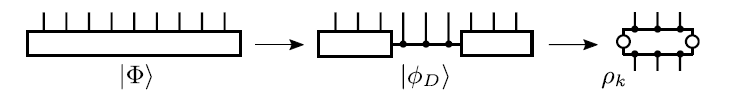}
\caption{Construction of a translation invariant MPO approximation to infinite translation invariant ground states by truncating the bond dimension in an intermediate region \cite{schuch2017matrix}; the resulting MPO $\rho_k$ for $k$ sites is then patched together and superimposed with its translations.}
\label{fig-note-Norbert-Frank}
\end{figure}

A corollary of this  result is the existence of efficient translationally invariant MPO desciptions for infinite translationally invariant gapped quantum spin chains, in the sense that any local expectation values  of the MPO approximation are $\epsilon$-close to the ones of the exact ground state. A direct proof was formulated by \citet{schuch2017matrix} and we sketch it here; building on it, similar results have been obtained for pure uniform MPS \cite{dalzell:local-MPS-approx,huang2019matrix}. The construction is as follows: define an MPO $\rho_k$ on $k$ sites by cutting the exact ground state $k+1$ times and then tracing the outer spins, Fig.~\ref{fig-note-Norbert-Frank}. This yields an MPO with bond dimension $D^2$ so that the trace distance to the true ground state reduced density matrix $\sigma_k$ is $\|\rho_k-\sigma_k\|_1\le 2\sqrt{2\epsilon(D)(k+1)}$.
In order to obtain a translationally invariant MPO for the infinite chain, take an infinite tensor product of this MPO with itself  and sum over all $k$ translations. The resulting MPO has bond dimension $k D^2$ and its reduced state on $\ell$ contiguous sites approximates  $\sigma_\ell$ up to error
$$\epsilon\le 2\sqrt{2k\epsilon(D)} + \frac{2(\ell-1)}{k}\;.$$
Choosing an appropriate $k$ and using \eqref{eq-AGSP} one gets that the bond dimension required to give an $\epsilon$-approximation on $\ell$ sites scales as $D\leq\frac{\ell}{\epsilon}$.  This result demonstrates that the MPS compression is still faithful in the thermodynamic limit, and hence provides clear evidence for the falling of the exponential many-body wall.

Finally, approximation results of thermal states by MPOs can also be obtained from area laws for thermal states, see \citet{kuwahara:thermal-area-law} and \citet{jarkovsky:mpdo-arealaw}.

\paragraph{Many-body Localization}

Matrix product states also pop up for the description of eigenstates of quantum spin systems subject to randomness. It has been argued that all eigenstates, not just the ground state, of many-body localized (MBL) Hamiltonians exhibit an area law for the entanglement entropy, and furthermore that all have an efficient description in terms of matrix product states \cite{bauer2013area,nandkishore2015many,wahl2017efficient}. This follows from a result of \citet{imbrie2016many}, proving that there is a low depth quantum circuit which completely diagonalizes any MBL Hamiltonian.

\subsubsection{Manifold of MPS and TDVP}

\paragraph{Manifold and TDVP}
As demonstrated in the previous paragraphs, ground states of local gapped quantum spin chains can efficiently be parameterized as matrix product states, and furthermore any such MPS is the ground state of a local Hamiltonian. This implies that the manifold of MPS is in one to one correspondence with all possible ground states, and this opens a large number of perspectives, such as the classification of all phases of matter of 1D spin chains. 

The manifold of infinite uniform matrix product states has been studied in detail by \citet{haegeman2014geometry}. MPS have the mathematical structure of a (principal) fiber bundle; this follows from the parameter redundancy corresponding to the gauge transforms. The total bundle space corresponds to the parameter space, i.e.\ the space of tensors associated to a physical site. The base manifold is embedded in the Hilbert space and can be given the structure of a K\"ahler manifold by inducing the Hilbert space metric. A similar construction holds in the finite case of MPS with open boundary conditions. The metric is governed by the Schmidt numbers, and is singular when the MPS is not normal/injective. 

Given a specific MPS on the manifold, we can associate a linear subspace to it, namely the tangent space on the manifold \cite{haegeman:varprinciple-lattices,vanderstraeten2018tangent}. The dimension of that space is $(d-1)D^2$, and is spanned by the vectors 
\[
|\psi_{\alpha\beta i}(A)\rangle=\frac{\partial}{\partial A^i_{\alpha\beta}}|\psi(A)\rangle
\ .
\]

\begin{figure}
\includegraphics[width=3cm]{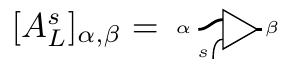} \hspace{0.3cm} \includegraphics[width=3cm]{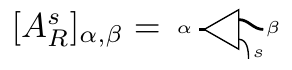} 

 \includegraphics[width=1.5cm]{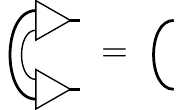} \hspace{1.8cm} \includegraphics[width=1.5cm]{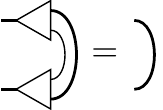} 

\vspace{0.4cm}

\includegraphics[width=7cm]{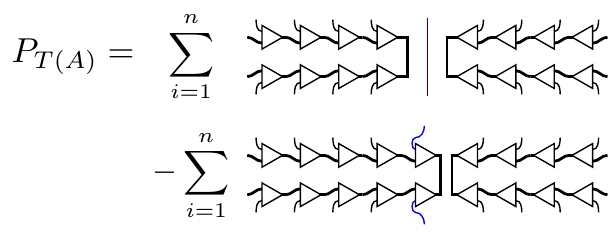}
\caption{For a given MPS, one considers the two possible canonical forms, given by tensors $A_L$ (resp. $A_R$),  where the right (resp. left) fixed point of the transfer operator is the identity (see Section \ref{sec_4} for more details). They are distinguished in the figure by the direction of the triangle. They allow to express the tangent space projector $P_{T(A)}$ as a sum of MPOs. The variable $i$ in the sum corresponds to the position of the blue index. }
\label{fig-tangent-plane}
\end{figure}

Due to the product rule of differentiation, these vectors correspond to plane waves and are hence normalized as delta-functions.  $D^2$ of these are linearly dependent due to the gauge freedom.  These degrees of freedom can be used to make the Gram matrix or metric $g_{xy}=\langle\psi_x(A)|\psi_y(A)\rangle$ locally Euclidean ($g_{xy}=\delta_{xy}$); the gauge transformations needed to achieve this are precisely the ones which bring the MPS in left and right canonical form. 
As shown in Figure~\ref{fig-tangent-plane}, the projector on the tangent space $P_{T(A)}$ in a specific point $|\psi(A)\rangle$ can be constructed in terms of a sum of matrix product operators which are expressed in terms of the tensors in left ($A_L$) and right ($A_R$) canonical form.

This expression is of great practical interest, as it allows to write down evolution equations within the manifold of MPS. Let us for example assume that we aim to simulate the time evolution generated by a Hamiltonian $H$ of a quantum state $|\psi(A)\rangle$ which is initially an MPS. The evolution described by the Schr\"odinger equation $i\partial_t\ket\psi=H\ket\psi$ will take the state outside of the manifold and hence make the problem seemingly intractable. The projector on the tangent plane however allows to pull the state back on the manifold, in a way which maximizes the overlap to any state on the manifold:
\[i\frac{\partial}{\partial t}|\psi(A)\rangle=P_{T(A)}H|\psi(A)\rangle\ .\]
The corresponding equations are the MPS equivalents of the time dependent variational principle (TDVP) as originally derived in the context of Hartree-Fock theory \cite{dirac1930note}: 
\[\partial_t A^i_{\alpha\beta}=F\left(A^i_{\alpha\beta}\right)\ .\]
This equation is non-linear due to the fact that the projector $P_{T(A)}$ depends itself on $A$. It conserves the energy and all symmetries that can be represented within the tangent plane, and it is possible to associate symplectic structure with a  Poisson bracket to it \cite{haegeman:varprinciple-lattices}. MPS therefore yields novel semi-classical descriptions of the dynamics of quantum spin chains. 

There are many methods to solve these differential equations in practice, and the most interesting case consists of evolving the equations in imaginary time such as to converge to the ground state. The density matrix renornalization group (DMRG) \cite{white1993density} can be understood in terms of splitting the differential equation into the $2N-1$ MPO terms, and then evolving each one of them for an infinitely large imaginary time step \cite{haegeman2016unifying}. Increasing the bond dimension can be understood in terms of a projection on the 2-site tangent plane, and time-dependent DMRG can completely be formulated within the time dependent variational principle. We refer the interested reader to a review by  \citet{vanderstraeten2018tangent} on tangent space methods for MPS.

\paragraph{Excitations}

The tangent plane of an MPS representing the ground state of a quantum spin system reveals another very intriguing aspect of MPS, namely the fact that the projection of the full many-body Hamiltonian on its linear subspace gives rise to an effective Hamiltonian whose spectrum reveals the dispersion relations of the true elementary excitations of the many-body Hamiltonian \cite{rommer1997class,porras2006renormalization,pirvu2012matrix,haegeman:mps-ansatz-excitations}. More precisely, the tangent plane yields a method for parameterizing plane waves of the form
$\psi_k(A,B)\rangle=$
\[\frac{1}{\sqrt{N}}\sum_x e^{2\pi ikx}{\rm Tr}\left[A^{i_1}\cdots A^{i_{x-1}} B^{i_x} A^{i_{x+1}}\cdots\right]|i_1\rangle|i_2\rangle\cdots\]
which correspond to Bloch wave-like excitations; the tensors $B^i_{\alpha\beta}$ can easily be determined by solving a linear eigenvalue problem of the effective Hamiltonian.  By making use of Lieb-Robinson techniques \cite{nachtergaele2010lieb,bravyi2006lieb, hastings2004locality,hastings:gap-and-expdecay}, it has been proven by \citet{haegeman2013elementary} that such an ansatz provides a faithful representation of the exact excitations in the system if these are part of an isolated band with a gap $\Delta(k)$ above it in the momentum sector $k$; by allowing the tensor $B$ to act on $l$ sites, the fidelity to the exact excited state is lower bounded by $1-p(l)\exp\left(-\Delta(k)l/2v_{LR}\right)$ with $p(l)$ a polynomial in $l$ and $v_{LR}$ the Lieb Robinson velocity.

This is a beautiful manifestation of the fact that ground state correlations reveal plenty of information about the excitation spectrum. In fact, as commented already, the spectrum of the transfer matrix $E$ of the MPS itself already contains this information: the correlation length can be extracted from the gap in eigenvalues of $E$, and the phases of all eigenvalues reveal information about the full dispersion relation \cite{zauner:tm-dispersion}.

\subsection{Bulk-Boundary Correspondences \label{bulk-boundary}}

One of the most intriguing consequences of tensor network descriptions is the fact that single tensors can represent entire many-body states if translational invariance is imposed. Once we know the geometry of the lattice, which tells us how to contract the local tensor with copies of itself, the whole state is completely determined. Thus, all physical properties (i.e., expectation values of observables) are contained in the tensor $A$, as they are all functions of the coefficients of that tensor. In a sense, one could establish a map between all physical properties of  many-body states (given a certain geometry in $d$ spatial dimensions), and the space of tensors corresponding to that geometry. However, this map is highly nonlinear, as the expectation values of observables in Hilbert space will be complex functions of the tensor.

The special form of MPS/PEPS allows one to establish other maps that are linear. In particular, if the PEPS is defined in a Hilbert space, $\mathcal H_d$, corresponding to a certain geometry in $d$ spatial dimensions, it is possible to map all physical properties to these of a different space, $\mathcal H_{d-1}$, corresponding to $d-1$ spatial dimensions \cite{cirac:peps-boundaries}. Furthermore, the map takes the form of a linear isometry. More explicitly, given a region $R$, it is possible to find an isometry, ${\cal U}_R$, that maps the reduced state and all operators acting on that region, to a state and operators acting on its boundary, $\partial R$, so that the expectation values computed in $R$ coincide with these computed in $\partial R$.

The existence of this holographic principle is not surprising, at least for ground states of local Hamiltonians fulfilling the area law (see Section \ref{sec2arealaws}). In fact, the area law states that the entropy of (the reduced state in) a region $R$ scales with the number of particles at its boundary. As the entropy counts degrees of freedom, it is natural to think that one can map the reduced state to a space that lives in the boundary $\partial R$. What is special about tensor networks is that the Hilbert space of this boundary is dictated by the bond dimension of the tensors, and one can thus talk about geometrical notions there too. For instance, the state in the boundary may be a mixed state that can be described as the Gibbs state of a local Hamiltonian, where the notion of locality refers to that geometry. All this will become clearer when we show how to explicitly construct the bulk-boundary correspondence. 

There exists another way of constructing a bulk-boundary correspondence which may be more ``physical'' \cite{yang:peps-edgetheories}. The one mentioned above relates the physical properties in a region of the bulk to those of another theory living in the boundary of such region, but in thermal equilibrium. The boundary Hamiltonian characterizing such a Gibbs state in the boundary does not have a dynamical meaning (i.e.\ does not generate the evolution), but just a statistical one. In contrast, it is possible to derive a Hamiltonian that describes the dynamics of the physical boundary (i.e.\ the edges) of a many-body system with respect to perturbations in the bulk. One can build an isometry to map this Hamiltonian to one that acts on the auxiliary indices associated to the edges of the system. Thus, one can describe the dynamics in the bulk by just transferring the dynamics to these auxiliary indices by using the isometry. For PEPS with finite correlation length, this isometry will only affect the lattice sites that are close to the physical boundary, and thus will be describing edge excitations. All this is very reminiscent of the physics of the quantum Hall effect in 2D, where there exists a 1D Hamiltonian that describes the dynamics of the edge states. What we will discuss is that something similar occurs with generic PEPS.

Very recently, this kind of bulk-boundary relations appearing in tensor
networks has been used to construct toy models of the AdS/CFT
correspondence \cite{maldacena:AdSCFT}, as it appears in holographic
principles proposed in the context of high-energy physics and string
theory. First of all, the construction of MERAs (and TTNs) can be
associated to a coarse tesselation of an Anti-de Sitter geometry, where
the renormalization direction coincides with the radial coordinate
\cite{swingle:mera-ads-cft, evenbly:tn-geometry}. The 1D  MERA
construction can be interpreted as a quantum circuit which implements a
conformal mapping between the physical Hilbert space and the
(renormalized) one in scale space \cite{czech2016tensor};  the
entanglement spectrum of the MERA can be identified with the one of a MPO
representing a thermal state, hence relating the bond dimension of the
MERA approximation to the bond dimension needed to represent thermal
states using MPOs \cite{van2020entanglement}. For MERA in 1D it is
straightforward to show that they display the logarithmic correction to
the area law associated to CFTs by simply finding the shortest path in the
MERA embracing the region in which one is interested \cite{vidal:mera} (on
the other hand, 2D MERA can be embedded in PEPS and thus obey an area law
\cite{barthel:peps-mera}). Swingle pointed out a very intriguing
connection between this way of determining the entropy and the
Ryu-Takayanagi formula relating the entropy of the ground state of a CFT
and the geometry of AdS space \cite{swingle:mera-ads-cft}. This indicated
that MERA could define geometries through that formula, and thus in a
sense relate the entanglement properties of many-body states with
geometries appearing in gravitational physics. Later on, it was realized
that by adding a physical index to the MERA tensors, one could build a
linear correspondence (an isometry) between these physical indices and the
auxiliary ones living in the boundary \cite{pastawski:holography}. This is
closely related to the bulk-boundary correspondence mentioned above. In
fact, if one builds PEPS in a (tesselated) hyperbolic geometry using the
quantum circuit construction, they are equivalent. In recent works
\cite{hayden:holographic-duality-from-tn,qi:holographic-coherent,qi:space-time-holographic,
kohler:holographic-toy-models}, it has been shown that by choosing appropriately the tensors which build the TNS, one can ensure certain desired properties of the AdS/CFT correspondence, and in this way explicitly build toy-models displaying such properties.

\subsubsection{Entanglement spectrum}

Let us consider a many-body state $\ket\Psi$ in $d$ dimensions and its reduced state, $\rho_R$ in a region $R$. The entanglement spectrum $\sigma_\Psi$ of the state with respect to region $R$ is defined as the spectrum of $H_R=-\log(\rho_R)$. It is clear that these dimensionless numbers are related to the entanglement of region $R$ and its complement $\bar R$. In fact, we can always write the Schmidt decomposition of $\ket\Psi$ as
 \be
 |\Psi\rangle = \sum_{n=1}^{d_R} \lambda_n |\varphi_n\rangle_R\otimes|\psi_n\rangle_{\bar R}\ ,
 \ee
where $\varphi_n,\psi_n$ are orthonormal sets in ${\cal H}_R$ and ${\cal H}_{\bar R}$, respectively. These are the Hilbert spaces corresponding to the lattice sites in region $R$ and $\bar R$, and have dimensions $d^{|R|},d^{|\bar R|}$, respectively, where as usual $|R|$ denotes the number of lattice sites in $R$. The $\lambda_n\in [0,1]$ are the Schmidt coefficients and completely characterize the entanglement of state $\Psi$ with respect to region $R$ and $\bar R$. As any Schmidt decomposition, the number of coefficients is $d_R\le \min(d^{|R|},d^{|\bar R|})$.  Given the normalization of $\ket\Psi$, their squares add up to one. By definition,
 \be
 \sigma_\Psi = \{ -\log(\lambda_n^2)\}_{n=1}^{d_R}\ ,
 \ee
which indicate that the entanglement spectrum is nothing but the Schmidt coefficients, which thus fully characterize the entanglement. The operator $H_R$ is usually referred to as the entanglement Hamiltonian.

It was argued by \citet{li:es-qhe-sphere} that for quantum Hall states (integer or fractional), the low-lying part of the entanglement spectrum coincides (up to a proportionality constant) with the Hamiltonian corresponding to the  CFT associated to its topological order. This has been verified for other states with topological order and proven for more general states whose wavefunction can be written in terms of correlators of a CFT under certain assumptions (see e.g.\ \onlinecite{qi:bulk-boundary-duality, dubail:entanglement-spectrum}, the references therein and Section~\ref{secESEM}). In models without topological order in $d=2$ dimensions it has also been found that the lower sector of the entanglement spectrum resembles that of local theories in $d=1$ dimensions \cite{lou:entanglement-spectra,cirac:peps-boundaries}. For instance, for the AKLT state in a square lattice it has been found that the entanglement spectrum also resembles that of a Wess-Zumino-Witten $SU(2)_1$ theory; this will be discussed in more detail in the following section.

\subsubsection{Boundary Theory}\label{sec:boundary-theory}

Both, the existence of an area law and the numerical results displaying a 1D-like spectrum in ground states of 2D theories indicate that it should be possible to compress the degrees of freedom corresponding to any region $R$ from $|R|$ to $|\delta R|$. More specifically, one could always find a $P_R$ such that $\tilde\rho_R=P_R \rho_R P_R \approx \rho_R$, where $P_R$ is a projector onto a subspace of ${\cal H}_R$ with dimensions $d_b^{|\partial R|}$, where $d_b$ is a constant integer. Note that this hints to an area law given the fact that the von Neumann entropy of a (mixed) state is upper bounded by the logarithm of the dimension of the Hilbert space on which it is supported; in this case, $S(\tilde \rho_R)\le |\partial R| \log(d_b)$. Furthermore, there should exist a Hamiltonian $H_{\partial R}$ acting on a different space corresponding to $d-1$ spatial dimensions, such that the lowest part of the entanglement spectrum coincides with the spectrum of $H_{\partial R}$. The latter should be somehow local (note that otherwise it does not make so much sense to talk about the spatial dimensions).

\begin{figure}
\includegraphics[width=7cm]{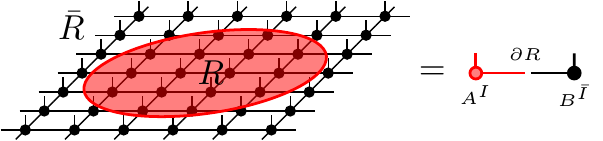}\\
\vspace{0.5cm}
\includegraphics[width=2.5cm]{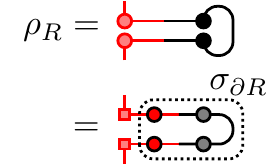}\hspace{1.5cm}
\includegraphics[width=2.5cm]{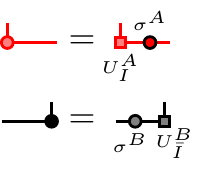}\\
\caption{Bulk-boundary correspondence: For any given region, $R$, $\rho_R$ can be isometrically mapped to a density operator, $\sigma_{\partial R}$, that is defined in the auxiliary indices of the boundary of $R$ via a polar decomposition (see main text).}    \label{Fig:boundary-theory}
\end{figure}

For PEPS it is possible to make the above conclusions rigorous \cite{cirac:peps-boundaries}. The projector $P_R$ comes automatically from the theory of tensor networks and, in fact, $\tilde \rho_R=\rho_R$, where $d_b=D$, the bond dimension. Furthermore, for any region, $R$, there exists an isometry
 \be
 U_R: {\cal H}_R \to {\cal H}_{\partial R}
 \ee
with $U_R^\dagger U_R=\Id_{\partial R}$ and $U_RU_R^\dagger=P_R$, such that
 \be
 \rho_R= P_R \rho_R P_R = U_R^\dagger \sigma_{\partial R} U_R\ ,
 \ee
where
 \be
 \label{sigmaR}
 \sigma_R= U_R \rho_R U_R^\dagger
 \ee
is an operator defined on the auxiliary indices at $\partial R$ (coming out of region $R$, see Fig.~\ref{Fig:boundary-theory}). In fact, for any observable $X_R$ acting on ${\cal H}_R$ we can define $X_{\partial R}=U_R X_R U_R^\dagger$, so that
 \be
 \langle \Psi_R| X_R|\Psi_R\rangle = {\rm tr}_R(X_R \rho_R)= {\rm tr}_{\partial R}(X_{\partial R} \sigma_{\partial R})\ .
 \ee
This expresses the fact that one can compute all physical properties of the bulk (in $R$) in terms of a state living in the boundary, $\sigma_{\partial R}$, and defines the bulk-boundary correspondence. Furthermore, we can identify the entanglement Hamiltonian $H_{\partial R}=-\log(\sigma_{\partial R})$ or equivalently write
 \be
 \sigma_{\partial R} = e^{-H_{\partial R}}\ .
 \ee
The isometric character of $U_R$ ensures that the entanglement spectrum $\sigma(H_R)=\sigma(H_{\partial R})$.

The derivation of the above statements is straightforward, and we reproduce it here. For that, let us consider the PEPS $\ket\Psi$, and denote by $A^I_i$ the tensor corresponding to the contraction of all the auxiliary indices in region $R$, where we have combined all physical indices of region $R$ into a single index $I$, and all the auxiliary indices which have not been contracted (and thus stick out of region $R$) into an index $i$. We do the same with region $\bar R$, and denote the corresponding tensor by $B^{\bar I}_{i}$. Note that the auxiliary indices connect $R$ and $\bar R$, and thus the auxiliary legs are characterized by the same combined index $i$.
Thus, we can write the coefficients of $\ket\Psi$ as
 \be
 \Psi_{I\bar I}  \equiv 
 \langle I,\bar I\ket\Psi =
 \sum_{i=1}^{D^{|\partial R|}} A^I_i B^{\bar I}_i\ .
 \ee
Considered as matrices, $A$ and $B$ can be expressed in terms of their polar decomposition
 \bea
 A^I_i &=& \sum_{r=1}^{D^{|\partial R|}} U^A_{I,r} \sigma^A_{r,i}\ ,
 \label{tensorAI}\\
 B^{\bar I}_i &=& \sum_{r=1}^{D^{|\partial R|}} U^B_{\bar I,r} \sigma^B_{r,i}\ .
 \eea
Here, $U^{A,B}$ are isometries and $\sigma^{A,B}$ positive semidefinite operators. From here, 
one directly obtains (\ref{sigmaR}) where $U_R=U^A$ and $\sigma_{\partial R}= 
\sigma^A  \big(\overline{\sigma^{B}}\big)^{2} \sigma^A$.

The possibility of explicitly building this bulk-boundary correspondence for PEPS is a direct consequence of its TN structure. Indeed, in this case there is a natural identification of the degrees of freedom of the boundary with these corresponding to the auxiliary indices. The boundary state, as well as the Hamiltonian ${\cal H}_{\partial R}$ act on that space. However, there is nothing which guarantees that this Hamiltonian is in any way local. In numerical studies, however, it has been always found that if the state $\ket\Psi$ has finite correlation length, then the boundary Hamiltonian is quasi-local \cite{cirac:peps-boundaries}; with a suitable treatment of sectors, this also holds for symmetry-breaking \cite{rispler:peps-symmetrybreaking,rispler:ZN-sym-breaking} and topological~\cite{schuch:topo-top} phases. More specifically, we can always expand it as
 \be
H_{\partial R} = \sum_{r=1}^{|\delta R|} h_r\ ,
 \ee
where $h_r$ combines operators with non-trivial support on exactly $r$ consecutive sites. In the studied examples, there is clear evidence that $\|h_r\|$ decays exponentially with $r$. Furthermore, in some cases where the many-body state is changed as a function of a parameter, $\Psi(g)$, and the corresponding correlation length, $\xi(g)$, diverges at some $g=g_c$, this exponential decay disappears, so that the boundary Hamiltonian displays long-range couplings. A first rigorous proof in this direction is provided in \cite{kastoryano2019locality,perez-hernandez:thermal-1D}: If the boundary Hamiltonian is sufficiently local, then the parent Hamiltonian of $\ket\Psi$ is gapped (and hence $\ket\Psi$ has finite correlation length).

As will be explained in Section~\ref{sec:3}, in case $|\Psi\rangle$ displays topological order and corresponds to a specific sector, $\sigma_{\partial R}$ itself is supported on a subspace of the boundary. In the presence of a finite correlation length, the boundary Hamiltonian is also expected to be local \cite{schuch:topo-top,haegeman:shadows} and the non-local projector related to the topological order can be understood as a superselection sector. 

A key point in the construction is that (global) symmetries in the bulk wavefunction $\ket\Psi$ will show up as (global) symmetries of the boundary state $\sigma_{\partial R}$ with specific representations, and correspondingly as symmetries of the entanglement Hamiltonian $H_{\partial R}$, which -- together with locality -- can significantly restrict the possible form of $H_{\partial R}=\sum h_r$.  This will be further discussed in  Section~\ref{secESEM}.

\subsubsection{Edge Theory}\label{edgetheory}

Let us consider a PEPS, $\ket\Psi$, defined on a 2D torus, with translational invariance along both directions. As explained above, it is always possible to find a 
 frustration-free parent  Hamiltonian $H$ for which $\ket\Psi$ is the ground state (possibly not unique). This Hamiltonian is itself translationally invariant, and thus can be written as a sum of translations of a projector, $H_r$, acting on few sites, which also annihilate $\ket\Psi$. Now, following \citet{yang:peps-edgetheories}, let us consider the same problem but with open boundary conditions. That is, we take the same Hamiltonian, $H_R$, but consider only the terms acting on some region $R$. The ground state of that Hamiltonian is now highly degenerate. In fact, by defining the tensor $A$ as in (\ref{tensorAI}) it is clear that for any
 \be
 |\varphi_r\rangle = \sum_I A^I_r |I\rangle\ ,
 \ee
we have
 \be
 H_R |\varphi_r\rangle =0\ .
 \ee

Let us denote by ${\cal H}_0\subseteq H_{\partial R}$ the subspace of the auxiliary indices spanned by all vectors $A^I_r$, with $I=1,\ldots, d^{|R|}$. In the generic case that $\ket\Psi$ is injective, all $\ket{\varphi_r}$ will be linearly independent and ${\cal H}_0= H_{\partial R}$. In case $\ket\Psi$ has topological order, 
${\cal H}_0\subsetneq \mathcal H_{\partial R}$ will be the support of an MPO projector (see Section \ref{sec:3}). In both cases, $\ket{\varphi_r}$ will span the whole ground space of $H_R$ by construction. We will denote by $P_R$ the projector onto that subspace.

Let us assume that $H_R$ is gapped, i.e.\ there is a $\gamma>0$ such that for any size $N$, the gap $\Delta_N$ above the ground-state subspace is $\Delta_N\ge\gamma>0$. If we add a small local perturbation to $H_R$ so that the new Hamiltonian reads
 \be
 H'_R = H_R + \epsilon V_R\ ,
 \ee
where $V_R$ is a sum of translations of a local operator acting on few neighboring sites, the degeneracy of the ground state subspace will be lifted. Assuming that $\epsilon$ is small enough such that perturbation theory applies, the effect of $V_R$ on the low energy sector can be determined with the help of degenerate perturbation theory, which yields an effective Hamiltonian
 \be
 H''_R = \epsilon P_R V_R P_R\ .
 \ee
Using the bulk-boundary correspondence of the previous subsection, we can map this Hamiltonian
to the auxiliary indices at the edge of the system:
 \be
 \label{edgeHam}
 h_{\partial R} = \epsilon U_R V_R U_R^\dagger\ .
 \ee
We can now use this Hamiltonian, which acts on the auxiliary indices corresponding to the edges of our system, in order to determine the low energy dynamics generated by the perturbation, and then map it back to the bulk. Note that, in contrast to the previous subsection, where the boundary Hamiltonian just had a statistical mechanics role, in the current scenario the edge Hamiltonian (\ref{edgeHam}) describes the real low-energy dynamics in the system.

In summary, we see that the isometry $U_R$ defined by the PEPS can be considered as a mapping between a theory that lives in the bulk and another one that lives in the boundary, and allows to relate the physics of the two.

As before, in practice the edge Hamiltonian turns out to be quasi-local for systems with finite correlation length. When mapping back the dynamical action of $H_{\partial R}$ from the boundary, only regions close to the edge (at a distance of about the correlation length) will be affected, so that perturbations only give rise to excitations at the edge. This is very reminiscent of the Quantum Hall effect, where the low energy excitations occur at the edge. Furthermore, the global symmetries of $\ket\Psi$ (and thus of $H$) will be inherited by $H_{\partial R}$ (see Section~\ref{sec:3}). In addition, by changing parameters in $V_R$, one can drive phase transitions in $H_{\partial R}$: This illustrates that it is possible to have phase transitions in the edge of a system, and in fact realize a range of different phases, without changing the phase of the gapped bulk~\cite{yang:peps-edgetheories}. Finally, as we will discuss later, the existence of topological order in $\ket\Psi$ is reflected in the fact that any $H_{\partial R}$ (resulting from a perturbation) has to commute with a non-local MPO projector,  which thus plays a role of a superselection rule at the boundary. Indeed this is the way the topological anomaly is revealed in this setup.

\subsection{Renormalization and phases of matter  \label{renorm_secII}}

As discussed in the previous sections, tensor networks give efficient representations of ground states, thermal states and elementary excitations in gapped locally interacting systems. Therefore, if one is interested in classifying the different possible features appearing in the low energy sector of gapped strongly correlated lattice models, one can restrict the attention to MPS and PEPS. 

In this review we will focus only on properties that are global (or topological), in the sense that they are stable under renormalization steps.  Since its conception by Kadanoff, Fisher and Wilson, the renormalization group (RG) has played a central role in many-body physics. From the conceptual point of view, the RG has clarified how simple toy Hamiltonians of spin systems can nevertheless exhibit the full spectrum of features of realistic Hamiltonians, as the universal properties of both theories at long length scales can be identified. In essence, RG provides a systematic method for integrating out UV degrees of freedom, thereby mapping a Hamiltonian to one for which the length scale is reduced. The correlation length of the ground state of a Hamiltonian which is a fixed point of an RG flow should hence be $0$ or $\infty$, the former case corresponding to trivial or topological phases, the latter to critical systems. 

Formally, a renormalization step can be understood as the composition of two processes:  blocking several sites, which coarse-grain the lattice, and acting with a reversible operation in the blocks which rearrange the local entanglement pattern. Reversibility is crucial since we want to guarantee an exact renormalization process, without discarding any relevant degree of freedom. 

Being interested in the topological content of a phase, arguably the best way to capture it is to restrict the attention to renormalization fixed points (RGFPs), where all local entanglement is integrated out and only the topological content remains.
It is however difficult to characterize such fixed points without a description of all possible renormalization flows. To circumvent that problem, we will define RGFPs in tensor networks intrinsically, from first principles, without referring to any concrete flow. 
For that we will identify some key properties that any RGFP of a gapped phase must have and conclude from there that there is indeed a renormalization flow for which the given state is a fixed point, together with a structural characterization of the tensor networks that are RGFPs.

We will first analyze the case of MPS, following \citet{verstraete:renorm-MPS} and \citet{cirac:mpdo-rgfp}, and comment on the implications for the classification of 1D phases. We will then analyze the case of MPOs, following \citet{cirac:mpdo-rgfp}, and show how a fusion category emerges from the RGFP conditions, which sheds light on the classification of 2D phases via the bulk-boundary correspondence analyzed in Section \ref{secESEM}. We will also  comment on RGFPs in higher dimensions and concrete renormalization flows in tensor networks.

\subsubsection{Renormalization Fixed Points in MPS \label{RFPinMPS}}

As just discussed, we should identify properties that any RGFP MPS must have. For that, given the ground state subspace $S$, of a local Hamiltonian $H$, we say that it has \emph{zero correlation length} if connected correlation functions vanish. That is, for any state $\Psi\in S$, and any two observables $A$ and $B$ acting on not neighboring regions (i.e.\ not directly connected by the action of $H$),
 \be
 \langle \Psi|A B|\Psi\rangle- \langle \Psi|A P_S B|\Psi\rangle=0
 \, 
 \ee
where $P_S$ is the projector onto $S$. Note that, according to this definition, a GHZ state has zero correlation length.

As normal MPS have a finite correlation length, 
and general MPS can be expressed as superpositions 
of normal MPS, it should happen that as we block, the correlation length decreases. Thus, RGFP must have zero correlation length. We will then say that a MPS is a RGFP if it has zero correlation length.
Since the transfer matrix $E$ of an MPS is the operator that mediates the correlations in the system (Section \ref{sec2TN3}), 
zero correlation length is equivalent to the condition
\be
 \label{E2=E}
 {E}^2={E}\ .
 \ee
 
There are other notions which are clearly connected to RGFP which can be shown to also be {\it equivalent} to zero correlation length in MPS \cite{cirac:mpdo-rgfp}, and can be taken then as alternative (but equivalent) definitions of RGFP for MPS.
The first one is the fact that the parent Hamiltonian can be written as sums of local terms which mutually commute. It has been recently shown by \citet{kastoryano:martingale} that if the parent Hamiltonian of a PEPS is gapped, then the norm of the commutator of the associated terms goes to zero as we coarse-grain the system (see Section~\ref{sec:4}). One can then expect that  RGFP PEPS (and in particular RGFP MPS) have commuting parent Hamiltonians. 
This property is indeed equivalent to RGFP,
as shown in \textcite{cirac:mpdo-rgfp}.

The second equivalent notion is the \emph{saturation of the area law}.
 Strong subadditivity of the von Neumann entropy \cite{lieb1973proof} implies that for a spin chain of size  $N$ and a region of size $L< N/2$,
 $S^{(N)}_{L+1}\ge S^{(N)}_L$,
 where $S^{(N)}_{L}$ denotes the entanglement entropy of that region. 
 Since MPS fulfill the area law, i.e.\ $S^{(N)}_{L}$ is upper bounded by a constant, independent of $L$ and $N$,  it follows that $\lim_{L\to \infty} S_L^{\infty} = c<\infty$. This implies that RGFPs must satisfy $S_L^{(N)}=c$ for all $L$ and thus in particular for $L=1$. That is, they must saturate the area law. As stated above, the converse is also true.

Let us now focus on property (\ref{E2=E}) and see what can be concluded from there. 

First of all, since different Kraus representations of a completely positive map must be related by an isometry \cite{stinespring1955positive,wolf:channel-notes}, a tensor $A$ corresponds to an MPS RGFP if and only if \cite{verstraete:renorm-MPS}
 \be
 \label{RFPPure}
 A^{i_1} A^{i_2} = \sum_{i_1,i_2} U_{(i_1,i_2),j} A^j
 \ee
for some isometry $U$. Graphically,
 \be
 \label{AA=UA}
 {\includegraphics[height=3.5em]{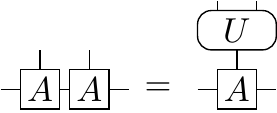}}
 \ee
That is, the MPS given by $A$ is the renormalization fixed point of a particular type of flow, obtained by acting with an isometry on the physical degrees of freedom.

Such a flow has a natural interpretation in the light of the usual definition of topological phases in quantum systems. Two ground states, or particularly two PEPS, are said to be in the same topological phase precisely if there is a low-depth local quantum circuit converting one state into the other. Via the quasi-adiabatic theorem \cite{hastings:quasi-adiabatic, bachmann:quasi-adiabatic}, this corresponds to the existence of a continuous gapped path of Hamiltonians connecting the two systems.  The intuition behind this definition is that in order to go to a different phase, one needs to  generate global (topological) correlations, which requires 
a time (i.e., circuit depth) which scales with the system size.

The flow described in Eq.~\eqref{AA=UA} above keeps a state in the same phase. If the unitary implemented in the renormalization flow aims to disentangle the left and right ends of a block, one expects only either nearest neighbor or purely global entanglement to remain in the limit. Indeed, if the
tensors are normal, the RGFP condition  ${E}^2={E}$ implies that ${E}$ is a rank one projector, and then (using the isometric relation between Kraus representations) we can split each spin at a given site $n$ into a left and a right system $n_{\ell}$ and $n_r$, such that the structure of the RGFP state 
up to local isometry is of the form
 \be
 \label{fixedMPS}
 |\Phi\rangle = \otimes_{n=1}^N|\varphi\rangle_{(n-1)_r,n_\ell},
 \ee
where $\ket\varphi$ is an entangled state defined on the right and left part of neighboring spins. If the state is not normal, then one has a direct sum of states of the form (\ref{fixedMPS}), where the terms in the sum are locally orthogonal (meaning that the corresponding $\ket\varphi$ are supported on orthogonal subspaces for each of the spins).

These states provide representatives for all possible phases of matter for closed 1D systems; see Section \ref{sec:3:1D-SPT}.

\subsubsection{MPDOs\label{sec:2:MPDO-RGFP}}

For the case of mixed states we follow a similar approach. Similarly to (\ref{E2=E}), in this context we say that a MPDO associated with tensor $M$ has zero correlation length if 
\be\label{Fig:ZCL-MPDO}
 {\includegraphics[height=2.5em]{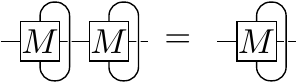}} 
\ee

As opposed to the pure state case, this is not enough to guarantee that a given MPDO has no length scale associated with it. In particular, this does not neccessarily imply that the MPDO fulfills a property analogous to the saturation of the area law. As mentioned earlier, in the context of mixed states the notion of an area law refers to the mutual information, instead of the entanglement entropy \cite{wolf:mutual-info-arealaw}. That is, a system is said to satisfy an area law for the mutual information if the mutual information between a region $R$ and its complement $\bar R$, $I(R:\bar R)=S(\rho_R)+S(\rho_{\bar R})-S(\rho_{R\bar R})$, can be bounded by the number of spins at the boundary of $R$ (up to a multiplicative constant). 
It has been shown that thermal states of short range Hamiltonians 
\cite{wolf:mutual-info-arealaw}
as well as fixed points of fast mixing Linbladians 
\cite{brandao:area-law-linbladians} 
fulfil an area law for the mutual information, which therefore 
characterizes the relevant corner of the Hilbert space for equilibrium states in analogy to the area law for the entanglement entropy for ground states.

Similar to the case of MPS, in any MPDO the mutual information between a region $R$ and its complement is upper bounded by a constant which only depends on the bond dimension, but not the system size or the size of $R$. Moreover, an analogous argument based on strong subadditivity implies that the mutual information increases monotonously with the size of the region $R$. Hence, just as before, we expect MPDO RGFPs to saturate the area law for the mutual information. As shown in \citet{cirac:mpdo-rgfp}, this condition is however not equivalent to zero correlation length.  In order to characterize RGFPs in MPDOs, we therefore need to impose both conditions independently:   We therefore say that an MPDO is an RGFP if it has both zero correlation length, and saturation of the area law for mutual information.

It can be proven that RGFP MPDOs are characterized (up to a technical condition) by the existence of two trace-preserving completely positive maps (quantum channels)  $\mathcal{T}$ and $\mathcal{S}$ such that
\be
\label{Fig:TandS}
 {\includegraphics[height=4.5em]{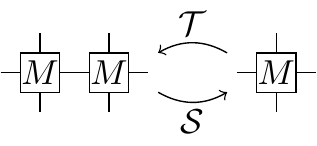}} 
\ee
It is immediate to see that taking traces in Eq.~(\ref{Fig:TandS}) implies zero correlation length. Moreover, since (\ref{Fig:TandS}) allows to grow or reduce a region by acting locally on it, it also implies  saturation of the area law. On the other hand, the fact that  both conditions together
imply (\ref{Fig:TandS}) is far less obvious \cite{cirac:mpdo-rgfp}.
We thus see that in analogy to the case of MPS, imposing RGFP conditions related to the absence of length scales gives rise to a particular type of RG flow for which the given MPDO is a fixed point. Here, the flow consists of blocking a finite number of sites and implementing a renormalizing quantum channel on the blocks whose action can be inverted. 
One can use this type of RG flow to define phases for mixed states, in analogy to the MPS case discussed above:
Two mixed states are said to be in the same phase if there exists a low-depth circuit of quantum channels that can map one state into the other \cite{coser:phases-mixed}. 

Just as in the MPS case, there is also a result which characterizes the structure of  RGFP MPDOs. Namely, it turns out that RGFP MPDOs generate a finite dimensional algebra of Matrix Product Operators, in the following sense:
Consider an MPDO generated by a tensor $M$ (obtained by contracting the tensors horizontally, as in Fig.~\ref{sec2fig5}), and consider the
MPO 
 \be
\label{Fig:MPDO-O_L}
{\includegraphics[height=6em]{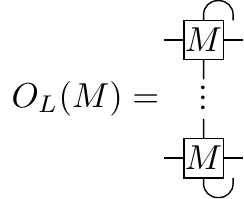}}  
 \ee
generated by $M$ by contracting the same tensors 
\emph{vertically} on a ring of length $L$. 
Assume w.l.o.g.\ that this MPO is in canonical form, 
i.e.\
 \be
  M  = \bigoplus_\alpha \mu_\alpha M_\alpha\ ,
 \ee
where the $M_\alpha$ are the different injective blocks (we do not include the case of multiple blocks, which can be treated in a similar way, see \onlinecite{cirac:mpdo-rgfp}).
Then, the given MPDO is an RGFP if and only if there exists 
a set of diagonal matrices $\chi_{\alpha,\beta,\gamma}$ with positive entries such that for each $L$, the operators $O_L(M_\alpha)$ linearly span an algebra with structure coefficient $c_{\alpha,\beta,\gamma}^{(L)}={\rm tr}(\chi^L_{\alpha,\beta,\gamma})$, i.e.
 \be
 \label{eq:algebra}O_L(M_\alpha)O_L(M_\beta)=\sum_\gamma c_{\alpha,\beta,\gamma}^{(L)}O_L(M_\gamma)
 \ee
 and
  \be
 \label{idempotent}
 \mu_\gamma = \sum_{\alpha,\beta} c^{(1)}_{\alpha,\beta,\gamma} \mu_\alpha \mu_\beta\ .
 \ee
That is, the vector $(\mu_\alpha)_\alpha$ is an idempotent for the ``multiplication'' induced by $c^{(1)}$. 

In case the structure coefficients $c_{\alpha,\beta,\gamma}^{(L)}={\rm tr}(\chi^L_{\alpha,\beta,\gamma})$ are independent of $L$, one can easily show that the $\chi_{\alpha,\beta,\gamma,k}\in \{0,1\}$  and therefore $c_{\alpha,\beta,\gamma}^{(L)}\in \mathbb{N}$. In this case, one can further show that the RGFP MPDOs generated by $M$ can be written as
 \be
 \rho^{(N)}(M)=\sum_{i=1}^d \lambda_i P^{(N)}_i e^{-H_N}
 \ee
where $d$ is the local Hilbert dimension of a single site, $P_i^{(N)}$ are projectors, $H_N=\sum_{i=1}^Nh_{i,i+1}$ is translationally invariant, nearest-neighbor and commuting $[h_{i-1,i},h_{i,i+1}]=0$, and $[P_i,e^{-H}]=0$ for all $i$.

This result establishes a connection with the boundary theories of topological PEPS in two dimensions, proving rigorously the desired structure for the boundary theory for RGFP (see Section \ref{bulk-boundary}): A global projector selecting the topological sector and a local boundary Hamiltonian commuting with it. At the same time,  it concludes -- based only on a natural RGFP condition -- the existence of an algebra of MPOs which, as will be explained in Section~\ref{sec:3:2D}, is the starting point to obtain the most general class of non-chiral topological models in two dimensions.

\subsubsection{Tree Tensor States and MERA} 
For a general quantum state describing a spin chain, it is possible to devise a multitude of renormalization processes. The simplest one is the real space renormalization, in which one joins a block of spins and truncates the corresponding Hilbert space to build a new one. It is straightforward to check that the concatenation of this procedure gives rise to a tree tensor state (TTN),  where each layer is characterized by the truncation map (see Section~\ref{TTNMERA}). As explained above, it is possible to choose these maps as isometries. The renormalization flow can be interpreted as the sequence of isometries corresponding to each level of the renormalization. The sequence may converge, so that one could define 
these TTN where all the isometries are the same
as fixed points of the RG flow.
These states give rise to a logarithmic violation of the area law, and have (averaged) correlation functions which in general decay as a power law.

A much more sophisticated  and comprehensive way of performing such a renormalization consists in including unitary operators (``disentanglers'') acting on neighboring spins before each step of the above procedure. The resulting states are the Multiscale Entanglement Renormalization Ansatz (MERA) \cite{vidal:mera}, which include more correlations than TTN as the tensor network now contains loops (see Section~\ref{TTNMERA}). Furthermore, the procedure can be applied to higher spatial dimensions with an area law scaling of the entanglement, something which is not possible with TTN.

The renormalization procedure in terms of MERA was introduced by \citet{vidal:mera}, where he interpreted the unitaries and isometries as ways of disentangling neighboring spins at each step. In fact, by reading the MERA or the TTN the other way round, one can see that they can be generated out of a product state by applying unitary operators.

As compared to previous procedures, MERA and TTN are more appropriate to describe critical states, where the correlation length diverges. In fact, given their structure one can extract properties of the Conformal Field Theory describing the critical behavior of the state, such as the form of the primary fields or their scaling dimensions \cite{pfeifer:conformal-MERA,giovannetti:conformal-MERA,milsted2017extraction,zou2018conformal}.

\subsubsection{RG in higher dimensions}

The renormalization procedures reviewed above can be extended to higher dimensions using PEPS. In principle, one can look for unitary operators acting on blocks of spins (in plaquettes, for instance) that disentangle some of them locally. This procedure will not work as well for PEPS as it does for MPS since in spatial dimension larger than one, blocking increases the bond dimension, as a direct reflection of the area law. Thus, one might want to choose different approaches. 

The most natural one is to truncate the states by replacing the unitary operator by an isometry to obtain a tree tensor network or adding disentangling unitaries to obtain a MERA \cite{evenbly:2d-mera}. If one considers tensor networks without physical degrees of freedom such as classical partition functions, another approach consists of replacing several tensors corresponding to neighboring spins by a single tensor but making sure that the tensor, in some way, generates a tensor network that is close to the original one \cite{evenbly:tensor-renormalization, gu:TERG, yang:loop-renormalization,bal2017renormalization}. Although these procedures may be useful as numerical tools, the state obtained at the end will in general not be the same as the original one. 

A way around is to look for fixed points of such types of renormalization procedures. The corresponding nontrivial fixed points turn out to form representative states for phases exhibiting topological quantum order \cite{wen2017colloquium}. As pioneered in \citet{dennis2002topological},  qubits in the toric code \cite{kitaev:toriccode} can be disentangled with local unitaries, and therefore the corresponding fixed point topological states can be represented in terms of a quantum circuit of isometries and unitaries. Essentially the same construction was used by  \citet{aguado:MERA-toric, konig:MERA-string-nets} to represent all quantum doubles \cite{kitaev:toriccode} and string nets  \cite{levin:stringnets}  as fixed points of renormalization flows in the form of a MERA. All those models have zero correlation length and are the ground states of frustration free Hamiltonians with local commuting terms. The ground states of those models can hence be obtained by projecting a product state on the ground subspaces of all those local Hamiltonian terms; such a construction generates a simple PEPS description for the ground states of such fixed point Hamiltonians \cite{verstraete:comp-power-of-peps}. This PEPS representation was worked out for string nets in \cite{gu2009tensor} and its emerging MPO symmetries were studied in \cite{schuch:peps-sym,sahinoglu:mpo-injectivity} and presented in Sec. \ref{sec:3:2D}.

\begin{figure}
\includegraphics[width=8cm]{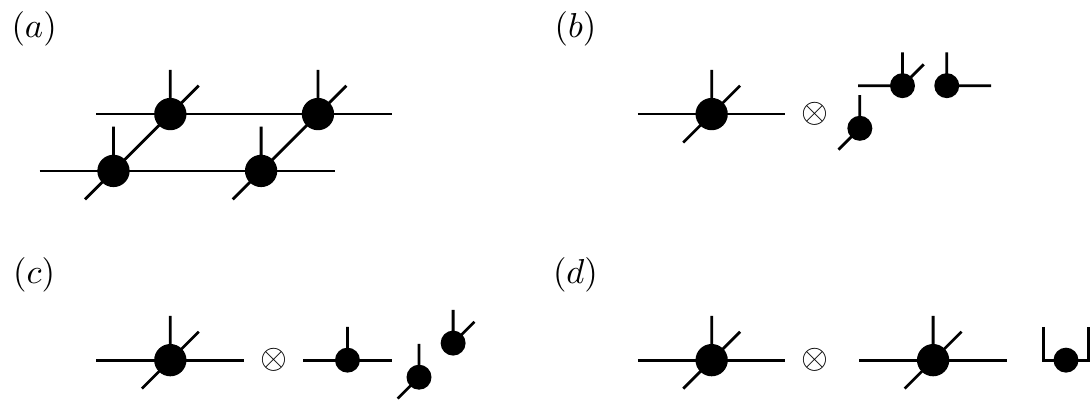}
\caption{Examples of different possibilities for RGFP in a 2D square configuration, where a unitary operation is applied to the physical indices of four spins (a) and invertible operations are applied to the auxiliary ones. (b) One can disentangle several spins, i.e. obtain the original tensor and other ones as  product states; (c) one can obtain the original tensor and other generating MPS; (d) one can obtain two copies of the original tensor}    \label{FigPEPSRG}
\end{figure}

From a more general perspective, one can consider renormalization fixed point equations for PEPS as, for instance, these shown in Fig.~\ref{FigPEPSRG}, and write down the corresponding (non-linear) stationary equations which fully characterize these PEPS which can be considered as RGFPs with respect to that property. Realizing such a program would involve nontrivial results from algebraic geometry, but has not been realized yet in full generality.

An alternative  method in two spatial dimensions consists of using the bulk-boundary correspondence reviewed in Section \ref{sec:boundary-theory}. Using this correspondence, the physical properties of a PEPS are characterized by a density operator that lives at the boundary. Let us next consider a PEPS on a cylinder. 
In the course of renormalizing the PEPS using any RG procedure, 
the boundary state itself will be renormalized as well, and as the RG fixed point is reached, one expects to obtain an RGFP MPDO 
\cite[][cf.~Sec.~\ref{sec:2:MPDO-RGFP}]{cirac:mpdo-rgfp}
at the boundary as well.
This RGFP condition at the boundary leads to an emerging  algebra of MPOs. We will discuss in Section \ref{secESEM} below how, from the existence of such algebra, one recovers in a direct way all known 2D non-chiral topologically ordered phases together with a description of their anyon excitations. The point of view of the characterization of renormalization group fixed point in terms of boundary density operators is therefore equivalent to the one in terms of disentangling circuits acting on the bulk, and both give rise to quantum double models and string nets.

\subsubsection{The limit to the continuum}

A natural question is whether one can also implement a process inverse to renormalization in the context of tensor networks.
That is, instead of coarse-graining the lattice to distill the global entanglement pattern, fine-graining it to obtain a meaningful continuum limit. 

Let us first discuss this for 1D, following  
\citet{verstraete2010continuous,delascuevas:mps-cont-limit}. 
As seen in Section \ref{sec:2:CorrsTOp}, MPS are characterized 
by a quantum channel $E$ (their transfer operator), up to a local basis change.
A blocking step of $r$ sites simply corresponds to taking the transfer operator to the $r$'th power, $E^r$,
as used in Section~\ref{RFPinMPS}.
A fine-graining step would therefore correspond  to taking integer roots of $E$. However, this is subtle as there exist quantum channels $T$ that cannot be divided \cite{wolf2008dividing}, in the sense that there does not exist any other quantum channel $R$ so that $R^2=T$. In order to guarantee a well defined continuum limit, one needs to require that the transfer operator is {\it infinitely divisible}, meaning that any posible integer root exists~\cite{delascuevas:mps-cont-limit}. This in turn is equivalent, up to a projector $P$ commuting with the given channel $E$, to the existence of an infinitesimal generator $\mathcal{L}$ which generates a semigroup $e^{t\mathcal{L}}, t\ge 0$, that interpolates the initial $E$ (which corresponds to $t=1$) all the way back to $t=0$
\cite{holevo:infdiv,denisov:infdiv}.
The state obtained by taking $t\rightarrow 0$ in this fine-graining
process is precisely the cMPS discussed in Section~\ref{FandCTN}.

\begin{figure}
\includegraphics[width=8.5cm]{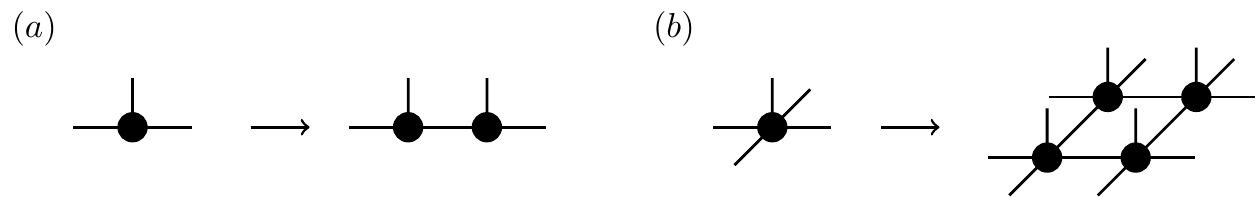}\\
\caption{The inverse renormalization procedure. (a) For MPS, in case $E$ is divisible, one can rewrite the tensor in terms of two tensors of the same bond dimension and iterate the procedure; (b) For PEPS in two dimensions, the same step would imply that the bond dimension of the new tensor has to be square rooted}    \label{fig:renorm2D}
\end{figure}

The above procedure cannot be easily extended to higher dimensions. The reason is that the inverse renormalization process should produce tensors with non-integer bond dimensions (as bond dimensions multiply when blocking), which is impossible. For instance, in 2D one should get from bond dimension $D$ to tensors of bond dimension $\sqrt{D}$ (see Fig.~\ref{fig:renorm2D}). At some point of the iteration, the square root will not be an integer, so that the procedure cannot work. The only way around is if in some sense, ``$D=\infty$''. In fact, continuous PEPS can be defined in this way, for instance in terms of path integrals where the discrete auxiliary indices of the tensors are replaced by functions which are integrated over when being contracted \cite{jennings2015continuum,tilloy2019continuous}.

\section{Symmetries and classification of phases \label{sec:3}}

Symmetries are a main guiding principle in quantum many-body physics, and the situation is no different for tensor networks. In fact, one of the main reasons for the success of tensor networks is precisely the fact that they make the role of symmetries in many-body systems so explicit: a quantum state described by an MPS or PEPS $|\psi_N\rangle$  will be invariant under a global symmetry $U^{\otimes N}|\psi_N\rangle$ if and only if all local tensors transform trivially under that symmetry. As a consequence, any global symmetry, including symmetries associated to topological order, will be reflected in the local symmetries of the tensors describing the many-body states. Phrased differently,  the entanglement spectrum acts like a signature of those symmetries. This yields a unifying principle for describing distinct gapped phases of matter, including topological ones for which there is no distinct local order parameter in the sense of \citet{landau1937theory}: distinct phases of matter can be distinguished by the different ways in which the local tensors transform under the global symmetries. The local tensors hence provide a non-trivial generalization of the notion of a local order parameter, and reduce the problem of classifying different gapped phases of matter to a problem in the representation theory of groups and algebras. It is a well known fact that there are certain topological obstructions to convert tensors, which transform according to different representations of the same group, into each other continuously. Those obstructions are precisely the ones responsible for the existence of topological quantum order.

One of the big success stories of many body physics has certainly been the realization that global symmetries can be lifted to local ones by introducing new ``gauge'' degrees of freedom. Such a procedure can also be carried out in the language of tensor networks, and gives rise to tensors with an increased intrinsic symmetry action on the entanglement degrees of freedom. The ensuing gauge theories exhibit fascinating properties such as excitations with anyonic statistics and non-trivial edge modes, and the fact that such features directly follow from the symmetry properties of the local tensors makes tensor networks a natural framework for describing and exploring quantum topological order. In fact, it can be argued that tensor networks implement the representation theory of braided fusion categories, which form the foundation of both topological and conformal field theories. 

This section is divided in two parts. The first part discusses symmetries of matrix product states, and is hence concerned with the classification of phases of quantum spin chains. The second part discusses symmetries of projected entangled pair states, including the case of quantum topological order. In both cases, we will limit the discussion to uniform (translationally invariant) systems.

\subsection{Symmetries in one dimension: MPS}
\subsubsection{Symmetric MPS \label{sec:3:sym}}

Normal (injective) uniform matrix product states exhibit the remarkable property that two states $|\psi(A)\rangle$ and $|\psi(B)\rangle$ are equal to each other if and only if there exists a gauge transform $X$ and a phase $\chi$ for which $A^i=e^{i\chi}X^{-1}B^iX$ \cite{perez-garcia:stringorder-1d}. If both $A^i$ and $B^i$ are in canonical form, then $X$ is guaranteed to be unitary due to the uniqueness of the fixed point. This is a consequence of the fundamental theorem of MPS, and will be discussed at large in Section~\ref{sec:4}. A useful feature is the following property of a normal MPS: 
\be X^{-1}A^iX=e^{i\chi}Y^{-1}A^iY\Rightarrow \chi=0\land\exists\phi: X=e^{i\phi}Y \ .
\label{eq:XAX=B}\ee 

Furthermore, any translationally invariant normal MPS has a uniform representation, i.e. the tensors $A^i$ do not depend on the site label. This property has very strong consequences for MPS that are invariant under global on-site, reflection and time-reversal symmetries: it implies that the tensors building up an MPS with global symmetries must themselves transform trivially up to a phase 
 under that symmetry. To illustrate this, let us consider the case of an MPS  in canonical form which  is invariant under a global on-site symmetry group $G : U(g)^{\otimes N}|\psi_N\rangle\simeq |\psi_N\rangle$. It follows from Eq.~(\ref{eq:XAX=B}) that
\be \sum_{j}U_{ij}(g)A^j=e^{i\phi(g)}X^{\dagger}(g)A^iX(g)\ . \ee
In other words, the 3-leg MPS tensor $A^i$ written as a vector in a vector
space of dimension $d.D^2$ transforms  trivially under the action of
$e^{-i\phi(g)}U(g)\otimes X(g)\otimes \bar{X}(g)$ with $\bar{X}(g)$ the
conjugate. This implies that the tensor $A^i$ can be written in terms of
Clebsch-Gordan coefficients of irreducible representations of the group
$G$ and the variational degrees of freedom can be incorporated by adding
multiplicities \cite{mcculloch2002non,singh:tns-decomp-symmetry,sanz:mps-syms,weichselbaum2012non,white1993density}.
 Such decompositions have been used since long and with great success in the context of DMRG. There are two non-trivial facts that one can conclude from that.  

First, the condition of injectivity in conjunction with translational invariance imposes constraints on the type of symmetries which can be realized in an MPS. This is best illustrated by an example. Let us consider a spin-$\tfrac12$ system with $\SU(2)$ symmetry. As the physical spin transforms according to a half integer representation, the Clebsch Gordan coefficients impose that the virtual irreps alternate between integer and half-integer representations \cite{sanz:mps-syms}. By blocking two sites, the MPS matrices $A^i$ therefore  exhibit two invariant subspaces, and hence the MPS cannot be normal, which shows that no uniform normal/injective  MPS can exhibit such a symmetry. As will be discussed later, this is the tensor network manifestation of the Lieb-Schultz-Mattis theorem.

Second, there is no need for the irreps on the virtual degrees of freedom to form representations of the group $G$: it is perfectly fine if they transform according to projective representations, that is, representations up to a phase
\be X_gX_h=e^{i\omega(g,h)}X_{gh}\ , \ee
as such phases leave $e^{i\phi(g)}U(g)\otimes X(g)\otimes \bar{X}(g)$ invariant \cite{pollmann:1d-sym-protection-prb,chen:1d-phases-rg,schuch:mps-phases}. As an authoritative example, if the physical system transforms according to $\SO(3)$, the virtual systems can either both transform according to half-integer or integer representations of $\SU(2)$, as readily seen from the Clebsch-Gordan coefficients; the half-integer representations of $\SU(2)$ form a projective representation of $\SO(3)$, with phases $\omega(g,h)=$ $0$ or $\pi$. 

For a given group $G$, it is a relatively easy task to find all possible projective representations, as the associativity of matrix multiplication heavily constrains the possible $\omega(g,h)$.  If the group is finite, this can be achieved by using the Smith normal form, which is similar to the Schmidt decomposition but with integer arithmetic. The following picture emerges \cite{chen:spt-order-and-cohomology}: for a given group $G$, the different projective representations fall into equivalence classes, where in a given equivalence class, the projective representations are related to each other by simple phases : $X_g=\exp(i\phi(g))\tilde{X}_g$. Those phases are clearly irrelevant from the point of view of MPS as they cancel, and hence only the equivalence classes count. Those classes are classified according to the second cohomology group $H^2_\alpha(G,U(1))$ with group action $\alpha$ (this action will be non-trivial when we  consider time-reversal and reflection symmetries), and are classified according to the solutions of the equation
\be \alpha_g(\omega(h,l))- \omega(g.h,l)+\omega(g,h.l)-\omega(g,h)=0 \mod 2\pi\ee
obtained by imposing associativity relations on the projective representations. The action $\alpha$ is a homomorphism from $G$ to the automorphism group $Z_2$ of $U(1)$, and hence consists of $\pm$. In the case of global symmetries excluding reflection and time-reversal, it is just the identity map: $\alpha_g(x)=x$ for all $g\in G$. As follows from the Smith normal form, there are only a finite number of such equivalence classes for a finite group, and these are labeled by integer and hence topological indices. 

In the case of a continuous (semisimple) Lie group $G$, irreducible projective representations are in one to one correspondence to irreducible linear representations of its universal covering group $C$, from which $G$ is then obtained by modding out a subgroup $Z_s$ of its center $Z$:   $G=C/Z_s$. Note that the relation between $\SU(2)$ and $\SO(3)$ discussed above is precisely of this form. In that example, half-integer representations of $\SU(2)$ correspond to projective non-linear representations of $\SO(3)$. For compact groups such as $\SO(3)$, this yields also a finite number of different inequivalent classes of projective representations (i.e. the second cohomology group is finite).

Before studying the remarkable physical implications of those projective representations, let us generalize the discussion to include time-reversal and/or reflection symmetries. A symmetry $\mathcal{G}$ of the system can be labeled by a subset of the tuples $x=(g,t,r)\in G\rtimes Z_2^T\times Z_2^R$ where $g$ denotes the physical group action and $t,r$ $\in$ $0,1$ denote the linear representation of time-reversal and reflection. From now on we will consider $\mathcal{G}$ of the form $G\rtimes H$ with $H$ a subgroup of $Z_2^T\times Z_2^R$.

Let us start with  discussing a  normal MPS in canonical form invariant under pure time-reversal symmetry represented by the tuple $x=(1,1,0)$ corresponding to the morphism $S_x(A^i)=\bar{A}^i$, it is elements-wise conjugation. As this is a symmetry of the system, the fundamental theorem imposes that there exists a $X_x$ and $\phi(x)$ such that $\bar{A}^i=e^{i\phi(x)}X^\dagger_xA^iX_x$. As conjugating a tensor twice yields the original tensor, we must have  
$A^i=S_x(S_x(A^i))=S_x(e^{i\phi(x)}X^\dagger_xA^iX_x)=e^{-i\phi(x)}\bar{X}_x^\dagger\bar{A}^i\bar{X}_x=(X_x\bar{X}_x)^\dagger A^i (X_x\bar{X}_x)$. By Eq.~\eqref{eq:XAX=B}, this is only possible for $X_x\bar{X}_x=\pm\Id$, and as discussed in the next section $\pm 1$ is a topological index. Note that if we had defined time-reversal symmetry in the form $x=(\sigma_y,1,0)$, as encountered in Wigner's discussion on time-reversal in spin-$\tfrac12$ systems, we would have obtained: $A^i=S_x(S_x(A^i))=(X_x\bar{X}_x)^\dagger A^i (X_x\bar{X}_x)=\sigma_y.\overline{\sigma_y\bar{A}^i}=-A^i$, which is in violation with Eq.~\eqref{eq:XAX=B}. This implies that ground states of systems exhibiting such a symmetry cannot be represented by a normal/injective MPS or, phrased differently, cannot be unique ground states of a local gapped Hamiltonian. This turns out to be the tensor network analogue of the famous Kramers theorem on time-reversal. From the mathematics point of view, the obstruction follows from the fact that the physical symmetry $\sigma_y$ acts as a projective representation under time-reversal. 

As has become clear by now, to discuss global symmetries labeled by $x_i=(g_i,t_i,r_i)$ and $x_3=x_1\circ x_2$, we need to define the actions of the symmetries on MPS tensors in the form $S_x\left(e^{i\phi}X^\dagger A^i X\right)=e^{i\chi^1_x(\phi)}\chi^2_x(X)^\dagger A^i \chi^2_x(X)$:  
here we introduced the functions $\chi^1_x(\phi)$ as the morphism on the phase $\phi$ and $\chi^2_x(X)$ the morphism on the gauge $X$ induced by element $x$.  As discussed before, if $x=(g,1,0)$ involves time-reversal, then $\chi^1_x(\phi)=-\phi+\phi(x)$ and $\chi^2_x(X)=X_x.\bar{X}$ with $\phi(x)$ and the gauge tensor $X_x$ depending only on the group element $x$. Similarly, a reflection $x=(g,0,1)$ in terms of MPS is implemented by taking the transpose of the matrices involved: $S_x\left(e^{i\phi}X^\dagger A^i X\right)=e^{i\phi}X^T (A^i)^T \bar{X}=e^{i(\phi+\phi(x))}X^T X_x^\dagger A^i X_x\bar{X}$. Hence in this case $\chi^1_x(\phi)=\phi+\phi(x)$ and $\chi^2_x(X)=X_x.\bar{X}$. If a group element involves simultaneous time-reversal and reflection $x=(g,1,1)$, then $\chi^1_x(\phi)=-\phi+\phi(x)$ and $\chi^2_x(X)=X_x.X$. The fundamental theorem of MPS imposes that $\chi^1_x$ forms a linear representation of the group $\chi^1_{x_1}\circ\chi^1_{x_2}=\chi^1_{x_1.x_2}$. This imposes a non-trivial constraint on the phases $\phi(x)$: 
\[(-1)^{t_1}.\phi(x_2)-\phi(x_1.x_2)+\phi(x_1)=0\mod 2\pi\]
with $t_1=1$ iff $x_1$ involves time-reversal. This is precisely the defining equation for a 1-cocycle of the first cohomology group $H^1_\beta(\mathcal{G},U(1))$ with group action $\beta_{x}(\phi)=(-1)^{t(x)}.\phi$. For finite groups and for compact semisimple Lie groups there are only a finite number of distinct cocycle solutions of this equation modulo the trivial co-boundary solutions. Indeed, the equivalence class of a cocycle is obtained by adding a co-boundary to it: $\phi(x)\rightarrow \phi(x)+(\beta_x(c)-c)$ for any constant $c$, and all such solutions are indistinguishable from the point of view of MPS. 

Similarly, the fundamental theorem implies that the morphism $\chi^2_x(.)$ must form a projective representation of $\mathcal{G}$: $\chi^2_{x_1}\circ \chi^2_{x_2}=\exp(i\omega(x_1,x_2))\chi^2_{x_1.x_2}$. Imposing associativity in the form $\chi^2_{x_1}\left(\exp(i\omega(x_2,x_3))\chi^2_{x_2.x_3}(X)\right)=\exp(i\omega(x_1,x_2))\chi^2_{x_1.x_2}\left(\chi^2_{x_3}(X)\right)$ leads to the following condition on the phases $\omega(x_1,x_2)$: 
\begin{align*}
(-1)^{(t_1+r_1)}\omega(x_2,x_3)-\omega(x_1.x_2,x_3)&\\
+\omega(x_1,x_2.x_3)-\omega(x_1,x_2)&=0\mod 2\pi
\end{align*}
This is precisely the defining equation for a 2-cocycle of the second cohomology group $H^2_\alpha(\mathcal{G},U(1))$ with group action $\alpha_x(\phi)=(-1)^{(t(x)+r(x))}\phi$. Given a finite group $G$, the finite number of equivalence classes of this equation can again be found explicitly by making use of the Smith normal form. In this case, the co-boundaries correspond to $\omega(x,y)\rightarrow \omega(x,y)+\alpha_x(\xi(y))-\xi(x.y)+\xi(x)$ for any function $\xi(x)$. $H^2_\alpha(\mathcal{G},U(1))$, as it is the case also for $H^1_\beta(\mathcal{G},U(1))$, is itself an (abelian) group and hence of the form $Z_n\times Z_m\times \dots$.

In summary, we have seen that all global symmetries of uniform normal (injective) MPS wavefunctions are reflected in the local tensors $A^i$ of the MPS; these symmetries can be represented projectively on the virtual level, and the classification of all possible ways in which this can occur can be obtained by solving the (integer) linear algebra problem of finding all 1- and 2-cocycles of the group of interest.

\subsubsection{SPT phases and edge modes\label{sec:3:1D-SPT}}
\paragraph{Symmetry protected topological order}

The way global symmetries are reflected on the local MPS tensors has very strong implications for the classification and description of phases of matter of 1D spin systems. The fundamental idea underlying this classification is the fact that the unique ground state of any local gapped quantum spin chain has an efficient representation in terms of a normal (injective) MPS.  This implies that the classification of gapped phases can be done  on the level of MPS as opposed to Hamiltonians, which is a huge simplification. 

Given two translationally invariant normal MPS parametrized by tensors $A^i$ and $B^i$, it turns out that there always exists an interpolation between them (even if they have a different bond dimension) for which all intermediate MPS are also injective: there is no topological obstruction for constructing such a path, and hence there only exists one phase for gapped quantum spin systems \cite{chen:1d-phases-rg,schuch:mps-phases} (note that the situation changes in the case of fermions, as will be discussed in \ref{sec:3:fermions}). This problem is equivalent to constructing an interpolating path $A^i(t)$ for which the transfer matrix $E=\sum_i A^i(t)\otimes \bar{A}^i(t)$ has a unique largest eigenvalue (which is guaranteed to be real). As demonstrated in \citet{schuch:mps-phases}, this can be achieved in three steps.  First, we block different sites until the tensor $A^{i_1}A^{i_2}...$ is injective, and then apply a filtering operation to bring this blocked MPS into the form of a renormalization group fixed point while keeping the unique largest eigenvalue property. Second, we can readily interpolate between any two of such dimer-type wavefunctions by a local quantum circuit without closing the gap. In the third step, we apply another filtering operation to obtain the tensor $B^i$. The corresponding parent Hamiltonian $H(t)$ is guaranteed to be gapped along the path, demonstrating that any two injective MPS are in the same phase. For a generalization of this theorem without the blocking step, see \cite{szehr2016connected}.

The situation changes drastically when symmetry constraints are imposed on the adiabatic path and hence on the MPS: a much smaller dimensional manifold can then be traversed during the interpolation, and topological obstructions might occur. Colloquially speaking, the submanifold of all normal MPS with a given symmetry decomposes into disconnected components. As a consequence, any symmetry preserving interpolating path between two states living on different components has to pass through a phase transition, and at this point the interpolating MPS will not be injective anymore \cite{chen:1d-phases-rg,schuch:mps-phases}. These different components correspond to different symmetry protected topological phases of matter (SPT phases). They are protected by translational invariance, on-site symmetries, time-reversal symmetry and/or reflection invariance.

The necessary mathematical framework for demonstrating this has been developed in Section~\ref{sec:3:sym}. By studying how an MPS $A^i$ transformed under one or a combination of the above symmetries $S_x(A^i)=\exp(i\phi(x))X_x^\dagger A^i X_x$, it was shown that $\phi(x)$ had to be a 1-cocycle of $H^1_\beta(\mathcal{G},U(1))$ and that we could associate to the gauge matrices $X_x$ a map (morphism) $\chi^2_x(X)=X_x.X$ or $\chi^2_x(X)=X_x.\bar{X}$ (dependent on whether time-reversal and/or reflection invariance is involved)  which itself forms a projective representation of the physical symmetry group: $\chi^2_{x_1}.\left(\chi^2_{x_2}(.)\right)=\exp(i\omega(x_1,x_2))\chi^2_{x_1.x_2}(.)$. The corresponding phases are characterized by the topologically distinct 2-cocycles $\omega(x_1,x_2)$ of the second cohomology group $H^2_\alpha(\mathcal{G},U(1))$.

A remarkable point is that the opposite is also true: whenever two injective MPS $A^i$ and $B^i$ exhibit equivalent 1- and 2-cocycles $\phi$ and $\omega$ when transforming under the group involving on-site, time-reversal and/or reflection symmetry, then there exists an adiabatic path of injective MPS which interpolates between them. The proof of this is basically equivalent to the one sketched when no symmetries are involved.  Within each of these phases, a representative MPS with zero correlation length can be constructed starting from any solution of the 1- and 2-cocycle condition. The translationally invariant SPT phases of gapped spin systems for a given symmetry group $G$ are therefore completely classified by $H^2_\alpha(\mathcal{G},U(1))\times H^1_\beta(\mathcal{G},U(1))$  \cite{chen:spt-order-and-cohomology}. The $H^2$ part has a strong influence on the entanglement spectrum. The $H^1$ part is related to the translational invariance of the system, and is not stable under blocking (note that the first cohomology group indeed plays a central role in the description of the space groups).

Let us explicitly construct the RG fixed point MPS which transforms according to a given 1- and 2-cocycle $\phi$ and $\omega$. The local physical Hilbert space will have the dimension squared of the number of elements in the symmetry group and can be parameterized as a tensor product $(a,b)$, while the virtual indices are labeled by the group elements. The MPS is defined as
\[ A^{ab}_{xy}=\left\{\begin{array}{cc} e^{i\alpha(a.x).\omega(a,x)+i\beta(b).\phi(b)}& {\rm if }\hspace{.1cm} y=a.x\\ 0 & {\rm otherwise}   \end{array}\right.\]
and the corresponding gauge matrices are of the form
\[ 
\left(X_g\right)_{xy}=\left\{\begin{array}{cc} e^{i\alpha(x).\omega(x,g)}& \mbox{\ if\ }\hspace{.1cm} y=x.g\\ 0 & 
\mbox{otherwise}  \end{array}\right.\ . 
\]

This works as the condition $S_q(A^{ab})=\exp(i\phi(q))X_q^\dagger A^{ab} X_q$ is equivalent to the 2-cocycle equation. This MPS has zero correlation length (which follows from the fact that the transfer matrix is a rank 1 projector) and hence represents the renormalization group fixed point in its corresponding phase.  In the particular case of e.g.\ an on-site $Z_2\times Z_2$ symmetry, the simplest group exhibiting non-trivial 2-cocycles, this construction precisely yields the 1D cluster state (Appendix \ref{app:examples}) when blocking pairs of adjacent sites.

The prime example of an MPS in a non-trivial SPT phase is the AKLT state \cite{affleck:aklt-cmp}, specified by the Pauli matrices $A^i=\sigma^i, i=x,y,z$. Its SPT character can be protected by multiple distinct physical symmetries: on-site $\SO(3)$ (with virtual symmetry given by a spin-$\tfrac12$ representation), on-site $Z_2\times Z_2$ (the smallest subgroup of $\SO(3)$ still exhibiting a non-trivial 2-cocycle), time-reversal, or reflection symmetry. In all these cases, the AKLT MPS tensor transforms projectively.

By making use of the Smith normal form, it is easy to solve for all cocycle conditions and determine the number of possible different SPT phases by combining those symmetries. If we consider the symmetries of the AKLT state (on-site $Z_2\times Z_2$, time-reversal $Z_2^T$ and reflection $Z_2^R$), we obtain the following classification: $H^2_\alpha(Z_2\times Z_2\times Z_2^T\times Z_2^R,U(1))=Z_2^{\times 7}$, while $H^1_\beta(Z_2\times Z_2\times Z_2^T\times Z_2^R,U(1))=Z_2^{\times 3}$. This means that there are 1024 distinct topological phases protected by this (large) symmetry group \cite{chen2011complete}. A simpler example is obtained when considering an on-site $Z_2$ in combination with time-reversal, leading to  
$H^2_\alpha(Z_2\times Z_2^T,U(1))=Z_2^{\times 2}$ and $H^1_\beta(Z_2\times Z_2^T,U(1))=Z_2$. Similarly, $H^2_\alpha(Z_2\times Z_2^R,U(1))=Z_2^{\times 2}$,  $H^2_\alpha(Z_2^T\times Z_2^R,U(1))=Z_2^{\times 2}$ and $H^2_\alpha(Z_2\times Z_2\times Z_2^T,U(1))=Z_2^{\times 4}$. The submanifolds of normal MPS subject to global symmetries clearly exhibit an incredibly rich structure.

\paragraph{Entanglement spectrum and edge modes}

In Section~\ref{bulk-boundary}, we have seen how PEPS provide a natural way to access the entanglement spectrum, compute boundary Hamiltonians, and determine the edge physics of quantum many-body systems. In the context of non-trivial SPT phases, all of these exhibit characteristic fingerprints of the phase, which we discuss in the following.  

A distinct feature of normal MPS belonging to a non-trivial SPT phase is the fact that their entanglement spectrum exhibits a very clear pattern of degeneracies. The fact that topological order is reflected in the entanglement spectrum was first observed by \citet*{li:es-qhe-sphere}. \citet{pollmann:symprot-1d} connected these ideas to the dangling spin-$\tfrac12$ edge modes in the AKLT chain, and by making use of the fundamental theorem of MPS they realized that these edge modes were protected. The consecutive work of \citet{chen:spt-order-and-cohomology} revealed that the right mathematical formalism to deal with this phenomenon is cohomology theory. 

The degeneracy of the entanglement spectrum and the existence of edge modes follows from the following  property of projective representations $X_g$:  they cannot be Abelian and cannot be reduced to 1-dimensional representations,  and hence are only reducible to matrices of dimension strictly larger than one. Let us first consider the case of on-site group symmetries. If a given injective MPS is in  canonical form and exhibits non-trivial SPT order, its entanglement spectrum is obtained by looking at the leading eigenvector of its transfer matrix $\sum_i A^i\otimes \bar{A}^i|\rho\rangle=|\rho\rangle$. Because of the uniqueness of the corresponding eigenvalue, $\rho$ has to inherit all symmetries of the MPS and will hence be invariant under the transformation $X_g\rho X_g^\dagger=\rho$. As $X_g$  has no 1-dimensional invariant subspaces, $\rho$ must necessarily have a spectrum in which all eigenvalues have a degenerate multiplicity. For the case of time-reversal and/or reflection symmetries, the situation is slightly more complicated. The virtual degrees of freedom of the MPS in the non-trivial SPT phase then transform according to an anti-symmetric gauge transform $X=-X^T$ if the MPS is in canonical form. The entanglement spectrum is then determined by the eigenvalues of the matrix $\rho=X\bar{\rho}X^\dagger$, with $\rho$ the leading eigenvector of the transfer matrix. All eigenvalues of $\rho$ are guaranteed to be degenerate: given a right eigenvector $|x\rangle$ with (real) eigenvalue $\lambda$, $\langle\bar{x}|X^{-1}$ is guaranteed to be a left eigenvector with the same eigenvalue. But $\langle\bar{x}|X^{-1}|x\rangle=0$  as it is the trace of a product of a symmetric with an antisymmetric matrix; this implies that $\langle\bar{x}|X^{-1}$ is the left eigenvector corresponding to a different right eigenvector, implying a two-fold degeneracy.

Exactly the same feature is responsible for the fact that non-trivial SPT phases exhibit edge modes when defined on systems with open boundary conditions: the ground state degeneracy of an MPS with respect to its parent Hamiltonian with open boundary conditions will at least be the dimension of the irrep space squared, as we can define boundary vectors on both sides which will transform non-trivially under $X_g$ and will not change the energy. A beautiful feature of MPS is the fact that these gapless edge modes can actually be constructed by lifting operators acting on the virtual level to the physical level (which is always possible when the MPS is normal). For the case of the AKLT model, this indeed leads to a $4$-fold degeneracy. There is however only one state in this 4-dimensional space which will be in the spin $0$ sector, and that one will exhibit long-range entanglement between the two edges. This is a general feature of SPT phases.

\paragraph{String order parameters for SPT phases}
Symmetry breaking phases can be distinguished by their local order parameters. Since SPT phases do not break any symmetries, we will need non-local observables to distinguish them. Such non-local order parameters have since long been used to study the AKLT model under the name of string order parameters \cite{den1989preroughening,kennedy1992hidden}. In general, if we consider a gapped quantum spin system with unique ground state exhibiting a physical symmetry group $U_g$, we can construct a family of observables acting on $L+2$ sites of the form
\[ O_{\alpha,g}(L):=R_\alpha\otimes U_g^{\otimes L}\otimes R_\alpha\ ,\]
where $R_\alpha$ is an observable that transforms according to a {\em non-trivial} linear representation $\alpha(g)$ of $G$: $U_g^\dagger R_\alpha U_g=\exp(i\alpha(g))R_\alpha$  (the {\em non-triviality} ensures that the local expectation value of $R_\alpha$ is zero). We say that a spin system exhibits string order iff for some $R_\alpha$ and $g$ the limit $\lim_{L\rightarrow\infty}\langle\psi| O_{\alpha,g}(L)|\psi\rangle\neq 0$. Using the language of MPS, we can readily prove that the existence of such string order implies the fact that the system is in a non-trivial SPT phase when the group under consideration is Abelian \cite{pollmann:spt-detection-1d}. This can most easily be shown by demonstrating the fact that it has to be equal to zero in the trivial phase, i.e. in a phase in which the virtual symmetries $X_g$ form a group representation of $G$. As $U_g$ represents a physical symmetry, the expectation value of $\langle O_{\alpha,g}\rangle$ (in the thermodynamic limit) is equivalent to evaluating  the product $X^L_{\alpha,g}.X^R_{\alpha,g}$
\[
\langle S_{\alpha,g}\rangle = 
\raisebox{-.8cm}{
	\includegraphics[width=5cm]{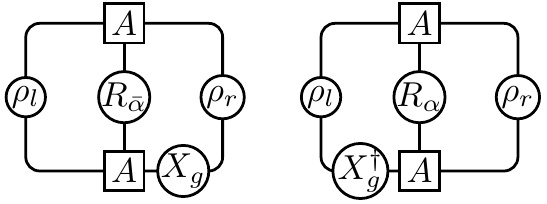}
}
\]
We could also have calculated this expectation value starting from the state $U_h^{\otimes N}|\psi\rangle$. By pulling this symmetry to the virtual level, a simple manipulation of the tensors then shows that $X^L_{\alpha,g}=e^{i\alpha(h)}X^L_{\alpha,hgh^{-1}}$. As we assumed  the group to be Abelian and the one-dimensional irrep $\alpha(h)$ to be non-trivial, this implies that the expectation value of the string order parameter has to be equal to zero. 

The expectation value of the string order parameter can certainly be non-zero in the case of a non-trivial SPT phase; in that case (and again assuming an Abelian group), we obtain $X^L_{\alpha,g}=e^{i\alpha(h)+\omega(h,g)+\omega(h.g,h^{-1})}X^L_{\alpha,hgh^{-1}}$. As the group is Abelian, $\omega(h.g,h^{-1})=-\omega(g,h)$, and hence the expectation value does not have to vanish if $\alpha(h)=\omega(g,h)-\omega(h,g)$. This is precisely what happens in the case of the AKLT chain and considering the string order parameter for the  $Z_2\times Z_2$ symmetry.  Note that not all SPT phases are detectable using this idea \cite{pollmann:spt-detection-1d}; this happens when all the commuting pairs of elements $g_1$ and $g_2$ also commute in the projective representation. 

A related idea which directly measures an observable targeting the commutator $\exp(i(\omega(g,h)-\omega(h,g)))\simeq {\rm Tr}(X_gX_hX_g^\dagger X_h^\dagger)$ can be defined in terms of swaps between distant regions, see
\citet{haegeman:1d-spt-orderparameter} for details.

\subsubsection{Symmetry breaking: virtual symmetries, Lieb-Schultz-Mattis, Kramers theorem, topological excitations \label{sec3:SB}}
\paragraph{Virtual symmetries}
Let us now consider the case of non-injective MPS which is invariant under a global symmetry. After blocking, the tensor $A^i$ corresponding to a non-injective MPS can always be written as a direct sum of injective MPS blocks $A^i=\bigoplus_\alpha A^i_\alpha$. The full MPS is hence a sum of these injective ones: $|\psi(A^i)\rangle=\sum_\alpha|\psi(A^i_\alpha)\rangle$. W.l.o.g. we will assume that the non-injective MPS $A^i$ cannot be decomposed into smaller blocks that are themselves invariant under the symmetry group $G$ under consideration; the global symmetry then permutes all the blocks into each other.  This situation happens precisely when considering the uniform superposition of all ground states of a symmetry broken system.  The paradigmatic example of such a system is given by the ferromagnetic Ising model $H=-\sum_i Z_iZ_{i+1}$ with global symmetry $X^{\otimes N}$, where the symmetric ground state is the superposition of all spins up and all spins down. This state can readily be represented as an MPS with bond dimension 2: $A^{\uparrow}=|0\rangle\langle 0|,A^{\downarrow}=|1\rangle\langle 1|$. This so-called GHZ state exhibits long-range order. Because the corresponding MPS is not injective, the tensors $A^i$ exhibit a non-trivial symmetry $G$ on the virtual level represented by the matrices that commute simultaneously with all $A^i$ ($\sigma_z$ in this GHZ case).  MPS with this property are called $G$-injective. Note that this symmetry is dual to the physical symmetry which permutes the blocks (represented by $\sigma_x$ in the case under consideration). 

For a given group $G$, one can construct a particularly simple $G$-injective MPS which is itself a renormalization group fixed point (all nonzero eigenvalues of the corresponding transfer matrix are equal to $1$) and hence provides a natural generalization of the GHZ to arbitrary groups. The physical symmetry action is represented by the regular representation.  It is defined by the MPS tensor with physical dimension $|G|^2$ and virtual dimension~$|G|$:

\be
\label{GinjMPS}
A^{ij}=\frac{1}{|G|}\sum_{g\in G}\, \sum_{\alpha, \beta}\left(L_g\right)_{\alpha i}\,\left(\bar{L}_g\right)_{\beta j} \, |\alpha)(\beta|\ ,
\ee
here $L_g$ denotes the left regular representation of the group $G$ \cite{schuch:peps-sym}. For the case of $Z_2$ symmetry, this can be written in the GHZ form by going to the dual basis (related by the discrete Fourier transform) both on physical and virtual level and with a blocking of two sites. Such a construction will be shown to be especially useful to construct topological phases in higher dimensions.

\paragraph{SET phases} 
In many interesting physical applications, only part of the total symmetry of the system is broken, while another part is unbroken. From the point of view of MPS, the broken symmetry leads to permutations of the different MPS blocks, while the unbroken one exhibits symmetries on the virtual level as encountered in the discussion of injective MPS. To accomodate both of these symmetries, the corresponding gauge representations $X_g$  acquires  the form of an induced representation. Induced representations are a technique in representation theory to lift representations $X_h$ of  a subgroup $H$ to representations of the full group $G$, that we will assume to be finite. Given a subgroup $H$, we can consider one representative $\tilde{g}_\alpha$ out of every left coset $\alpha=g.H$; any element of $G$ can then uniquely be written as $\tilde{g}_\alpha.h_\beta$. This defines functions $\phi_1$, $\phi_2$ such that  $g_a.\tilde{g}_\alpha=\tilde{g}_{\phi_1(a,\alpha)}.h_{\phi_2(a,\alpha)}$. The induced representation is then given by 

\[V_a=\sum_{\alpha}|\phi_1(a,\alpha)\rangle\langle \alpha|\otimes X_{h_{\phi_2(a,\alpha)}}\]

It acts as the regular representation between the blocks, but as an irrep within each individual block. This is indeed precisely the symmetry exhibited by symmetry-enriched topological phases (SET phases): the $X_h$ matrices form (possibly projective) representations of the unbroken symmetry group, while the broken symmetries yield permutations of the different blocks of the non-injective MPS. Along the lines of the discussion on SPT phases, it can now be established that two systems are in the same phase if and only if the permutation representations $\phi_1(a,\alpha)$ are equal to each other and furthermore that the $X_h$ must belong to the same equivalence class of $H^2(H,U(1))$ \cite{schuch:mps-phases}. An alternative approach based on a 1D version of anyon condensation is explained in \citet{garre2017symmetry}.

\paragraph{Kramers theorem and Lieb-Schultz-Mattis}
One of the most fascinating aspects of quantum spin chains is the beautiful interplay between symmetry and degeneracy. This has led to a wealth of theorems, of which Kramers' Theorem \cite{kramers1930theorie}, the Lieb-Schultz-Mattis theorem \cite*{lieb1961two}, and  the Mermin-Wagner theorem \cite{mermin1966absence} are certainly the most famous and useful ones. Let us discuss these in the light of MPS.

Kramers' theorem dictates that eigenstates of time-reversal invariant systems must be degenerate if the total spin of the system is half-integer. More precisely and in the case of spin-$\tfrac12$, the theorem is valid whenever time-reversal is implemented as an anti-unitary transformation in the form $|\psi\rangle\rightarrow (\sigma_y)^{\otimes N}|\overline{\psi}\rangle$ as originally derived by Wigner (see  \onlinecite{bargmann1954unitary}). In Sec.~\ref{sec:3:sym}, it was shown that no normal uniform MPS can exhibit such a symmetry, as there is a topological obstruction involving the global phase making it impossible for a normal uniform MPS to accomodate an on-site symmetry which acts projectively. 

To deal with such a situation, we hence have to modify the uniform MPS ansatz. In the case of spin-$\tfrac12$, this is most easily done by introducing an ansatz with a two-site unit cell: 
\be |\psi(A,B)\rangle=\sum {\rm Tr}\left(A^{i_1}B^{i_2}A^{i_3}B^{i_4}\cdots\right)|i_1\rangle|i_2\rangle\cdots\label{mpsAB}\ee
A careful analysis of the possibilities leads to the conclusion that there always exists a gauge such that $B^i$ can be chosen equal to one of the following : $\bar{A}^i$, $(\sigma_y\otimes\Id)\bar{A}^i$  or 
$(\sigma_y\otimes\Id)\bar{A}^i(\sigma_y\otimes\Id)$. The other (degenerate) ground state is then obtained by either shifting the MPS over 1 site, or by acting with the time-reversal symmetry on the state.  The total time-reversal invariant uniform MPS can then be written as the superposition of both, yielding a non-injective MPS with tensors

\be \mat{cc}{0& A^i\\B^i & 0} \label{nonijAB}\ee

Note that this MPS has no invariant subspaces, but exhibits the structure of a limit cycle (see Sec.\ref{Sec:MPV}). Only by blocking two sites do we get a $G$-injective structure. Note also that a translational-invariant breaking term in the Hamiltonian can distinguish the two blocks: the blocked MPS $\tilde{A}^{ij}=A^iB^j$ is injective and is hence the unique ground state of a local Hamiltonian. This is precisely the mechanism in the Su-Schrieffer-Heeger model \cite{su1979solitons}, where a staggered field opens a gap and leads to a unique ground state.

Similarly, the Lieb-Schultz-Mattis theorem dictates that the singlet ground state of an $\SU(2)$ invariant quantum spin chain whose unit cell transforms according to a half-integer representation of $\SU(2)$ must be gapless or symmetry broken. This theorem has been extended to the case of any half-integer charge, for example for the case of a $U(1)$ symmetry with one charge per two unit cells \cite{oshikawa1997magnetization}. From the point of view of MPS, this situation is very similar to the case of Kramers' theorem. As discussed in Section~\ref{sec:3:sym}, the structure of Clebsch-Gordan coefficients imposes that a half charge couples to one half-integer and one integer charge. As a consequence, the MPS will be exactly of the form of Eq.~\eqref{nonijAB} with $B^i$ related to $A^i$ by e.g.\ a transpose. Just as in the case of Kramers theorem, it follows that any MPS with such a symmetry will be non-injective, and will become $G$-injective after blocking two sites. In the case of a spin-$\tfrac12$ antiferromagnetic Heisenberg model, the ground state is critical, and can hence not be represented exactly as an MPS. Nevertheless, DMRG  methods are able to reproduce the ground state energy to an astounding precision with an MPS exhibiting $\SU(2)$ symmetry. What happens is that the DMRG algorithm is artificially introducing a tiny staggering in the antiferromagnetic strength, a relevant perturbation opening up a gap, breaking the translational invariance. In the uniform case, we will get an MPS of the form Eq.~\eqref{mpsAB}, and this representation will be of central importance to capture the topological non-trivial spinon excitations. 

Finally, let us say a few words about the Mermin-Wagner theorem, which states that a continuous symmetry in a quantum spin chain cannot be broken with an order parameter (observable) that does not commute with the Hamiltonian. As the unique ground state of a local gapped Hamiltonian can be represented as an injective MPS, the impossibility of defining an MPS with the relevant symmetries of the Hamiltonian (including translational invariance) implies that the ground state of that Hamiltonian has to be critical or has to break translational invariance.  If the ground state is critical, any good variational (hence injective) MPS approximation of that ground state will either have to break the continuous symmetry or the translational invariance;  which one leads to a better approximation depends on the respective scaling exponents of both perturbations.

\paragraph{Topological excitations: domain walls and spinons}
A direct implication of symmetry breaking is the emergence of topological excitations. In the case of $G$-injective MPS such as the Ising model in the ferromagnetic phase, these are domain wall excitations that tunnel between the different blocks $A^i_\alpha$ of the MPS \cite{haegeman:mps-ansatz-excitations}: 
\be 
\vert\psi(X)\rangle=\sum \cdots
A_{1}^{i_{x-2}}A_1^{i_{x-1}}
    e^{ikx}X^{i_{x}}A_2^{i_{x+1}}\cdots
\bigotimes\vert i_y\rangle\ ,
\ee
where $X^{i_x}$ is a ``tunneling'' tensor which couples the different blocks $A^i_1$ and $A^i_2$, 
and the sum runs over all $i_y$ and all positions $x$ of $X$.
Note that such excitations only make sense for an open infinite system, and that the momentum $k$ is only defined up to a constant shift. This ansatz can readily be used to simulate dispersion relations of the elementary excitations of symmetry broken quantum spin chains, where the variational parameters of these excitations are encoded in $X^i$, giving rise to an effective Hamiltonian for the quasi-particle excitations. The topological trivial excitations can then be understood as scattering states of such domain walls. This structure also emerges when studying excitations of critical systems using a variational MPS approach: the MPS will slightly break the symmetry, and the elementary excitations will then tunnel from one ground state to the other one. This was e.g.\ observed when studying the elementary excitations of the Lieb-Liniger model using cMPS \cite{draxler2013particles}.

A similar  situation occurs when the translation symmetry is broken instead of the on-site symmetry, as discussed in relation to Kramers and Lieb-Schultz-Mattis. The MPS description then acquires a $\cdots ABAB\cdots$  or $\cdots ABCABC\cdots$ etc. structure. The elementary excitations become (topological) dislocations of the form  $\cdots ABABXAB\cdots$. If we consider the Heisenberg spin-$\tfrac12$ antiferromagnet and its MPS description with such a 2-site unit cell, the emerging topological excitations are spinons: the corresponding tensor $X^i$ transforms according to a half-integer object, as it intertwines between two MPS tensors which have the same half-integer or integer spin index \cite{zauner2018topological}. This is a topological feature as there is clearly no local operator which can create such an excitation. The MPS picture hence gives a very precise meaning to the spin in a spin wave, as originally coined by \citet{faddeev1981spin}.

In general, the framework of MPS makes it very clear how elementary excitations can acquire fractionalized quantum numbers. This will even be more pronounced in the case of 2 dimensions, where PEPS provides a natural framework for describing anyons.

\subsubsection{Fermions and the Majorana chain\label{sec:3:fermions}}

The case of virtual symmetries becomes particularly intriguing when symmetry breaking is prohibited due to the existence of a superselection rule. This happens in the case of a chain of fermions, where superpositions between states with an even and odd number of fermions are ruled out. In contrast to the situation of symmetry breaking discussed in the previous section, the fermion superselection rule has the power to stabilize a $G$-injective GHZ state,  as all corresponding symmetry broken states $|\psi\left(A^\alpha\right)\rangle$ would violate the superselection rule.

This scenario was first discussed by \citet{kitaev:majorana-chain}, and the corresponding non-trivial Hamiltonian is called the Kitaev or Majorana chain. He demonstrated that there are 2 distinct phases for interacting fermionic spin chains, and hence that there is no adiabatic gapped path between the trivial phase and the Majorana phase. The ground state of the Kitaev chain is a fermionic MPS with bond dimension 2. As discussed in Section~\ref{FandCTN}, the natural language for fermionic MPS is the one in terms of $Z_2$ graded algebras. Just as happened in the case of SPT phases, the Kitaev chain has edge modes which are exponentially localized around the boundary, and this can be understood in terms of the entanglement degrees of freedom which exhibit the virtual symmetry.  When considering the Kitaev chain on a ring with periodic boundary conditions, the ground state is unique and has odd parity (see Eq.~\eqref{Majchain}):
\begin{equation}\label{eq:kitaev-chain-mps}
|\psi\rangle=\sum_{i_1\cdots i_N}\tr\left[{YA^{i_1}A^{i_2}\cdots A^{i_N}}\right]|i_1\rangle\otimes_g|i_2\rangle\otimes_g\cdots
\ .
\end{equation}
The tensors $A^0=\openone$ and $A^1=Y=\sigma_y$ both commute with  $Y$
which forms the representation of the virtual symmetry, and hence the MPS
is non-injective and can be written as a sum of two injective MPS
$|\psi_1\rangle+|\psi_2\rangle$. The symmetry broken states
$|\psi_i\rangle$ however contain  superpositions of an even and odd number
of fermions, and are hence unphysical.  As demonstrated in
\citet{bultinck:fermionic-mps-phases}, the uniqueness of the ground state is guaranteed by the $Z_2$ graded version of injectivity. For a system with open boundary conditions, we get a 2-fold degeneracy as opposed to a 4-fold one as in the case of the AKLT model; the Hilbert space "dimension" of a Majorana fermion is indeed $\sqrt{2}$ as opposed to $2$. As a consequence, the Majorna chain has the unique feature that the system with periodic boundary conditions can be represented through an MPS with open boundary conditions and bond dimension $2$.

There is also an intriguing interplay between time-reversal symmetry and the $Z_2$ superselection rule. As first demonstrated by \citet{fidkowski:1d-fermions}, 8 different SPT phases emerge. This has to be contrasted to the spin case, where time-reversal gives rise to only 2 cases. These 8 phases can be distinguished by studying how the entanglement degrees of freedom transform under the time-reversal symmetry, and this gives rise to 3 different indices 
\cite{bultinck:fermionic-mps-phases}.
The first index distinguishes the Majorana case (with a virtual $Z_2$ symmetry) from the trivial non-Majorana case (without such a symmetry). A new $Z_2$ index $\kappa$ emerges according to the transformation rules under conjugation of the tensors $\bar{A}^i=e^{i\chi}XA^iX^{-1}$; just as in the case of spins, the index $\kappa$ is witnessed by $\bar{X}.X=(-1)^\kappa\openone$. A third $Z_2$ index $\mu$ characterizes how the matrix $Y$, which represents the center of the MPS algebra, transforms under the gauge $X$: $X.Y=(-1)^\mu Y.X$. Those three $Z_2$ indices give rise to the celebrated $Z_8$ classification of interacting fermionic spin chains, and a representative of each of the 8 classes can be constructed by taking tensor products of Kitaev chains.  

A wide variety of such fermionic SPT phases can be constructed by repeating this construction for other groups and symmetries. In analogy to the discussion on SET phases, this can be achieved by making use of induced representations, where the physical and purely virtual symmetries are combined in a natural way.

\subsubsection{Gauge symmetries}

The idea of lifting global symmetries to local ones by introducing new gauge degrees of freedom has proven to be of fundamental importance in the field of particle physics. It turns out that a procedure similar to the minimal coupling prescription can be implemented on the level of wavefunctions whenever these are expressed in terms of MPS \cite{buyens2014matrix,kull2017classification}: starting with an MPS describing matter fields with a global symmetry $U_g$ implemented (projectively) on the virtual degrees of freedom as $X_g$, it is straightforward to introduce new (gauge) degrees of freedom and tensors on the edges which will lift the global symmetries to local ones. To achieve this, let us consider a tensor with physical degrees of freedom corresponding to the group elements, and define it as:
\be A^{a^{-1}.b}=\sum_{ab}|a)|a^{-1}.b\rangle(b| \label{gaugemps} \ee

Acting with the left regular representation $L_g$ on the physical level is equivalent to acting with the right regular representation $R_g$ on $|a)\rightarrow |ag)$, and acting with the right regular representation $R_g$ on the physical level amounts to acting with $R_{g^{-1}}$: $(b|\rightarrow (bg^{-1}|$. Note that this tensor provides the natural generalization of the GHZ-state in the dual basis for any group.

\subsubsection{Critical spin systems:  MPO symmetries \label{CSS}}
In Section~\ref{sec3:SB}, we have discussed the difficulty of representing ground states of critical quantum spin systems using injective MPS, and hinted to the fact that there are topological obstructions to do so. Whenever the (continuous) symmetries of a translationally invariant quantum spin Hamiltonian cannot be represented by a uniform injective MPS, then the ground state of that Hamiltonian has to either be critical and exhibit power law decay of its correlations, or exhibit symmetry breaking. An important question is the characterization of the nonlocal symmetries emerging for such  critical systems which prevent the exact description of their ground states with finite bond dimension MPS. The formalism of matrix product operators (MPO) provides exactly that. Additionally, the MPO formalism provides a constructive way of writing down Hamiltonians with such a symmetry. It turns out that the same symmetries are the ones responsible for the existence of non-chiral topological order in $2+1$ dimensions. This is of course not surprising, as there is a very intimate connection between topological quantum field theory in $2+1$ and conformal field theory in $2+0$ or $1+1$ dimensions \cite{witten1989quantum,elitzur1989remarks,fuchs2002tft,moore1989classical}. This section therefore also provides the mathematical background for studying topological gapped systems in 2 dimensions.

The symmetries under consideration have been studied at great length in the fields of quantum groups, integrability and conformal field theory. The picture that has emerged is that critical spin systems exhibit "topological symmetries" or anomalies that can be seen as lattice remnants of the full conformal group \cite{aasen2016topological,vanhove2018mapping}. In the example of the critical quantum transverse Ising model,  this "symmetry" corresponds to the famous Kramers-Wannier duality, and the scale-invariant symmetry operations form a closed algebra as opposed to a group.  Matrix product operators are precisely the right framework to provide representations of these algebras \cite{sahinoglu:mpo-injectivity,bultinck:mpo-anyons,williamson:SET,lootens2020matrix,molnar:WHA}.

\paragraph{MPO algebras}

Given a tensor 
\[
\includegraphics[width=1cm]{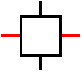}
\]
in canonical form (in each direction) we can define the algebras
\[
\mathcal{A}^{(N)} = \left \{ \; \parbox[c]{0.18\textwidth}{ \includegraphics[scale=0.75]{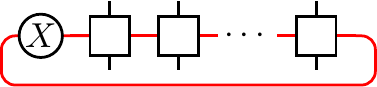} }: X  \right \} 
\]
with $N$ the system size. We can identify all $\mathcal{A}^{(N)}$ if the product is independent of $N$. In such a case we say that the algebra 
\[
\mathcal{A} = \left \{ \; \parbox[c]{0.10\textwidth}{ \includegraphics[scale=0.75]{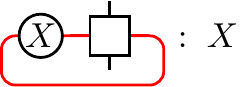} } \right \} 
\]
is an {\it MPO algebra} \cite{bultinck:mpo-anyons,molnar:WHA}. 

A series of non-trivial consequences can be derived from this definition; for the precise mathematical statements, we refer to \citet{lootens2020matrix}:

\begin{enumerate}
    \item Any MPO algebra can be decomposed into a finite set of injective MPOs represented as $O_a$ with $a$ taken from a set of labels $a\in\mathcal{C}$; these $O_a$ form a ring with nonnegative integer fusion coefficients $N_{ab}^c$: $O_a O_b=\sum_c N^c_{ab}O_c$. We use the notation $^2\!F_a$ for describing the MPO tensors.

    \item The Fundamental theorem of MPS implies the existence of a \emph{fusion} tensor $^1\!F$ which satisfies the following \emph{zipper} equation:
    \[\parbox[c]{0.07\textwidth}{ \includegraphics[scale=0.75]{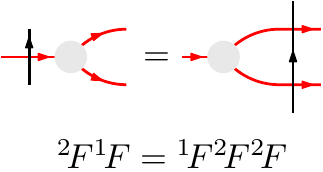} } \]
     If the labels $a\in\mathcal{C}$ correspond to the irreps of a group, then the $^1\!F$ can be identified with  the Clebsch-Gordan coefficients.

    \item Associativity of the zipper equation requires the existence of a recoupling tensor $^0\!F$ which solely depends on the set of labels $a\in\mathcal{C}$ satisfying the following equation:
    \[\parbox[c]{0.07\textwidth}{ \includegraphics[scale=0.75]{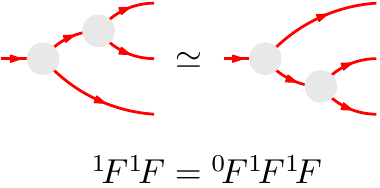} } \]
     If the labels $a\in\mathcal{C}$ correspond to the irreps of a group and the MPO tensors encode these irreps, then the $^0\!F$ are equal to the Racah or Wigner 6j symbols.  

    \item Any recoupling tensor $^0\!F$ has to satisfy the ubiquitous algebraic pentagon equation. For given MPO fusion rules $N_{ab}^c$,  there only exists a finite number of possible inequivalent solutions to the pentagon equation. This puts a huge restriction on the possible MPO algebras, and puts its study squarely in the realm of fusion categories.

    \item The scale invariance of the MPO algebra implies the existence of a different set of fusion tensors $^3\!F$ acting on the physical degrees of freedom satisfying the following \emph{pulling through} equation:
        \[\parbox[c]{0.07\textwidth}{ \includegraphics[scale=0.75]{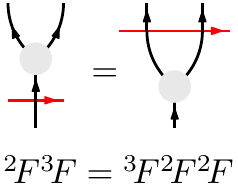} } \]
          It follows that there also exists a \emph{dual} algebra obtained by switching the role of physical and virtual indices of the MPO tensor $^2\!F$; note that this is precisely the same duality as the one described in Sec. \ref{sec:2:MPDO-RGFP} describing renormalization group fixed points. This \emph{vertical} MPO algebra can now again be decomposed into its injective blocks, giving rise to a new set of discrete labels $\alpha\in\mathcal{D}$ and fusion coefficients $\tilde{N}_{\alpha\beta}^\gamma$. This $\mathcal{D}$ will also form a fusion category. As will become clear from Sec. \ref{sec:3:2D}, we also call $^3\!F$ the PEPS tensor.
          
    \item Recoupling of the fusion tensors $^3\!F$ implies the existence of a set of fusion tensors $^4F$ satisfying:
        \[\parbox[c]{0.07\textwidth}{ \includegraphics[scale=0.75]{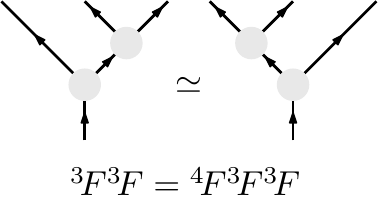} } \]

    \item The tensor $^4\!F$  satisfies the pentagon equation, with solutions completely determined by $\tilde{N}_{ij}^k$.

\end{enumerate}

The 5 objects $^i\!F$ and accompanying six consistency equations appear in the field of tensor categories and form the defining equations of a bimodule category. Such categories have been studied extensively in the context of describing boundaries between systems exhibiting topological quantum order \cite{kitaev:gapped-boundaries}, but have a completely different meaning here. 

A $(\mathcal{C},\mathcal{D})$ bimodule category $\mathcal{M}$ has a new set of labels $A,B,\ldots \in\mathcal{M}$ which will represent the \emph{entanglement degrees of freedom}  and therefore the choice of this bimodule category determines the explicit representation of the MPO/fusion/PEPS tensor.  We can then identify the fusion, MPO and PEPS tensors as follows (note that all tensor legs are labelled by triple indices belonging to either $\mathcal{C}$, $\mathcal{M}$ or $\mathcal{D}$, some of which might be trivial):
\begin{eqnarray*}
\left(^1\!F^{abC}_A\right)^{B,kj}_{c,mn} &=&  \parbox[c]{0.07\textwidth}{ \includegraphics[scale=0.75]{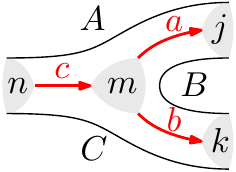} } \\
\left(^2\!F^{aC\alpha}_B\right)^{D,nk}_{A,jm} &=& \parbox[c]{0.07\textwidth}{ \includegraphics[scale=0.75]{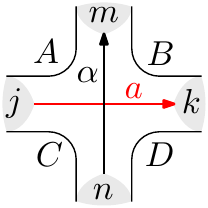} } \\
\left(^3\!F^{A\alpha\beta}_{B}\right)^{\gamma,km}_{C,jn} &=& \parbox[c]{0.07\textwidth}{ \includegraphics[scale=0.75]{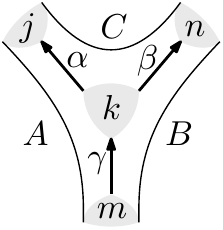} }
\end{eqnarray*}

For bimodule categories that are invertible, the categories $\mathcal{C}$ and $\mathcal{D}$ are Morita equivalent; this requires that their Drinfeld doubles or Drinfeld centers are equivalent. As described in Section~\ref{tsaa}, this Drinfeld center has a very tangible physical meaning as it represents the (output) fusion category of the anyon excitations in topological phases of matter described by string nets defined by the (input) category $\mathcal{D}$; equivalently, it describes the primary fields for lattice realizations of CFTs. A particularly simple choice of an invertible bimodule category is obtained by choosing $\mathcal{C}=\mathcal{D}=\mathcal{M}$; in that case, all $^i\!F$ symbols are equal and the six pentagon equations are all equivalent \cite{bultinck:mpo-anyons}. 

Instead of the categorical description which arises naturally from the fundamental theorem of MPS, an equivalent formulation of MPO algebras in terms of weak Hopf algebras appears in \citet{molnar:WHA}, where the size independence of the MPO algebra is formalized in the form of a co-multiplication
\[
	\includegraphics[width=5cm]{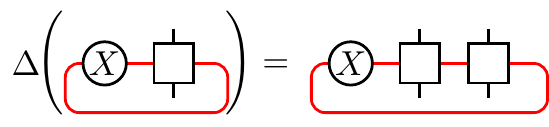}
\]
This equivalence makes the seminal result of \cite{Hayashi:WHA} connecting fusion categories with weak Hopf algebras very explicit. The additional known connection with subfactor theory is also made explicit for MPO algebras in \cite{kawahigashi2020remark}.

The central \emph{pulling through} equation can also be rephrased completely in terms of the full MPO algebra \cite{sahinoglu:mpo-injectivity,molnar:WHA}:
\begin{equation}\label{fig:pull-through-wha}
	\raisebox{-1cm}{\includegraphics[width=5cm]{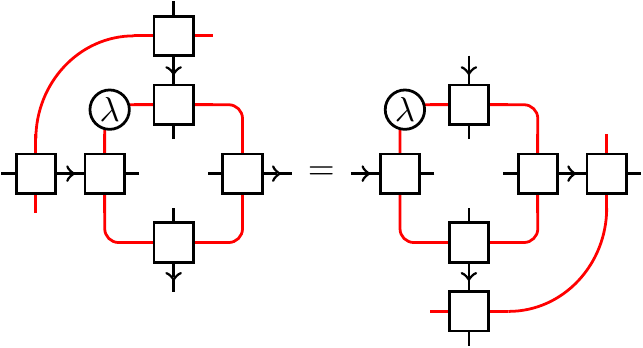}}
	\end{equation}

Here $\lambda$ is an extra block diagonal tensor which is a direct sum of identities acting on the invariant subspaces of the MPO tensors, and weighted with the quantum dimensions of categorical objects corresponding to related MPO injective blocks. Moreover, reversing the arrows, as done in the right and bottom tensors, means taking the {\it inverse representation} \cite{bultinck:mpo-anyons}.

Let us illustrate this with an example, for which $\mathcal{C}$ represents the labels of a group $g\in G$ and $\alpha\in \mathcal{D}$ the labels of its irreps $D^\alpha$, each $\alpha$ appearing as many times as its dimension. $\mathcal{M}$ can then be chosen to be trivial, and the injective MPOs $O_g$ is  represented by the MPO tensors 

\[A_g =\sum_{\alpha ij}D^\alpha(g)_{ij}\, |g)(g|\otimes |i,\alpha\rangle\langle j,\alpha| \]
where we keep using the notation already established in Section~\ref{MPSandPEPS}: {\it curved} ket/bras $|\,\cdot\,)$ correspond to the virtual level and standard ones $|\,\cdot\,\rangle$ to the physical one.

Up to a unitary transformation on the physical indices, this is equal to  $A_g =|g)(g|\otimes L_g$, with $L_g$ the regular representation, and leads to the representation for quantum doubles used  by \citet{schuch:peps-sym}. 
The $G$-injective MPO is then defined as (compare with \eqref{GinjMPS})
\be \label{eq:G-injective-MPO}
\frac{1}{|G|}\sum_{g\in G} L_g^{\otimes N}
\ee
The corresponding pulling through equation \eqref{fig:pull-through-wha} then becomes equal to 
\begin{align*}
&L_h^{\otimes 2}\otimes (\Id^{\otimes 2})\sum_{g}L_g^{\otimes 2}\otimes L_{g^{-1}}^{\otimes 2}=
\\
&\hspace*{2cm}=\left(\sum_{g}L_g^{\otimes 2}\otimes L_{g^{-1}}^{\otimes 2}\right) (\Id^{\otimes 2})\otimes L_h^{\otimes 2}
\end{align*}
which is trivially true. 

Stepping up one level of sophistication, MPO algebras can accommodate 3-cocycles, corresponding to non-trivial solutions to the pentagon equations and requiring non-trivial entanglement degrees of freedom $\in\mathcal{M}$ \cite{buerschaper:twisted-injectivity,williamson:mpo-spt}. The most general MPO algebras form representations of all bimodule categories with a spherical structure \cite{lootens2020matrix}.

\paragraph{MPO symmetries}

The pulling through equation for the $^2\!F$ tensors can be used to define operators that commute with the full MPO algebra:
\[\includegraphics[width=8cm]{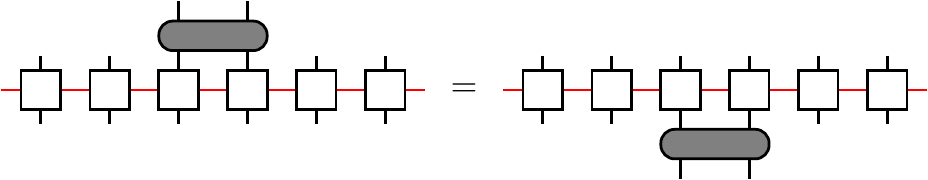}\]
If these tensors can be made Hermitian, they define a Hamiltonian on a one-dimensional lattice by taking the sum of their translations;  the corresponding full Hamiltonian will commute with the complete MPO algebra. All so-called \emph{anyonic spin chains} \cite{gils2009collective} can be constructed in that way, and the MPO symmetries hence realize the corresponding topological symmetries.

Let us illustrate this with two simple examples. If one considers the two unitary solutions of the pentagon equations for the fusion rules corresponding to the group $Z_2$, one of them corresponds to a non-trivial 3-cocycle. The local Hamiltonian commuting with the corresponding MPO is precisely the cluster state Hamiltonian with critical magnetic field \cite{bridgeman2017anomalies}, which is equivalent to the XY model \cite{lahtinen2015realizing}. Similarly, if one starts from the Ising fusion rules, one obtains the critical Ising model in transverse magnetic field, and for the Fibonacci fusion rules,  the critical ``golden chain'' Hamiltonian emerges \cite{vanhove2018mapping,lootens2019cardy}. 

In a similar vein, it is possible to construct classical statistical mechanical lattice models using this construction; the construction gives rise to the RSOS models of Andrews, Baxter and Forester \cite{andrews1984eight,aasen2016topological}, which are known to yield lattice critical systems corresponding to all CFTs in the minimal series. 

This clearly suggests that the MPO symmetry is exactly the one that is responsible for criticality; it emerges at the critical point,  and provides a topological obstruction for a normal/injective MPS to have such a symmetry. Strictly speaking, it is not a symmetry as the MPOs do not have to be invertible; in the case of the Ising model for example, the MPO implementing the Kramers-Wannier duality is not invertible, and this is a consequence of the fact that the critical theory must both be symmetric with respect to the local $Z_2$ symmetry and the dual disorder symmetry \cite{kadanoff1971determination} which anticommute. Furthermore, it can be shown that a perturbation which commutes with the full MPO algebra and keeps translational invariance is generically prohibited from opening a gap \cite{buican2017anyonic}. However, MPO symmetries are not enough to guarantee criticality, as MPOs only encode the topological features of critical systems; extra constraints related to the geometry, the so-called discrete holomorphicity, have to be imposed and lead to a lattice version of the conformal symmetry \cite{fendley2021integrability}.

For the present discussion, let us now demonstrate that a uniform normal MPS cannot exhibit an MPO symmetry obtained from a non-trivial solution of the pentagon equations. For the group case of 3-cocycles, this was originally proven by  \citet{chen:spt-order-and-cohomology} \cite[see also][]{williamson:mpo-spt,molnar:quasi-injective-peps}, and their proof can readily be extended to the general case of MPO algebras. It is proven by contradiction, and is based on the fundamental theorem. Let us sketch the proof. 

First, if we act with an MPO $O_\alpha$ on an injective MPS $|\psi(A)\rangle$ and it is proportional to $|\psi(A)\rangle$, the fundamental theorem implies the existence of an intertwiner $V_\alpha$ which reduces the local tensors of the  MPS $O_\alpha |\psi(A)\rangle$ back those of the original one:
\[
	\includegraphics[width=4cm]{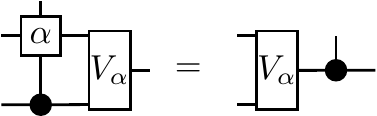}
\]
If we act with 2 MPOs, then there are two inequivalent ways of reducing to the original one: either we first reduce the MPS and $O_\alpha$ via $V_\alpha$, followed by $V_\beta$, or we can first reduce the two MPOs with the intertwiner $F$: 
\[
	\includegraphics[width=4cm]{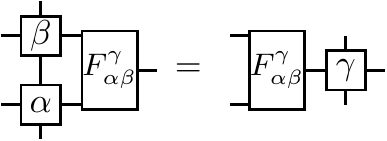}
\]
Note that in the case of multiple fusion channels $\gamma$, it is immaterial which fusion channel is taken as long as it does not give zero.  It has been proven by \citet[][Theorem 22]{molnar:quasi-injective-peps} that these 2 ways of reducing to an injective MPS must be equivalent up to a scalar $\lambda(\alpha,\beta;\gamma)$. 

\[
	\includegraphics[width=6cm]{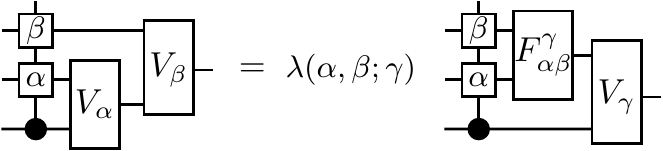}
\]

We now act with three MPOs on the supposed injective MPS for which all of them are symmetry operators. Just as in the defining equation of the pentagon equation, there are two different ways of changing the order of the reductions, and both have to be equivalent. If follows that 
\[\frac{\lambda(\alpha,\beta;a).\lambda(a,\gamma;b)}{\lambda(\beta,\gamma;c).\lambda(\alpha,c;b)}=F^{\alpha\beta\gamma}_{bac}\]
There is however a topological obstruction to achieve this: a non-trivial solution $F$  of the pentagon equation can never be a simple product of functions which act exactly as gauge transforms on these same F-symbols. In the case of groups and 3-cocycles, the left hand side corresponds exactly to the co-boundary and enforces the F-symbol to be trivial. This is the contradiction, and proves that the MPS cannot be injective. As a consequence, any Hamiltonian with an MPO symmetry will either be critical or have a symmetry broken groundstate space. 

This theorem is very powerful and is a clear demonstration of the fact that MPO symmetries yield a Lieb-Schultz-Mattis-like  proof for ``topological'' symmetries as opposed to continuous ones. Note that the proof assumed no fusion multiplicities, it is all $N_{ab}^c\leq 1$, and counterexamples can be constructed if this is not the case.  Note also that a non-trivial MPO symmetry does not prevent a tensor network description for density matrices, as the renormalization group fixed points discussed in Section~\ref{renorm_secII} are exactly of that form. Similarly, it is possible to construct MERA with non-trivial MPO symmetries, and hence allows for the description of critical phases with exact topological symmetries \cite{bridgeman2017anomalies}. Those two facts turn out to be intimitaly related to each other, as the entanglement Hamiltonian of a MERA is exactly of the MPO form \cite{van2020entanglement}. 

The fact that local Hamiltonians which commute with non-trivial MPO symmetries necessarily have to be critical or symmetry broken has a crucial influence on the edge modes of systems exhibiting topological quantum order in 2 dimensions, and is the origin of the CFT-TQFT correspondence and anomalies on these edges. This will be discussed in Section~\ref{sec:3:2D}. The MPO picture for describing critical quantum spin systems transcends to the statistical mechanical world, in which the MPOs become Wilson loop type operators which generalize the disorder operators introduced in the context of the 2D Ising model by \citet{kadanoff1971determination}. A systematic study of these MPO symmetries allows to represent all fields, including the chiral ones, in terms of the so-called tube algebra, which is an MPO algebra representing the Drinfeld Center of the input category. Many of the intriguing properties of conformal field theories can as such be transmuted to the lattice, including notions like orbifolding and the coset construction \cite{lootens2019cardy}.

\subsection{Symmetries in two dimensions: PEPS\label{sec:3:2D}}

The full power of symmetries in tensor networks is revealed in the description of quantum many body systems in two dimensions by PEPS, in which different  phases of matter can be distinguished by the different representations of these symmetries acting on  the local PEPS tensors. As such, tensor networks provide \emph{local} order parameters for topological phases of matter as a direct manifestation of the entanglement properties of such systems. The most general language for describing these symmetries is in terms of matrix product operators, which play the role of Wilson loops on the virtual level. These MPOs provide a unifying framework for describing both the entanglement structure of the ground state manifold and for characterizing the elementary excitations in these systems. 

\subsubsection{Symmetric PEPS \label{sec:3:MPO-injective}\label{sec:3:2D-SPT}}

The arena of topological quantum phases in two dimensions is much richer than the one for quantum spin chains. From the tensor network point of view, this can be understood from the fact that tensors have a much richer but also more complicated structure than matrices. The most general form of the fundamental theorem of MPS, on which much of the previous section was built upon, does not have an easy generalization to two dimensions, and we will therefore concentrate on the cases for which the symmetries on the virtual level can be described in terms of tensor products of local operators or of matrix product operators. 

We will distinguish five different situations: a. the case of injective PEPS, describing gapped systems with a unique ground state; b. non-injective PEPS as unique ground states of gapped SPT phases; c. non-injective PEPS as ground states of systems exhibiting genuine topological order; d. PEPS descriptions of SET phases; e. chiral phases.

\paragraph{Injective PEPS}
Global symmetries of injective PEPS behave essentially equivalently to global symmetries in the MPS case: the fundamental theorem of injective PEPS (Section~\ref{sec:4:fund-thm-peps}) dictates that two uniform PEPS are equal if and only if they are related by a gauge transform on the virtual level:

\be \sum_j U_{ij}(g)A^j_{\alpha\beta\gamma\delta}=
\ee
$$ e^{i\phi}\sum_{\alpha'\beta'\gamma'\delta'} X(g)_{\alpha\alpha'}X(g)^{-1}_{\beta\beta'}Y(g)_{\gamma\gamma'}Y(g)^{-1})_{\delta\delta'}A^i_{\alpha'\beta'\gamma'\delta'} $$

Just as in the 1D case, $X(g)$ and $Y(g)$ form possibly projective representations of the group $G$ characterizing the global symmetries of the system. The fact that a PEPS tensor exhibiting this feature has a global symmetry follows immediately from the fact that all these gauge transformations cancel each other pairwise. The surprising content of the fundamental theorem is that it is also a necessary condition.  This again implies that the PEPS tensor can be decomposed as a product of Clebsch-Gordan coefficients, as used already extensively in numerical PEPS algorithms.  Note that different projective representations do not necessarily lead to different phases if one does not impose translational invariance, as blocking several sites together allows to relate different projective representations to each other.

The canonical example for an injective PEPS with global symmetries is the cluster state \cite{raussendorf:cluster-short} on the honeycomb lattice, which is the unique ground state of the commuting parent Hamiltonian $\sum_{ijkl}X_iZ_jZ_kZ_l$ of qubits where $j,k,l$ are the nearest neigbours of site $i$. This state has a very simple PEPS representation \cite{verstraete:mbc-peps}:
\[\includegraphics[width=5cm]{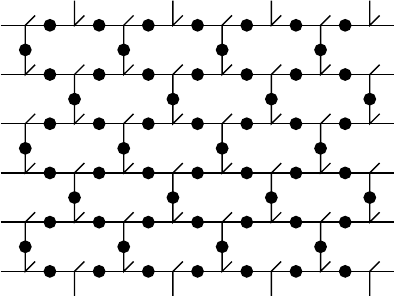}\]
 with 
\[ 
\raisebox{-.5cm}{\includegraphics[width=1.2cm]{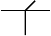}}= \ket{0000}+\ket{1111} \; , \quad \raisebox{-.1cm}{\includegraphics[width=0.7cm]{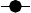}}= \frac{1}{\sqrt{2}}\left(\begin{matrix} 1 & 1\\ 1 & -1\end{matrix}\right) \]

It exhibits the global symmetry $Y^{\otimes N}$, as this is just the product of all (commuting) terms of the Hamiltonian. To use the fundamental theorem of PEPS, we first block two sites of this PEPS such as to make it uniform, and it can then readily be checked that the local physical symmetry is equivalent to acting on all virtual legs with the same $Y$ on all four legs. This cluster state is interesting from the point of view of quantum information theory, as it allows to do universal quantum computation by implementing local measurements on its qubits \cite{raussendorf:cluster-short}. The underlying mechanism which allows for this remarkable feature is the fact that local measurements on the physical qubits effectively teleport the virtual degrees of freedom, and in that process implement quantum gates \cite{verstraete:mbc-peps}. A related mechanism underlies the concept of topological quantum computation by braiding anyons, which can be understood in terms of quantum circuits on the entanglement degrees of freedom of the PEPS describing the topological phase. 

Note that we had to block sites of the cluster state to get a uniform PEPS description. From the point of view of space group symmetries, this is not wholly satisfactory as this leads to a loss of symmetry in the system. It turns out that the full space group symmetry can be done justice for general PEPS by including matrices which just act on the virtual edges connecting the vertices of the PEPS. By imposing translational symmetry, it will then follow that this decorated PEPS will be uniform and exhibit all lattice symmetries \cite{jiang:spt-tns-anycond}.

\paragraph{Non-injective PEPS: SPT phases}
Injective PEPS on a square lattice are rare, as the injectivity condition is typically violated at the corners of the region of interest. Unlike the MPS case however, non-injective PEPS can still be unique ground states of local gapped Hamiltonians. The 2D AKLT model on the square lattice is such an example. The non-injectivity gives rise to a much more interesting algebraic structure in the form of a family of  matrix product operator symmetries $O_g$ labelled by the group elements of the global symmetry \cite{williamson:mpo-spt,molnar:quasi-injective-peps}:
\be \raisebox{-.5cm}{\includegraphics[width=4cm]{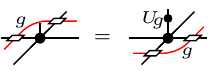}} \label{pulth} \ee

It is easy to see that this local condition is sufficient for the complete PEPS to be invariant under the global symmetry $U^{\otimes N}(g)$ in the thermodynamic limit. 

To be consistent, the MPOs $O_g$ should form a representation of the group $G$: $O_g.O_h=O_{gh}$. Remarkably, the fundamental theorem of MPS allows to translate this condition into a local condition for the tensors defining these MPOs, as two MPOs are equal to each other if and only if there exists an intertwiner (or fusion tensor) connecting them to each other:

\[\includegraphics[width=6.5cm]{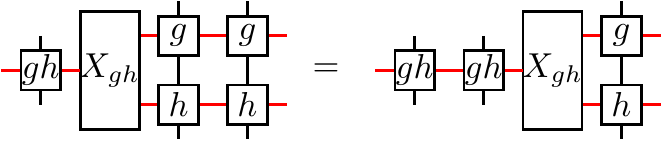}\]

The associativity condition for these fusion tensors then leads to the condition that both the elements of these MPOs and of the intertwiners can be identified with the elements of a 3-cocycle, determined by the third cohomology group $H^3(G,U(1))$.  This situation is completely equivalent to the one discussed in Section~\ref{CSS}, but for the special case of the fusion algebra being a group. The third cohomology group is well known to classify symmetry protected topological phases in 2 dimensions \cite{chen:2d-spt-phases-peps-ghz}, and PEPS hence provides a natural realization of such phases.

Just as in the case of 2-cocycles, there is a systematic way of writing down PEPS tensors which exhibit such MPO symmetries \cite{williamson:mpo-spt}. Indeed, the pulling through equation (\ref{pulth}) can componentswise be identified with the 3-cocycle condition. 

The canonical example of a non-trivial SPT PEPS was derived in \cite{chen:2d-spt-phases-peps-ghz} as the CZX state with global $Z_2$ symmetry, for which the virtual MPO symmetry is represented by the following two matrix product unitaries acting on qubits: $O_1=\openone$, $O_Z=\bigotimes_{i}CZ_{i,i+1}\otimes\bigotimes_i X_i$.  Here the commuting matrices $CZ_{i,i+1}=\sum_{ij}(-1)^{i.j}|ij\rangle\langle ij|$ represent diagonal controlled Z-gates. $O_Z$ has bond dimension $2$, but the square of it is not in canoncial form and has a 1-dimensional invariant subspace equal to $O_1=\openone$.  

These ideas were worked out in the papers \cite{buerschaper:twisted-injectivity,williamson:mpo-spt,molnar:quasi-injective-peps}. It was demonstrated  that this notion of MPO-injectivity - also called semi-injective - is sufficient for guaranteeing the uniqueness of the ground state of the corresponding parent Hamiltonian, and that the corresponding PEPS fully characterize short-range entangled SPT phases.

\paragraph{Virtual Symmetries}
\label{virtualsym}

One of the most striking features of two-dimensional quantum spin systems is the fact that there exist topological phases of matter which are stable under any perturbations \cite{klich:stability,bravyi:tqo-long}. This robustness is a consequence of its nontrivial entanglement structure, which is reflected in the behaviour of the topological entanglement entropy, of its edge modes, and in the anyonic statistics of its elementary excitations. Tensor networks provide a natural language for describing all those features in terms of the local symmetries of the tensors involved.

The fact that there is a connection between topological phases of matter and symmetries in the tensors has its roots in the pioneering work of \citet{gu:TERG}, and was described in \citet{schuch:peps-sym}. There it was shown that the symmetries in the virtual indices of a PEPS characterize its topological order. Specifically, a PEPS is constructed for each finite group $G$, where the local tensor is given by the G-injective MPO  \eqref{eq:G-injective-MPO}:
\be \includegraphics[width=6.5cm]{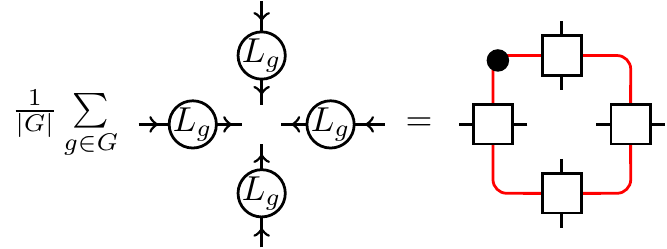} \label{fig:G-injective-tensor-Fig} \ee
where input indices correspond to the virtual degrees of freedom and output indices to the physical Hilbert space. Strictly speaking the arrows on the right and bottom tensors must be reversed to make arrows match in the PEPS construction. In this case reversing the arrows is nothing but taking the inverse representation. The PEPS constructed this way is called $G$-injective. 

The symmetries of the tensor, Eq.~(\ref{fig:pull-through-wha}),
allow to recover easily all topological invariants (topological entanglement entropy, ground state degeneracy, anyonic statistics,\ldots), and correspond to the phase of the quantum double models of \citet{kitaev:toriccode}.
It is important to notice that by acting with an invertible operator on the physical index, one can perturb the tensor and induce  finite  correlation  lengths and, for large perturbations, topological phase transitions. In this sense, $G$-injective PEPS allow to recover not only the quantum double model, but all the associated phases (see Section~\ref{sec:H-to-states}) and the phase transitions between them. They also allow to study particularly relevant models which are not renormalization fixed points, as e.g. the nearest neighbour resonating valence bond (RVB) state \cite{anderson:rvb}, which is an $\SU(2)$ invariant spin liquid when defined on a frustrated 2D lattice. As explained by \citet{verstraete:comp-power-of-peps}, this RVB has a very simple description in terms of a PEPS with bond dimension $3$. As shown by \citet{schuch:rvb-kagome}, it exhibits a non-trivial purely virtual $Z_2$ symmetry, and can adiabatically be continued to the toric code phase without crossing a phase transition. Indeed, thanks to the PEPS approach, a parent Hamiltonian was derived for it \cite{schuch:rvb-kagome,zhou:rvb-parent-onestar}.

The RVB state is an example where the symmetries in the virtual indices arise naturally from physical symmetry constrains. For example, if a global SU(2) spin-$\tfrac12$ symmetry is encoded locally in the tensor
\begin{equation}\label{fig:product-encoding-symmetry}
\includegraphics[width=5cm]{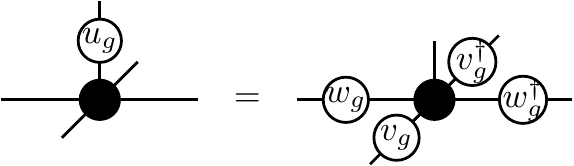}
\end{equation}
then the virtual symmetries must necessarily  be reducible and contain both integer and half-integer representations, e.g $\tfrac12\oplus 0$, and the tensor must be supported on the sector containing an odd number of half-integer ones. This enforces a $\mathbb{Z}_2$ virtual symmetry in the tensor.  This effect is nothing but a version in the context of PEPS of the celebrated Hastings' 2D version of the Lieb-Schultz-Mattis theorem \cite{hastings:liebschultzmattis-2d}.

The $G$-injective construction was generalized by \citet{buerschaper:twisted-injectivity} to twisted quantum double models (twisted $G$-injective PEPS), to the case of an arbitrary fusion category by \citet{sahinoglu:mpo-injectivity}, under the name MPO-injective PEPS, and those MPO constructions were unified by invoking the structure of a bimodule category by \citet{lootens2020matrix}. 

For the case of topologicsl phases, the pulling through equation \eqref{fig:pull-through-wha} becomes very similar to the case of SPT phases but without a physical action:
\be
\label{pulth-2}
\includegraphics[width=5cm]{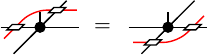}\ee

As discussed in Sec. \ref{CSS}, the bimodule categories define MPO algebras and PEPS tensors satisfying the pulling through equations through a set of 6 coupled pentagon equations. This construction starts from two Morita-equivalent fusion categories  $\mathcal{C}$, $\mathcal{D}$ labelling the different MPO tensors respectively the degrees of freedom of the physical Hilbert space.  The $(\mathcal{C},\mathcal{D})$-bimodule category $\mathcal{M}$ then represents the entanglement degrees of freedom of the PEPS tensor depending on $\mathcal{D}$ and $\mathcal{M}$.  The PEPS tensor is represented by $^3\!F$ of Sec. \ref{CSS}, while the MPO tensor corresponds to $^2\!F$. For the case of quantum doubles, $\mathcal{D}$ is given by the group $\mathcal{G}$, and $\mathcal{C}$ can either be chosen to be equal to the irreps of that group, for which $\mathcal{M}$ is trivial, or equal to $\mathcal{G}$, in which case $\mathcal{C}=\mathcal{D}=\mathcal{M}$ and nontrivial 3-cocycles become possible.

The same bimodule categorical objects can also be used to define intertwiners between different PEPS realizations of the same state $|\psi\rangle$ \cite{lootens2020matrix}; the fact that such different realizations exist is a consequence of the fact that the category $\mathcal{D}$ corresponding to the physical degrees of freedom can have several Morita-equivalent categories $\mathcal{C}_i$, each with a compatible $\mathcal{M}_i$; different $\mathcal{M}$ will lead to completely different but locally equivalent PEPS descriptions as $PEPS_{\mathcal{D},\mathcal{M}_1}$ and $PEPS_{\mathcal{D},\mathcal{M}_2}$. Translating the seminal work of  \citet{kitaev:gapped-boundaries} in the tensor network language, these intertwiners can  again be described in terms of MPOs. It is then possible to construct intertwiners relating different physical and/or virtual tensor network representations of quantum doubles and string nets with each other. In particular, the mapping of quantum double models to string net tensor network descriptions can readily be completed  using such MPOs \cite[see also][]{buerschaper2009mapping,kadar2010microscopic}.

\paragraph{SET phases}
Just as in the one dimensional case, virtual and physical symmetries can be combined in a non-trivial way. The corresponding phases are called SET phases. Such systems have at length been studied by \cite{barkeshli:SET}, and the mathematical framework to describe the possible phases is given by graded unitary fusion category theory. Such graded fusion categories can again be realized within the context of matrix product operators, and the non-chiral case gives rise to a PEPS desciption where the MPO symmetry reflects this grading \cite{williamson:SET}. Analogously to SPT phases, string order parameters can be defined to detect the symmetry fractionalization pattern of SET phases that also involve swaps between distant regions, see \citet{Garre_Rubio_2019} for details.

Graded fusion categories also allow to characterize topological phases for fermionic systems in terms of superpivotal categories \cite{aasen2019fermion}. By adopting the language of graded tensor networks, it is possible to realize these phases in terms of fermionic PEPS \cite{bultinck:fermionicPEPS}, for which the Majorana defects can explicitly be constructed. Also the generalization of the toric code to the fermionic case \cite{gu2014lattice}  can readily be understood in terms of these graded tensor networks.  

\paragraph{Chiral phases}

Phases with chiral order~\cite{bernevig:book-topo-ins} exhibit a number of phenomena which distinguish them from the non-chiral topologically ordered states discussed above. In particular, they exhibit protected gapless edge modes characterized by a chiral CFT, whose spectrum is also matched by the entanglement spectrum~\cite{li:es-qhe-sphere}. These phases can either be protected by the fermionic parity superselection rule or by an additional symmetry, such as time-reversal, and can show up both in free fermion and in interacting models, most notably Kitaev's honeycomb model~\cite{kitaev:honeycomb-model}. 
However, some of the properties of chiral systems -- such as the gapless nature of the entanglement spectrum which is suggestive of some kind of ``non-renormalizability'', or the fact that their Wannier functions cannot be locally supported~\cite{kohn:wannier-functions} -- suggest that it might not be possible to describe them as PEPS.

As it turns out, PEPS can describe non-interacting fermionic systems with chiral order exactly~\cite[][an example is given in the Appendix]{dubail:chiral-fpeps,wahl:chiral-fPEPS}.  The resulting free fermion PEPS are ground states of a flat-band Hamiltonian with algebraically decaying interactions, whose bands have non-zero Chern number, i.e., they exhibit non-trivial chiral order. The simplest of these examples is a topological superconductor (protected by fermionic parity); more complex examples such as topological insulators can be e.g.\ constructed from two or more copies thereof.  
The flat-band Hamiltonian exhibits gapless chiral edge modes and a matching chiral entanglement spectrum, and it exhibits correlations (and thus interactions) which decay as $1/r^3$~\cite{wahl:chiral-boundaries}.  The PEPS tensor exhibits a virtual symmetry, characterized by an unoccupied mode.  As shown by~\citet{wahl:chiral-boundaries}, this unoccupied mode is stable under blocking, and thus gives rise to an empty mode formed jointly by one Majorana mode on the left and right boundary each at a given momentum $k$, analogous to the two Majorana edge modes in the Kitaev chain; these edge modes match exactly the point where the edge mode is absorbed into the bulk.  Systems with a higher Chern number (and a higher number of gapless edge modes) correspondingly exhibit a larger number of such symmetries. Moreover, the same symmetry is needed to construct the ground states on the torus -- just as in the case of the Majorana chain [Eq.~\eqref{eq:kitaev-chain-mps}], anti-periodic boundary conditions are required to describe the chiral PEPS wavefunction.

Numerical findings suggest that PEPS can also describe interacting chiral phases. One possibility to construct such interacting models is by Gutzwiller-projecting several copies of a non-interacting topological superconductor or insulator, a construction for which field theory predicts an interacting topological model; numerical study of two copies indeed reveals entanglement spectra consistent with a chiral Kalmeyer-Laughlin state, but at the same time is suggestive of critical correlations~\cite{yang:chiral-topo-fpeps}. An alternative approach put forth by \citet{poilblanc:kl-peps-1,poilblanc:kl-peps-2} is to construct a PEPS from tensors $A$ which themselves possess a ``chiral symmetry'' (that is, which are invariant under combined reflection and conjugation); it is found that the resulting PEPS exhibit entanglement spectra consistent with chiral theories (depending on the additional symmetry such as $\mathrm{SU}(2)$ or $\mathrm{SU}(3)$ encoded in the tensor), as well as algebraic correlations~\cite{poilblanc:kl-peps-1,poilblanc:kl-peps-2,hackenbroich:chiral-su2,chen:peps_su2_2_csl,chen:su3_peps_csl,haegeman:medley}.
In all these cases, a limitation to their analytical study is in the fact that both entanglement spectra and correlations can only be determined numerically, leaving an uncertainty 
in distinguishing a truly chiral entanglement spectrum from a non-chiral one with a very small gap in the entanglement spectrum (and very different velocities of the counterpropagating chiral theories), as well as critical correlations from a very large but finite correlation length~\cite{hackenbroich:chiral-su2}.

While PEPS can represent states with chiral order, there are limitations to their ability to exactly capture chiral systems: As shown by \citet{dubail:chiral-fpeps} and \citet{read:chiral-peps-wannier-functions}, PEPS cannot capture non-interacting 
(both intrinsic and symmetry-protected) chiral states with exponentially decaying correlations exactly; and \citet{mozgunov:knabe-bound-chiral} have shown that the existence of a gapped parent Hamiltonian with periodic boundaries implies an open-boundary spectrum inconsistent with chiral edge modes with linear dispersion (see Sec.~\ref{sec:4:hams-gaps} for a precise statement), providing a partial no-go result also for interacting chiral phases. 

Despite these no-go theorems, the approximation results for the faithful approximations of low-energy states of gapped Hamiltonians (with a suitable density of states) still apply~\cite{hastings:locally,molnar:thermal-peps}, and it has been found numerically that PEPS are well suited to approximate ground and thermal states of both non-interacting
 and interacting Hamiltonians, and allow to simultaneously approximate chiral entanglement spectra and exponentially decaying correlations on the relevant scales   \cite{wahl:chiral-fPEPS,poilblanc:chiral_hafm_squarelattice,chen:peps_su2_2_csl,chen:su3_peps_csl}.

\subsubsection{Entanglement Spectrum and Edge Modes \label{secESEM}}

\paragraph{Entanglement Hamiltonians}
One of the defining properties of PEPS is the fact that the tensors describe how entanglement is routed throughout the system. In practice, tensor networks implement an effective holographic dimensional reduction of the physical degrees of freedom to a 1-dimensional system of entanglement degrees of freedom: all correlations in the PEPS are determined by the fixed points / entanglement Hamiltonians of the 1D transfer matrices of the PEPS. The local MPO symmetries of the tensors immediately translate to MPO symmetries of these transfer matrices. As we have discussed, the symmetries on the virtual level can be much richer than the ones on the physical level, and for topological ordered systems amount to non-trivial MPOs. The results reported in Section~\ref{CSS} then immediately imply that the corresponding entanglement Hamiltonians ought to be critical, hence realizing an explicit tensor network analogue of the fingerprints of conformal field theory in the entanglement spectrum \cite{dubail:entanglement-spectrum,qi:bulk-boundary-duality}.  

The situation is slightly different for SPT phases as for topological phases. In the former case, the transfer matrix has the symmetry $[O_g\otimes \bar{O}_g,T]=0$ and has a unique fixed point $|\rho\rangle$ which hence inherits this symmetry $O_g\rho O_g^\dagger=\rho$. As first observed by \citet{chen:2d-spt-phases-peps-ghz}, whenever the symmetry on the entanglement degrees of freedom is realized through a non-trivial co-cycle, then the corresponding entanglement Hamiltonian $\rho=\exp(-\beta H_E)$ will be critical or symmetry broken. Note that whenever the SPT ground state is a renormalization group fixed point with zero correlation length, the corresponding temperature will be $\beta=0$ and hence one has to perturb the system to witness $H_E$ \cite{bultinck2018global}.

For the case of a genuine topological phase, the MPO symmetries do not have to come in pairs, and any off-diagonal combination is also allowed: $[O_a\otimes \bar{O}_b,T]=0$. This has interesting consequences for the fixed point structure of the corresponding transfer matrix $T$: whenever $\rho$ is a fixed point, then $O_a\rho O_b^\dagger$ will also be a fixed point for any choice of $a,b$. This degeneracy of the fixed points follows from the non-injectivity of the PEPS and is a clear signature
of topological order \cite{schuch:topo-top}. Given $N$ independent MPO-symmetries $O_a$, one would naively think that there will be $N^2$ distinct fixed points. This is however not the case, as there are exactly $N$ fixed points, implying that the entanglement structure exhibits a subtle type of symmetry breaking. Following \cite{haegeman:shadows}, this can be understood from the necessity of having anyons in the system: if the degeneracy would be $N^2$, then the anyons would be confined, and if the degeneracy would be $1$, all anyons would be condensed. Only the case of $N$ different fixed points provides the perfect balance and leads to genuine topological order. A quantum phase transition occurs whenever this fixed point structure changes. 

This result is in complete accordance with the famous formula for the topological entanglement entropy \cite{kitaev:topological-entropy,levin:topological-entropy}, which states that the entanglement entropy of a certain region in the bulk has a $\log(D_q)$ correction with $D_q$ the total quantum dimension of the underlying fusion algebra. This is a direct consquence of the fact that the edges of the block under consideration do not live in the full Hilbert space but are constrained to the MPO-invariant subspace, whose dimension is a constant factor $D_q$ lower than the full one.

\paragraph{Edge Modes}
A remarkable feature of PEPS is the fact that it provides a Hilbert space with a tensor product structure on the edge of a PEPS with open boundary conditions: these are precisely the bonds/entanglement degrees of freedom which are unconnected, and hence span a Hilbert space $D^{L}$ with $D$ the bond dimension and $L$ the number of uncoupled bonds (see Section~\ref{edgetheory}). Note that those degrees of freedom cannot directly be accessed, as they are virtual, but that Hamiltonian terms acting on the physical boundary of the system will induce an effective Hamiltonian on that Hilbert space. Assuming that the bulk is gapped, low-energy modes can emerge on that virtual Hilbert space, and an important question is to study the spectrum and features of these edge excitations. If the system exhibits non-trivial symmetries such as in the case of topological order or SPT, then a fascinating matter occurs: when the effective Hilbert space of the spin chain is projected onto the symmetric or invariant subspace of the MPO symmetries, non-local features emerge which are impossible for normal spin chains. In the language of quantum field theory, the MPOs induce anomalies in the boundary.

Let us first discuss the case of SPT phases. Acting with the physical symmetry $U_g^{\otimes N}$ on the wavefunction induces a non-trivial MPO symmetry action on the boundary. The effective Hamiltonian of the edge modes 
is hence MPO symmetric, and, as discussed in Section~\ref{CSS}, must therefore be either symmetry breaking or critical. Even  in the case of symmetry breaking, there will be a ground state with all symmetries, but it will be realized as a highly entangled GHZ or cat state. This is equivalent to the situation in the 1D AKLT model, where the only $SO(3)$ invariant state is one in which the dangling spins at the ends form a singlet. The ground state of any SPT state with open boundary conditions can therefore not be unique, and from the physics point of view, symmetry breaking will happen.

The situation is very different for the topological case, where the bulk symmetry is purely virtual. This implies that the Hilbert space of the edge modes has to be projected on the MPO symmetric subspace. If $P_s$ is the projector on the $O_g$-invariant subspace, then the Hamiltonian $P_sH_{edge}P_s$ is gappable and has a unique ground state. If the MPO symmetry is non-trivial, that ground state must be MPO-symmetric and hence has to be a GHZ-state. The different components in this GHZ are related to each other by acting with $O_g$ in them. This is a fascinating phenomenon: the physics on the edge induces stable highly entangled states which would be impossible to create in a genuine 1-dimensional system. For the case of quantum doubles, twisted quantum doubles and string nets, the boundaries are always gappable, a fact that is known to follow from the feature that the bulk theory of these systems always yields a Lagrangian subgroup of the anyons \cite{kitaev:gapped-boundaries,levin:Lagrangian}:  the set of  anyons in the bulk can be divided into two sets, one in which every anyon has trivial statistics with respect to each other one, and a second subset for which every one exhibits non-trivial statistics with at least one of the first set. As we will show in the next paragraph, anyons are described by idempotents of an extended MPO algebra, and always satisfy this criterion. The corresponding ground state will be of GHZ-type or not depending on whether the underlying MPO algebra is trivial (such as e.g.\ for the case of quantum doubles) or not (such as for e.g.\ the twisted quantum doubles and string nets).

\subsubsection{Topological sectors and anyons} \label{tsaa}
Two defining features of topological phases are the fact that the ground state degeneracy depends on the genus of the manifold on which the spins are defined, and the fact that the elementary excitations are anyons with non-trivial statistics. The fact that both of these features are intimitly connected to each other is made explicit when studying them from the point of view of PEPS: both are defined by exactly the same tensors \cite{schuch:peps-sym,bultinck:mpo-anyons}.  

\paragraph{Topological sectors}
Let us first discuss the different ground states for a topological spin system on a torus. Consider on the one hand a uniform PEPS, and on the other hand the same uniform PEPS but with a virtual non-trivial MPO winding around one of the directions of the torus. As the PEPS exhibits the MPO symmetry, the location of the MPO is immaterial, and this second state turns out to be orthogonal to the first one. Note that winding two MPOs $O_a$ and $O_b$ along the same direction is equivalent to the sum of $\sum_c N_{ab}^cO_c$, as we can always pull them through the lattice until they touch each other. We could of course also wind an MPO along the other direction, and get another different state. A problem arises however when we try to wind two MPOs along the two different directions: this inevitably leads to a crossing of the two MPOs, and we have to define a new (blue) tensor object defining this crossing:

\be\label{fig:crossing-tensor-topo-sectors}
\includegraphics[width=3cm]{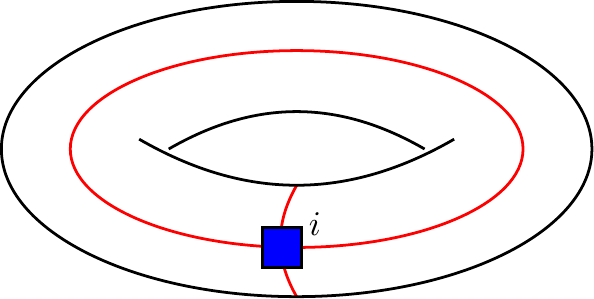}
\ee

By varying this blue tensor, and requiring that it can be pulled through the PEPS tensors, we are able to completely characterize all ground states of the topological theory. Physically, the pulling through property implies that the topological sector is invisible by local measurements on the physical Hilbert space.

It is of course possible to put more MPOs in the PEPS, and doing so induces more and more crossings. However, it should not lead to new ground states, and hence the {\it enlarged} MPOs

\be\label{fig:enlarged-MPO}
\includegraphics[width=3cm]{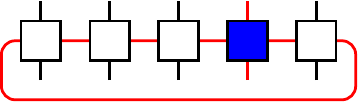}\ee
themselves should form an algebra. It was proven that they form a $C^*$ algebra, which means that they form a closed algebra when multiplying with each other and under conjugation \cite{bultinck:mpo-anyons}.

Such a $C^*$ algebra has a natural decomposition into minimal central idempotents $P_i$, $P_iP_j=\delta_{ij}P_i$, and these different blocks  provide the full set of orthogonal ground states or topological sectors on the torus. From the categorical  point of view, this construction is called the Drinfeld Center \cite{drinfel1987quantum,muger2003subfactors}, and the $C^*$ algebra is called the Ocneanu tube algebra \cite{Evans:ocnenau-tube}. The  states corresponding to the idempotents are the ground states with minimal entanglement \cite{zhang:multigrover_2bit}, which indeed provide a natural partitioning of the ground state sector. From a practical point of view, these idempotents can be calculated from the structure factors of this enlarged MPO algebra. This program can be realized in a straightforward way for all PEPS described in the language of bimodule categories. Similarly, it can be worked out in terms of weak Hopf algebras:  using that physical and virtual indices in the associated MPOs correspond respectively to the algebras $\mathcal{A}$ and $\mathcal{A}^*$, the enlarged MPO gives a representation, as vector space, of $\mathcal{A}\otimes \mathcal{A^*}$. Apart from the algebra structure already commented, there is a natural coproduct, given by $\Delta_\mathcal{A}\otimes \Delta_{\mathcal{A}^*}$, that makes it a weak Hopf algebra. This is called the Drinfeld double and it is precisely the weak Hopf algebra associated to the Drinfeld center category (see Section~\ref{CSS}).

Let us illustrate in the particular case of the G-injective PEPS defined by tensor \eqref{fig:G-injective-tensor-Fig} the process just described that obtains the Drinfeld double of a finite group $G$, just by imposing a pulling through condition on the enlarged MPOs of the associated G-injective PEPS \cite{schuch:peps-sym}.

For that, let us start analyzing the conditions of the crossing tensor needed to pull it through the lattice.  When we move a horizontal virtual MPO $O_g=L_g^{\otimes N}$ one row down, if there is also a vertical virtual MPO $O_h=L_h^{\otimes M}$, the group element of the vertical MPO associated to the position in between the two rows get conjugated by $g$.

\be\label{fig:crossing-tensor-G}
\includegraphics[width=7.5cm]{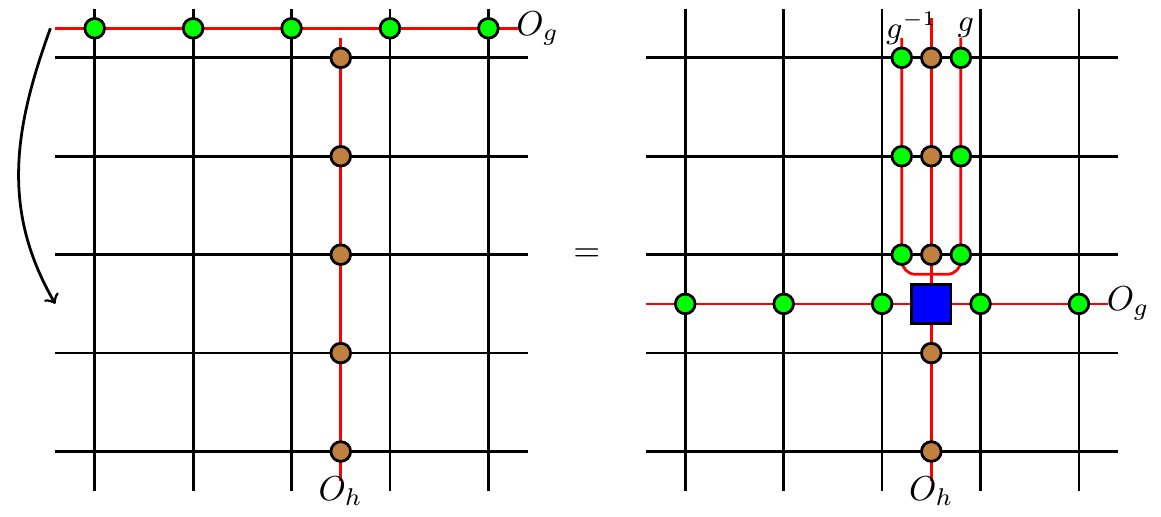}
\ee

 Therefore the crossing tensor needed for that action is precisely a linear combination of tensors of the form
\be\label{eq:elementary-crossing} 
|g)(g|\otimes |ghg^{-1}\rangle\langle h |.
\ee

The enlarged MPO associated to the crossing tensor \eqref{eq:elementary-crossing} is precisely $L_g^{\otimes N}\otimes |ghg^{-1}\rangle\langle h|$, which corresponds to the element $g\otimes \delta_h \in \mathbb{C}(G)\otimes \mathbb{C}^G$, where $\mathbb{C}^G$ denotes the set of functions $G\rightarrow \mathbb{C}$ and $\delta_h$ is the function defined by $\delta_h(k)=1$ for $k=h$ and $0$ otherwise.

Given two pairs $(g,\delta_h)$, $(g',\delta_{h'})$ in $\mathbb{C}(G)\otimes \mathbb{C}^G$, the multiplication induced by their associated enlarged MPOs is given by 
$$(g,\delta_h)\cdot(g',\delta_{h'})= (gg',\delta_{h}(g'h'g'^{-1})\delta_{h'} )$$
since trivially
$$\left(L_g^{\otimes N}\otimes |ghg^{-1}\rangle\langle h|\right)
\cdot \left(L_{g'}^{\otimes N}\otimes |g'h'g'^{-1}\rangle\langle h'|\right)=
$$
$$=L_{gg'}^{\otimes N}\otimes \delta_{h}(g'h'g'^{-1}) |ghg^{-1}\rangle\langle h'|=$$
$$=L_{gg'}^{\otimes N}\otimes \delta_{h}(g'h'g'^{-1}) |(gg')h'(gg')^{-1}\rangle\langle h'|$$

The algebra $\mathbb{C}(G)\otimes \mathbb{C}^G$ with such multiplication is precisely the definition of the Drinfeld double of the group $G$. 

It is shown in \cite{gould1993quantum} that the generating idempotents in this case are given by fixing a conjugacy class, a representative $h\in G$ for it, and an irrep $\alpha$ of the centralizer of $h$, $Z(h)=\{g\in G: gh=hg\}$.
The associated central idempotent is the one given by a crossing tensor proportional to:
$$T_{h,\alpha}= \sum_{k\in G}\sum_{g\in Z(h)} \chi_\alpha(g^{-1})|k^{-1}gk)(k^{-1}gk|\otimes |k^{-1}hk\rangle\langle k^{-1}hk| $$
which gives an enlarged MPO proportional to
$$P_{h,\alpha}=\sum_{k\in G}\sum_{g\in Z(h)} \chi_\alpha(g^{-1}) L_{k^{-1}gk}^{\otimes N}\otimes |k^{-1}hk\rangle\langle k^{-1}hk|$$

Note that the conjugation by $k$ can be absorbed in the PEPS, due to the virtual symmetry, which gives finally the following simpler crossing tensor and enlarged MPO:
$$\sum_{g\in Z(h)} \chi_\alpha(g^{-1})|g)(g|\otimes |h\rangle\langle h| $$
$$P_{h,\alpha}=\sum_{g\in Z(h)} \chi_\alpha(g^{-1}) L_g^{\otimes N}\otimes |h\rangle\langle h|$$

The topological sectors are then indexed by a conjugacy class and an irrep of its centralizer, as expected \cite{kitaev:toriccode}. For the particular case of the toric code, we obtain the projectors $(\Id^{\otimes N} \pm X^{\otimes N}) \otimes |h\rangle\langle h|$ with $h\in \{0,1\}$.
For the general case of bimodule categories, the idempotents can readily be found by diagonalizing linear combinations of the adjoint representation of the $C^*$ algebra \cite{lootens2020matrix}.

\paragraph{Anyons}
In an amazing twist, it turns out that the idempotents defining the ground state manifold on the torus also define the elementary bulk excitations. The orthogonality of the idempotents provides a natural decomposition of the Hilbert space into topological sectors, and some of them have a  non-trivial MPO string attached to them. The corresponding anyons have non-trivial self-statistics, and non-trivial braiding. All of these features can succinctly be understood  from the fact that virtual MPO strings are attached to them, which are immaterial as they can be moved at will through the lattice \cite{schuch:peps-sym,bultinck:mpo-anyons}.  A pair of anyons in the PEPS picture hence has the following form:

\[\includegraphics[width=4cm]{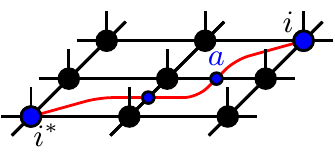}\]

Here the blue tensor projects onto one of the idempotents defined above:

\[\includegraphics[width=7cm]{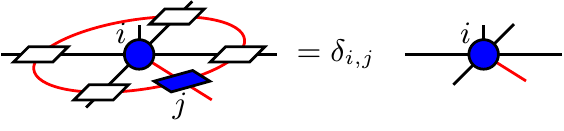}\]

In the particular case of $G$-injective PEPS, the blue tensor takes the form:

\be\label{fig:blue-G}
\includegraphics[width=3.5cm]{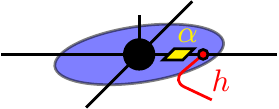}
\ee

The topological spin of an anyon characterizes the phase that the wavefunction acquires when rotating the anyon over $2\pi$:

\[\includegraphics[width=8cm]{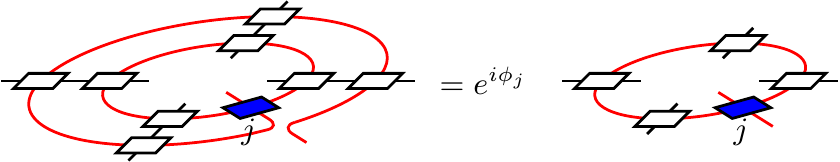}\]

For the case of the toric code, one sees that the anyon characterized by the idempotent $(I-Z^{\otimes N})/2$ with a string attached to it has toplogical spin-$\tfrac12$, and is hence  called a fermion.

Anyons have new fusion rules generating the output category. Those are precisely the ones associated to the Drinfeld center. Crucially, anyons exhibit also braiding properties, encoded in a tensor $\mathcal{R}$, that gives a {\it generalized pulling through equation}:

\[\raisebox{-1.2cm}{\includegraphics[width=3.5cm]{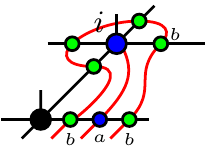}} = 
\raisebox{-1.2cm}{\includegraphics[width=3.5cm]{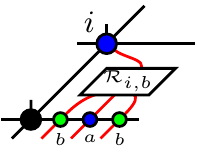}}
\]

With this one can compute the braiding of an anyon around another by doing the following four steps:

\[(1)\includegraphics[width=3.5cm]{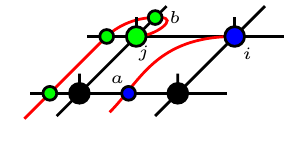} \hspace{.1cm}
(2) \includegraphics[width=3.5cm]{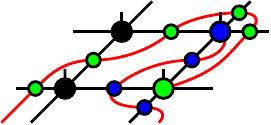}
\]

\[(3)\includegraphics[width=3.5cm]{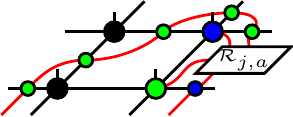} \hspace{.1cm}
(4) \includegraphics[width=3.5cm]{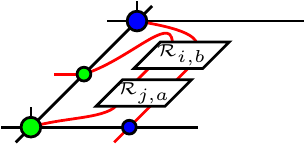}
\]

This idea can readily be exploited to see that the concept of topological quantum computation \cite{kitaev:toriccode,freedman:topological_computation}, in which anyons are braided, can be described in terms of the usual quantum circuit model for quantum computation applied to the entanglement degrees of freedom (see \cite{bultinck:mpo-anyons} for the details).

Braiding takes a simpler and explicit form in the group case. For instance, in case of a trivial irrep $\alpha$, braiding corresponds to conjugation (adjoint action) as it is illustrated in \eqref{fig:crossing-tensor-G}.

Note that distinct input categories can lead to equivalent Drinfeld doubles; this e.g.\ happens for the case of a twisted $Z_2\times Z_2$ and the $Z_4$ quantum double. Two categories with this feature are called Morita equivalent \cite{muger:Morita,kitaev:gapped-boundaries}, and as discussed in Sec. \ref{virtualsym} it is then possible to construct an intertwiner -- in the form of an MPO - between the two corresponding PEPS which preserves all topological features and effectively implements an automorphism of the corresponding topological sectors. This idea has been extended to a systematic study of boundaries between different theories -- e.g.\ between a topological phase and a trivial phase such as with open boundary conditions -- and allows to construct all possible boundary conditions which are compatible with the anyonic bulk physics. For the case of the toric code, this leads to the concept of rough and smooth edges - absorbing electric vs.\ magnetic excitations on the boundary \cite{bravyi:roughandsmooth, lootens2020matrix}.

\paragraph{Anyon condensation}

Although the topological phases described by quantum doubles and string nets are stable with respect to any perturbation \cite{klich:stability,bravyi:tqo-long}, a topological phase transition will occur whenever the perturbation becomes large. From the point of view of PEPS, such a transition is characterized by a change in the MPO symmetry-breaking pattern of the eigenvectors of  the transfer matrix \cite{schuch:topo-top}. In the topological phase described by an input category with $N$ labels (and hence $N$ corresponding MPOs), there are $N$ linearly independent eigenvectors $\sigma_i$ of the transfer matrix with eigenvalues of modulus 1, obtained by acting with the MPO's on a fiducial one $\rho$:  $\sigma_i=O_i\rho$. Note that for e.g.\ the toric code, $\rho=\openone$. A crucial property is that the MPOs $O_i\rho O_j^\dagger$ do not yield new fixed points, and furthermore that $O_i\rho O_i^\dagger$ is not linearly independent of $\rho$. As demonstrated  by \citet{haegeman:shadows}, this can easily be seen to be a necessary condition for the possibility of having anyons, as otherwise the expectation value of a PEPS with a pair of anyons connected with an MPO string would be zero, as illustrated in the following figure:
\[
\includegraphics[height=4.5cm]{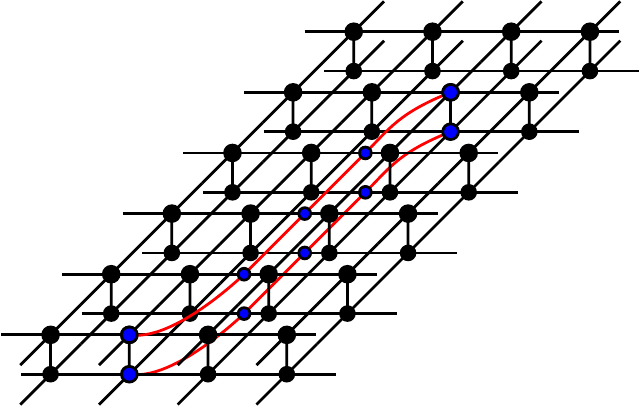}
\]
If the fixed point $O_i\rho O_i^\dagger$ were orthogonal to $\rho$, the norm of the state with 2 anyons connected by a string would decay exponentially in the distance between the anyons, as it would be obtained by raising a mixed transfer matrix with spectral radius strictly smaller than one to the distance between the anyons.  This explains why the eigenvectors of the transfer matrix  must strike a delicate balance of the symmetry breaking pattern. 

This immediately clarifies that a (large) perturbation increasing the number of fixed points of the transfer matrix will lead to confinement of the anyons - and hence to a topological phase transition.  Conversely, the situation where the number of distinct fixed points decreases makes the existence of strings irrelevant -- this corresponds to the condensation of anyons. As shown  by \citet{duivenvoorden:anyon-condensation}, the PEPS description is fully compatible with the standard rules of anyon condensation \cite{bais:quantum-sym-breaking}: 1.\ Only particles with trivial self-statistics can condense; 2.\ anyons become condensed iff they have mutual non-bosonic statistics with some condensed anyon; 3.\ non-condensed anyons which differ by a condensed anyons become indistinguishable; 4.\ anyons which can fuse to two different condensed anyons split up into two distinguishable anyons. Not surprisingly, very similar rules apply in the context of orbifolding in conformal field theory. This connection can indeed be fully established by making use of the formalism of strange correlators \cite{vanhove2018mapping} and of non-invertible bimodule categories \cite{lootens2020matrix} realizing all possible ways in which anyons can condense.

When restricting to the group case, this leads to a natural connection between anyon condensation and SET phases  \cite{garre2017symmetry}. Given a group $G$ and a normal subgroup $H$, there is a natural way of condensing anyons which is restricting the $G$-injective tensor to the subgroup $H$. The strings of the anyons then fullfil the pulling through equation \eqref{pulth} if they come from $H$. If not, they get confined and the pulling through equation gets degraded to one of the form of Eq.~\eqref{pulth-2}, where a unitary must be applied in the physical level. These unitaries form precisely a representation of the quotient group $Q=G/H$. That is, condensing anyons make some global symmetry emerge under which the new condensed model is in an SET phase. Interestingly, one can show that all SET phases appear in this way. This bimodule MPO point of view of anyon condensation is a generalization of this idea.

\section{Formal results: Fundamental Theorems, Hamiltonians%
\label{sec_4}\label{sec:4}}

In this section, we provide formal statements for two types of mathematical questions linked to tensor networks.  In Section~\ref{sec:4:fund-thm-mpv} and Section~\ref{sec:4:fund-thm-peps}, we enunciate the Fundamental Theorems for Matrix Product Vectors and PEPS, respectively. These fundamental theorems relate different MPS representations of the same MPS, MPO, or PEPS, and are essential in the classification of phases under symmetries, renormalization fixed points, and MPO algebras describing topological order.  In Section~\ref{sec:4:hamiltonians}, we discuss the relation of MPS/PEPS and Hamiltonians. In particular, we explain how MPS and PEPS appear as ground states of parent Hamiltonians, and review the known Theorems about their ground space structure, their gap, and their robustness against perturbations.

\subsection{The Fundamental Theorem of Matrix Product Vectors\label{Sec:MPV}\label{sec:4:fund-thm-mpv}}

\subsubsection{Overview}

In the preceding sections, we have encountered different kind of tensor networks: MPS, MPOs, MPDOs, and MPUs.  All these have in common that they live on a sequence of spaces $\mathcal H_d^{\otimes N}$, $N\in\mathbb N$, where $\mathcal H_d$ is the corresponding $d$-dimensional local Hilbert space of states or operators
Therefore, all these cases can be seen as a special case of \emph{Matrix Product Vectors} (MPVs)
\begin{equation}
 \label{MPV}
 |V^{(N)}(A)\rangle = \!\!\!\!\!\sum_{i_1,\ldots,i_N=1}^d\!\! \mathrm{tr}\left( A^{i_1}\ldots  A^{i_N}\right) |i_1\cdots i_N\rangle \in \mathcal H_d^{\otimes N}\ ,
\end{equation}
where the $A^i$ are $D\times D$ matrices. We will denote the family of MPVs generated by $A$ by $\mathcal V(A):=\big\{\ket{V^{(N)}(A)}, N\in\mathbb N\big\}$.

The map $A \mapsto {\cal V}(A)$
is not one-to-one -- that is, different tensors $B$ and $C$ can generate the same family of vectors,  $|V^{(N)}(A)\rangle = |V^{(N)}(B)\rangle\ \forall N$.
This is for instance the case if $A$ and $B$ are related by a similarity transformation (or \emph{gauge transformation}), $B^i=YA^{i}Y^{-1}\
\forall i$, as $Y$ cancels out in Eq.~\eqref{MPV}.
The goal of the Fundamental Theorem of MPVs is to characterize the most general way in which two MPV representations $A$ and $B$ of the same family $\mathcal V(A)=\mathcal V(B)$ are related.

The relevance of the Fundamental Theorem is manifold. It is the basic tool in 
the classification of phases in 1D systems under symmetries, $U^{\otimes N}\ket{V^{N}(A)}=\ket{V^{N}(A)}$, where $B^i=\sum u_{ij}A^j$ and $A^i$ describe the same MPV~(Sec.~\ref{sec:3:1D-SPT}). 
Its application to MPO algebras is relevant in the study of topological order in MPO-injective PEPS~(Sec.~\ref{sec:3:MPO-injective}), as well as the characterization of SPT phases in 2D~(\ref{sec:3:2D-SPT}).  It is also being applied in the characterization of RG fixed points through bulk-boundary correspondence~(Sec.~\ref{sec:2:MPDO-RGFP}), or in the classification of MPUs and Quantum Cellular Automata~(Sec.~\ref{sec:2:MPU}).

The derivation of the Fundamental Theorem consists of two steps.
First, we will show that any MPV tensor $D$ can be brought into a \emph{canonical form} $A$ such that they describe the same MPV family, $\ket{V^{(N)}(A)}=\ket{V^{(N)}(D)}$.
Second, we will present the Fundamental Theorem, which in essence states that given any two tensors $A$ and $B$ in canonical form with  $\ket{V^{(N)}(A)}=\ket{V^{(N)}(B)}$, they are related by a gauge transformation,
\begin{equation}
\label{eq:4:fund-gaugetrafo}
B^i=YA^iY^{-1}\mbox{\ for all $i$}\ .
\end{equation}

The following discussion follows closely \textcite{cirac:mpdo-rgfp}, where further details can be found.

\subsubsection{Canonical form and normal tensors}

In the following, we will introduce the canonical form of MPVs and show how to get a tensor into its canonical form. The reason we require a canonical form is that the similarity transform \eqref{eq:4:fund-gaugetrafo} is not the only way in which two tensors  can generate the same MPV. Consider, for instance, the case where the $B^i$ are upper triangular, e.g. 
 \be \label{eq:upper-triangular}
 B^i = \left(\begin{array}{cc} B^i_1 & B^i_o\\ 0 & B^i_2 \end{array}\right)
 \ ,
 \ee
where $B^i_k$ are $D_k\times D_k$ matrices, and $B^i_o$ is a $D_1\times D_2$ matrix. 
As the off-diagonal block drops out in the trace, the MPV generated by $B$ 
is the same for any choice of $B^i_o$.
The goal of the canonical form is exactly to get rid of those unphysical off-diagonal blocks.

The upper triangular form of $B^i$ can be abstractly characterized as the presence of a subspace $\mathcal S_1$ of dimension $D_1$ which is left invariant under the action of all $B^i$. That is, $B^i \mathcal S_1\subset \mathcal S_1$, or equivalently, denoting by $P_1$ ($Q_1=\Id-P_1$) the orthogonal projector onto $\mathcal S_1$ ($\mathcal S_1^\perp$),
 \be
 B^i P_1 = P_1 B^i P_1,\quad  Q_1 B^i = Q_1 B^i Q_1.
 \ee
Note that this shows that there exist a nontrivial left invariant subspace if and only if there exists a nontrivial right invariant one.

Arguably the easiest way to lift the redundancy  in (\ref{eq:upper-triangular}) is to fix $B^i_o=0$. This corresponds to changing $B^i$ to $P_1B^iP_1+ Q_1B^iQ_1$, which yields the same family of MPVs.  Moreover, there is no loss of generality in assuming that $\mathcal S_1$ does not contain any smaller invariant subspace (if it would, we could choose $\mathcal S_1$ to be that smaller invariant subspace instead).  We thus replace 
\[
B^i \to P_1B^iP_1+ Q_1B^iQ_1
\]
and repeat the argument with the block $Q_1B^iQ_1$ (yielding projections $P_2$), and so on. 
After a finite number of steps, there will be no (non-trivial) invariant subspace anymore. The matrices $\{A^i\}$, defined as 
 \be
 \label{eq:II_Aiplusk1}
 A^i = \sum_{k=1}^r P_k B^i P_k = \bigoplus_{k=1}^r \mu_k A^i_k\ ,
 \ee
generate the same family $\mathcal V(A)=\mathcal V(B)$ of MPV as the initial $B^i$. Note that in a suitable basis, the $A^i$ are all block-diagonal with $r$ blocks.  The positive numbers $\mu_k$ are scaled such that the CP map ${\cal E}_k$, defined through
 \be
 \label{Ek}
 {\cal E}_k (X) = \sum_{i=1}^d A_k^i X A_k^{i\dagger},
 \ee
has spectral radius equal to 1. Note that ${\cal E}_k$ is just the transfer operator of  the tensor $A_k$  associated to the $k$-th block. 

As shown by \textcite{fannes:FCS} and \textcite{perez-garcia:mps-reps} (see also \textcite{wolf:channel-notes}), each CP map ${\cal E}_k$ has a unique eigenvalue $\lambda=1$ and the corresponding left and right eigenvectors are positive and full rank. A CP map with these properties is called {\it irreducible} \cite{wolf:channel-notes}.

However, irreducible CP maps can have other eigenvalues of magnitude one,
always of the form $e^{i 2\pi q/p}$, with $p,q$ integers, $\mathrm{gcd}(q,p)=1$,
and where $p$ is a divisor of $D$. In order to remove them, we block
$p$ spins. This blocking procedure results in a new tensor $C^{i_1,\ldots, i_p}=A^{i_1}\cdots A^{i_p}$. As has been shown by \textcite{cadarso:fractionalization}, the blocked matrices are still block-diagonal, such that the  corresponding transfer operators have a unique eigenvalue of magnitude (and value)
equal to one, and the corresponding left and right eigenvectors are
positive and full rank (as there are no invariant subspaces). A CP map with theses properties is called {\it primitive} \cite{wolf:channel-notes}.

One can now state the main definition of the section.

\begin{defn}
\label{def:4:normal-tensor-mps}
A tensor $A_k$ is called \emph{normal} if its transfer operator $\mathcal E_k$, Eq.~\eqref{Ek}, is a primitive channel. The  corresponding MPV $|V^{N}(A_k)\rangle$ is called a \emph{normal MPV}.
 
We say that a tensor $A$ is in \emph{canonical form} if
\begin{equation}
 \label{eq:II_CF1}
    A^i = \bigoplus_{k=1}^r \mu_k A^i_k\ ,
\end{equation}
and the tensors $A_k$ are normal tensors.
\end{defn}

The above provides an algorithm for transforming any tensor $B$ after blocking into  another tensor $A$ which is in canonical form, such that they both generate the same family of MPVs, $\mathcal V(A)=\mathcal V(B)$.

\subsubsection{Basis of normal tensors}

While the canonical form does no longer suffer from ambiguities due to off-diagonal blocks, there is still a source of ambiguity:
Among the different blocks there could be some that generate the same (or linearly dependent) vectors. In order to properly treat this case one needs to introduce the concept of a basis of normal tensors.

\begin{defn}
\label{def:4:BNT}
A \emph{basis of normal tensors} for $A$ is a set of normal tensors
 $A_j$ ($j=1,\ldots,g$) so that (i) for each $N$, $\ket{V^{(N)}(A)}$ can be written as a linear combination of $V^{(N)}(A_j)$, and (ii) there exists some $N_0$ such that for all $N>N_0$, the $\ket{V^{(N)}(A_j)}$ are linearly independent.
\end{defn}
The following result from \textcite{cirac:mpdo-rgfp} characterizes bases of normal tensors and in particular shows that such a basis always exists.
\begin{prop}\label{prop:char-BNT}
The tensors $A_j$ ($j=1,\ldots,g$) form a basis normal tensors for $A$ if and only if: (i) for all normal tensors $\tilde A_k$ appearing in the canonical form (\ref{eq:II_CF1}) of $A$, there exists a $j$, a non-singular matrix $X_k$, and a phase $\phi_k$ such that  
\be
 \label{eq:II:A=XAX}
 \tilde A_k = e^{i\phi_k} X_k A_j X_k^{-1}
 \ee
  holds; (ii) the set is minimal, in the sense that for any element $A_j$, there is no other $j'$ for which (\ref{eq:II:A=XAX}) is possible.
\end{prop}
Note that given a set of normal tensors, a basis of normal tensors can be constructed (and efficiently obtained numerically) by computing the largest eigenvalue $\lambda_{jk}$ of 
the \emph{mixed transfer operator}  $\mathcal F_{jk}(X)=\sum_i A_j^i X (A_k^i)^\dagger$ and choosing a maximal subset for which $|\lambda_{jk}|<1$ for all pairs $j$, $k$. The $X$ relating $A_j$ and $A_k=e^{i\phi}XA_jX^\dagger$ with $|\lambda_{jk}|=1$ is then obtained by comparing the largest eigenvector $\rho$ of $\mathcal F_{kk}$ with the eigenvector $X\rho_k$ of $\mathcal F_{jk}$; the phase can then be inferred immediately.

One can then write the matrices of any tensor, $A$, in canonical form in terms of a basis of normal tensors $A_j$, as
 \begin{subequations}
 \bea
 \label{eq:II_ABasicTensors}
 A^i &=& 
 \bigoplus_{j=1}^g \bigoplus_{q=1}^{r_j} \mu_{j,q} X_{j,q} A^i_j X_{j,q}^{-1}
 \\
 &=&
 X \left[\bigoplus_{j=1}^g \left(M_j \otimes A^i_j\right)\right] X^{-1} \ ,
 \eea
 \end{subequations}
where $M_j$ is a diagonal matrix with entries $\mu_{j,q}$,
and
 \be
 \label{eq:II_X}
 X= \bigoplus_{j=1}^g \bigoplus_{q=1}^{r_j} X_{j,q},
 \ee
so that
 \be
 \label{decBSV}
 |V^{N}(A)\rangle = \sum_{j=1}^g \left(\sum_{q=1}^{r_j} \mu_{j,q}^N \right) |V^{(N)}(A_j)\rangle\ .
 \ee
 
\subsubsection{Fundamental Theorem of MPVs}

We are now ready to state the \emph{Fundamental Theorem of Matrix Product Vectors} \cite{cirac:mpdo-rgfp}. It clarifies which degree of freedom is left for two $A$ and $B$ in canonical form which generate families of MPVs which are proportional -- or equal -- to each other:
\begin{thm}[Fundamental Theorem for proportional MPVs]
\label{thm1}
Let $A$ and $B$ be two tensors in canonical form, with bases of normal tensors $A^i_{k_a}$ and $B^i_{k_b}$ ($k_{a,b}=1,\dots,g_{a,b}$), respectively. If for all $N$, $A$ and $B$ generate MPV that are proportional to each other, then: (i) $g_a=g_b=:g$; (ii) for all $k$, there exists a $j_k$, phases $\phi_{k}$, and non-singular matrices $X_{k}$ such that $B^i_{k}=e^{i\phi_{k}}X_{k}A^i_{j_k}X_{k}^{-1}$.
\end{thm}

\begin{cor}[Fundamental Theorem for equal MPVs]
\label{II_cor2}
If two tensors $A$ and $B$ in canonical form generate the same MPV for all $N$ then: (i) the dimensions of the matrices $A^i$ and $B^i$ coincide; (ii) there exists an invertible matrix, $X$, such that
 $A^i = X B^i X^{-1}$.
\end{cor}
The $X_k$ and $\phi_k$ can again be obtained constructively and numerically efficiently following the procedure described after 
Proposition~\ref{prop:char-BNT}.

Note that one can select a gauge such that 
the CP maps $\mathcal{E}_k$ associated to the normal tensors are unital, i.e.\ $\mathcal E_k(\openone)=\openone$ (by replacing $A_k$ with $\rho^{-1/2}A_k \rho^{1/2}$, with $\rho$ the fixed point of $\mathcal E_k$). In that case, both Theorem \ref{thm1} and Corollary \ref{II_cor2} hold with the extra condition that $X$ and $X_k$ are unitary matrices.

The Fundamental Theorem of MPV can be generalized to hold without the need of blocking to remove $p$-periodic components \cite{delascuevas:fund-thm-periodic}, which allows to apply the Fundamental Theorem to analyze e.g.\ symmetries with regard to the original unit cell.
We state only the analogue of Corollary \ref{II_cor2}., see \textcite{delascuevas:fund-thm-periodic} for the analogue of Theorem \ref{thm1}. In \textcite{delascuevas:mps-cont-limit}, this result has been applied to the analysis of the existence of a continuum limit in the context of MPS.

\begin{thm}\label{thm:fundamental-general}
Let $A$ and $B$ be tensors in block diagonal form as in (\ref{eq:II_Aiplusk1}), that is, the CP map of each block is irreducible. If $|V^{(N)}(A)\rangle =|V^{(N)}(B)\rangle$ for all $N$, then there exists a diagonal matrix $Z$, and an invertible matrix $Y$ so that 
\begin{enumerate}
\item[(i)] $[Z,A^{i}]=0$ for all $i$,
\item[(ii)] $ZA^{i}=YB^i Y^{-1}$ for all $i$, and
\item[(iii)] $|V^{(N)}(A)\rangle= |V^{(N)}(ZA)\rangle$ for all $N$.
\end{enumerate}
\end{thm}
The idea behind the appearance of the diagonal matrix $Z$ is that the MPV associated to a block $A_k$ whose CP map has eigenvalues $e^{i 2\pi q/p}$ satisfy that $|V^{(N)}(A_k)\rangle =0$ unless $N$ is a multiple of $p$. Hence we can multiply this block by any complex $p$-th root of unity and the whole MPV $|V^{(N)}(A)\rangle$ will not be affected, giving an extra degree of freedom. What is proven in Theorem \ref{thm:fundamental-general} is that this is essentially the only extra freedom one gets in the general case without blocking.

\subsection{Fundamental Theorems for PEPS\label{sec:4:fund-thm-peps}}

Fundamental Theorems for PEPS only exist for a number of special cases.  Let us first discuss the case of  \emph{normal PEPS}.

\begin{defn}
\label{def:4:injectivity}
A PEPS tensor $A$  is called injective if it is injective as a linear map $A:
(\mathbb C^D)^{\otimes r}\to\mathbb C^d$ from the virtual to the physical system (with $r$ the coordination number), that is, if there exists a left-inverse $A^{-1}$,
\[
A^{-1}A=\openone_{(\mathbb{C}^D)^{\otimes r}}\ .
\]
\end{defn}

\begin{defn}
A tensor $A$ generating a 2D PEPS is called \emph{normal} if it becomes injective after blocking a sufficiently large rectangular region $H\times V$.
\end{defn}
Tensors which are injective on two regions are also injective on their union, and thus if a normal tensor $A$ becomes injective in a region of size $H\times V$, it is also injective for any region $\tilde{H}\times \tilde{V}$ with $\tilde{H}\ge H$ and $\tilde{V}\ge V$~\cite{perez-garcia:parent-ham-2d,molnar:normal-peps-fundamentalthm}. It has beeen shown that if $A$ is normal, the required $H$ and $V$ are upper bounded by a constant which only depends on the bond dimension $D$ and the graph, but not the specific $A$~\cite{michalek:2d-quantum-wielandt}.

For MPS, the notion of normality defined above is equivalent to the previously introduced notion (Definition~\ref{def:4:normal-tensor-mps})
of a normal tensor~\cite{sanz:wielandt}:
Injectivity implies  that the tensor $A$ must be normal, and conversely, the quantum version of Wieland's theorem states that after blocking at most 
$2D^2(6+\log_2D)$ sites, every normal tensor becomes injective (\textcite{michalek:wielandt-Dsquare}, see also \textcite{perez-garcia:inj-peps-syms,rahaman:wielandt-Dsquare}).

This yields the following version of the Fundamental Theorem for normal PEPS shown in \textcite{molnar:normal-peps-fundamentalthm} (an earlier version of which was proven in \textcite{perez-garcia:inj-peps-syms}):

\begin{thm}[Fundamental Theorem for normal PEPS]
\label{thm:canonical-2D}
Let $A$ and $B$ be two normal tensors such that every $H\times L$ region is injective, and let $A$ and $B$ generate the same PEPS for some system size $n\times m\ge (2H+1)\times (2L+1)$. Then, 
there exist invertible matrices $X,Y$ and $\lambda \in \mathbb{C}$ such that 
$A=\lambda\, B\,(X^{-1}\otimes Y^{-1} \otimes X\otimes Y)$, and $\lambda^{nm}=1$.
\end{thm}
     
Based on it we have the desired 2D analogue of Corollary \ref{II_cor2} for normal 2D PEPS.

\begin{cor}
Let $A$ and $B$ be two normal tensors generating 2D PEPSs. Then they define the same state for all sizes if and only if there exist invertible matrices $X,Y$ such that $ A^i= B^i(X^{-1}\otimes Y^{-1} \otimes X\otimes Y)$ for all $i$. Moreover $X$ and $Y$ are unique up to proportionality.
\end{cor} 

Just as in one dimension, this theorem in particular provides a local characterization of all normal PEPS having a global on-site symmetry or an spatial symmetry, see \textcite{perez-garcia:inj-peps-syms} for details.

Beyond the normal case, a fundamental theorem for PEPS has also been proven for so-called \emph{semi-injective PEPS} (see section \ref{sec:3:MPO-injective}), which in particular encompass 2D SPT phases such as the CZX model of \textcite{chen:2d-spt-phases-peps-ghz} and its generalizations \cite{chen:spt-order-and-cohomology,williamson:mpo-spt}; there, the relation between two PEPS tensors $A$ and $B$ generating the same state is given by Matrix Product Operators instead~\cite{molnar:quasi-injective-peps}. Using the theory of bimodule categories, it is also possible to construct such MPO intertwiners between equivalent PEPS with different bond dimensions \cite{lootens2020matrix}.

Let us note that one cannot hope for a Fundamental Theory for PEPS in the same generality as in 1D, since the corresponding problem in its full generality is undecidable~\cite{scarpa:peps-zero-testing}.

\subsection{Hamiltonians%
\label{sec:4:hamiltonians}}

In this section, we discuss the relation of MPS and PEPS with Hamiltonians.  In particular, we provide the construction of the parent Hamiltonian, the precise conditions under which it has a unique ground state or a ground space with a controlled degeneracy, and discuss the conditions under which these Hamiltonians can be proven to be gapped. We also discuss constructions which provide alternative Hamiltonians associated to an MPS. These results extend the seminal results of \citet*{affleck:aklt-prl,affleck:aklt-cmp} and \citet*{fannes:FCS} on exact parent Hamiltonians for the AKLT states and for finitely correlated states.

\subsubsection{Parent Hamiltonians and ground space}

\paragraph{Construction of the parent Hamiltonian} We start with the 1D case. Given an MPS, consider the space
\begin{equation}
\label{eq:4:GL-for-parent}
\mathcal{G}_{L}=\left\{
    \sum_{i_1,\ldots i_{L}} 
    \tr(A^{i_1}\cdots A^{i_{L}}X) |i_1\cdots i_{L}\rangle : X\in \mathcal{M}_D\right\}
\end{equation}
of all states spanned by $L$ consecutive sites of the MPS, given arbitrary boundary conditions $X$. 
Graphically, this corresponds to the states
\be
\raisebox{-2em}{\includegraphics[height=4em]{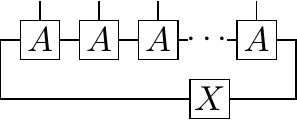}}\ .
\ee
In some cases, we will also write $\mathcal G_{i,\dots, j}$ to denote $\mathcal G_{i-j+1}$ on sites $i,\dots, j$.
\begin{defn}[Parent Hamiltonian]
 A parent interaction is any hermitian positive semidefinite operator $h\ge0$ acting on $L$ sites, whose kernel equals $\mathcal G_L$, Eq.~\eqref{eq:4:GL-for-parent}. The corresponding \emph{parent Hamiltonian} of the MPS with tensor $A$ on $N$ sites is then given by $H_N=\sum_{i=1}^N h_i$, where $h_i$ denotes $h$ acting on sites $i,\dots,i+L-1$ mod$(N)$.
\end{defn}
Since the dimension of $\mathcal G_L$ is at most $D^2$, the parent Hamiltonian will necessarily be non-trivial as soon as $d^L>D^2$ (as $\mathcal G_L$ cannot be the full space).

Similarly, we can define parent Hamiltonians in two dimensions (or on other graphs) by considering a sufficiently large region $R$, where $\mathcal G_R$ is the space spanned by the states 
\be
\label{fig:Parent-Ham-2D}
{\includegraphics[height=5em]{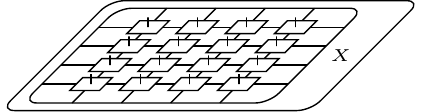}}
\ee
with arbitrary boundary conditions $X$, and the terms in the Hamiltonian are again positive semidefinite operators with kernel $\mathcal G_R$. Note that in two dimensions, the resulting parent Hamiltonian can have different types of terms if one considers more then one type of region (e.g.\ two rectangular regions of size $2\times 1$ and $1\times 2$). Note that the same construction can be carried out without translational invariance, and for general graphs.

It is clear that the parent Hamiltonian has the MPS/PEPS $\ket\psi$ as a ground state, as $H_N=\sum h_i\ge0$ and $H_N\ket\psi = \sum h_i\ket\psi = 0$. Hamiltonians with the property that the ground states minimize the local terms are called \emph{frustration free}. In the following, we will discuss the conditions under which $H_N$ has a unique ground state, or more generally a ground space with a controlled structure.

\paragraph{Normality, injectivity and unique ground states.}

Let us first recall the notion of injectivity from Sec.~\ref{sec:4:fund-thm-peps}, Definition~\ref{def:4:injectivity}: An MPS or PEPS tensor is injective if it has a left-inverse $A^{-1}$ when considered as a map from virtual to physical system, $A^{-1}A=\openone$. For MPS, this is equivalent to the property that the  matrices $A^i$ span the whole set of ${\cal M}_{D\times D}$ matrices. 
Let us also recall that any injective MPS tensor is normal, and any normal MPS tensor becomes injective after blocking at most $L_0\le 2D^2(6+\log_2D)$ sites.

For an MPS or PEPS with injective tensors, it can be easily be proven that the parent Hamiltonian defined on nearest neighbors has a unique ground state. To this end, it is most convenient to construct the PEPS as in section \ref{MPSandPEPS}  by applying local linear maps to maximally entangled pairs. 
 That is, given a regular graph $G$ of degree $r$ with vertex set $V$ and edge set $E$, one can consider the so called \emph{isometric PEPS} of bond dimension $D$,
\[
\ket{\Omega^G}:=\bigotimes_{e\in E} |\omega\rangle_e\ ,
\]
where $|\omega\rangle$ is the maximally entangled state of dimension $D$. A general MPS or PEPS $\ket{\Psi^G(A)}$ is then given by a linear map (the tensor) $A:  (\mathbb{C}^D)^{\otimes r}\rightarrow \mathbb{C}^d$ 
\be\label{eq:4:peps-as-peps} 
|\Psi^G(A)\rangle :=
    \Big(\bigotimes_{v\in V} A\Big)\bigotimes_{e\in E} |\omega\rangle_e
    =\Big(\bigotimes_{v\in V} A\Big)\ket{\Omega^G}\ ,
\ee
cf.~Fig.~\ref{sec2fig1}. We omit the superscript $G$ whenever the graph is unambiguous.

For the isometric PEPS $\ket\Omega$, a possible parent Hamiltonian is given by $h=1-\ket{\omega}\bra\omega$, and it is immediate to see that $H_N=\sum h_i$ has $\ket\psi$ as its unique ground state. We can now construct a parent Hamiltonian for $\ket{\Psi(A)}$ as
\begin{equation}
    \label{eq:4:deformed-parent-1}
h' = (A^{-1}\otimes A^{-1})^\dagger h (A^{-1}\otimes A^{-1})\ .
\end{equation}
Clearly, the kernel of $h'$ is $\mathcal G_L$, and thus, $\ket{\Psi(A)}$ is a ground state of $H'_N=\sum h'_i$. Now assume $H'_N$ had another ground state, $\ket\Phi$, i.e., $h_i'\ket\Phi=0$. Then, 
$h_i(A^{-1})^{\otimes N}\ket\Phi = Xh_i'\ket\Phi=0$ (here, $X$ acts as $A^\dagger$ on sites $i$, $i+1$, and $(A^{-1})$ on all others), i.e., $(A^{-1})^{\otimes N}$ would be another ground state of $H_N$, whose ground state is however unique. Thus, the ground state of $H_N'$ must be unique as well. (This technique can more generally be seen as establishing a one-to-one correspondence between ground states of the parent Hamiltonians of two PEPS $\ket\psi$ and $\ket{\psi'}=R^{\otimes N}\ket A$ which are related by an invertible map $R$ -- not necessarily the PEPS map $A$ -- and thus can also be applied e.g.\ to the cases with topological or otherwise degenerate ground space structure described in the next subsection.)

This leads to the following result~\cite{perez-garcia:parent-ham-2d}:
\begin{thm}
\label{thm:4:unique-gs-2L0}
Consider a PEPS where the tensors have been blocked such that all tensors are injective. Then, the $2$-body parent Hamiltonian constructed from all nearest neighbor sites has a unique ground state. This results holds independent of the graph and translational invariance.

In particular, given a square lattice, if injectivity is reached by blocking $H\times V$ sites, then the parent Hamiltonians containing all terms derived for patches of size $H\times (2V)$ 
and $(2H)\times V$ has a unique ground state.  In 1D, the result correspondingly applies for parent Hamiltonians acting on $2L_0$ sites, with $L_0$ the injectivity length ~\cite{perez-garcia:mps-reps}.
\end{thm}
The injectivity length of a normal MPS is the smallest number of sites which have to be blocked such that its tensors become injective. This result can be considerably strengthened -- yielding more local Hamiltonians -- by avoiding to block until injectivity is reached. For a normal MPS, consider two Hamiltonian terms $h$ and $h'$ acting on sites $1,\dots,L_0+1$ and $2,\dots, L_0+2$, where $L_0$ is the injectivity length, and let us denote by $A$, $B$, and $C$ the tensors on site $1$, the blocked tensor of site $2,\dots, L_0+1$, and the tensor at site $L_0+2$, respectively; note that $B$ is injective.  The joint ground space of $h$ and $h'$ is the intersection $\mathcal I=\mathcal G_{1,\dots,L_0+1}\cap \mathcal G_{2,\dots,L_0+2}$, 
i.e., all states of the form
\begin{equation}
    \label{eq:4:initiate-intersection-proof}
\raisebox{-1.6em}{\includegraphics[width=3cm]{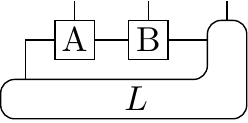}}\quad=\quad
\raisebox{-1.6em}{\includegraphics[width=3cm]{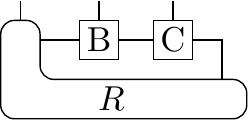}}
\end{equation}
for some boundary conditions $L$ and $R$. We can invert $A$, which yields 
\[
\raisebox{-1.6em}{\includegraphics[width=3cm]{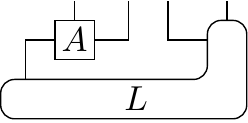}}\quad=\quad
\raisebox{-1.6em}{\includegraphics[width=3cm]{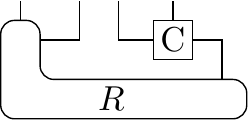}}
\]
We can now re-attach (``grow back'') $B$ to $A$:
\[
\raisebox{-1.6em}{\includegraphics[width=3cm]{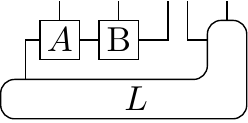}}\quad=\quad
\raisebox{-1.6em}{\includegraphics[width=3cm]{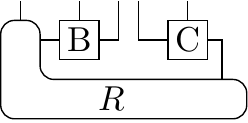}}
\]
Since $B$ is injective, so are $A$ and $B$ jointly, and we can invert them (calling the inverse $S$):
\[
\raisebox{-1.6em}{\includegraphics[width=3cm]{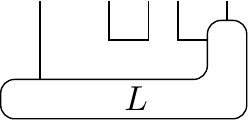}}\quad=\quad
\raisebox{-1.6em}{\includegraphics[width=3cm]{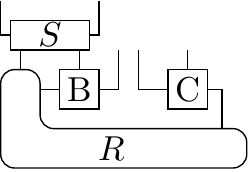}}
\]
We thus find that 
\[
\raisebox{-1.6em}{\includegraphics[width=3cm]{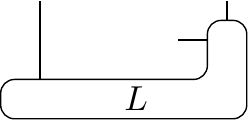}}\quad=\quad
\raisebox{-1.6em}{\includegraphics[width=3cm]{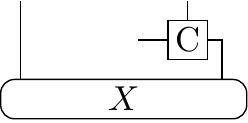}}
\]
i.e., any state in the intersection $\mathcal I$ is also contained in $\mathcal G_{1,\dots,L_0+2}$;
and the converse is trivially true. 
We can now iterate this argument to show that the ground space of a Hamiltonian with terms acting on $L_0+1$ sites is the same as for a Hamiltonian with terms acting on $L_0+k$, $k>1$, sites. Once we have reached $k=L_0$, we can resort to the above Theorem, or alternatively apply a similar argument when closing the boundaries.
(These two properties are called the \emph{intersection property} and \emph{closure property}, respectively. \cite{fannes:FCS,perez-garcia:mps-reps,schuch:peps-sym}) 

The same technique (inverting and growing back) also works in two dimensions; e.g., on the square lattice we would start from the equality 
\[
\raisebox{-1.6em}{\includegraphics[width=4cm]{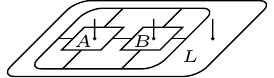}}\,=\,
\raisebox{-1.6em}{\includegraphics[width=4cm]{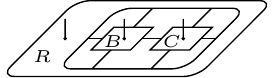}}
\]
and apply the same arguments as above.  This yields the following strengthened result~\cite{fannes:FCS,perez-garcia:mps-reps,schuch:rvb-kagome}:
\begin{thm}[Uniqueness of ground state]
\label{thm:4:unique-gs-L0_plus_1}
Consider a normal MPS which becomes injective upon blocking $L_0$ sites. Then, the parent Hamiltonian defined on $L_0+1$ sites has a unique ground state.

Consider a normal PEPS on the square lattice which becomes injective upon blocking $H_0\times V_0$ sites. Then, the parent Hamiltonian defined on $(H_0+1)\times V_0$ and $H_0\times (V_0+1)$ sites has a unique ground state.

Both results hold independent of translational invariance.
\end{thm}
Note that the locality bound need not be tight. E.g., the AKLT model has $L_0=2$, yet the two-site parent Hamiltonian is suffient to obtain a unique ground state. This can be seen e.g.\ by checking by hand that $\mathcal G_{1,2}\cap \mathcal G_{2,3}=\mathcal G_{1,2,3}$.

For MPS, this provides a full characterization of all MPS which appear as unique ground states of local Hamitonians: Non-normal MPS, which have more than one block in their  canonical form, exhibit long-range order, and we will see in the next section that their parent Hamiltonian exhibits a degenerate ground state subspace. 

For PEPS, there exist classes of states that even not being injective, the are unique ground states of a parent Hamiltonian. This holds for all states constructed analogously to \eqref{eq:4:peps-as-peps}, by replacing $\ket\Omega$ with any other state which is a unique ground state of a frustration free local Hamiltonian. For instance, this is the case for semi-injective PEPS (\textcite{molnar:quasi-injective-peps}, cf. Secs.~\ref{sec:3:2D-SPT} and~\ref{sec:4:fund-thm-peps}), where $\ket\Omega$ is a product of entangled states across plaquettes, and which is a unique ground state of a $4$-body Hamiltonian (acting on each plaquette).
Another class of PEPS with unique ground states are given by the family of MPO-injective PEPS that fulfill that the MPO has a single block in the canonical form (Sec.~\ref{sec:4:fund-thm-mpv}).

\paragraph{Block-injectivity, entanglement symmetries, and degenerate ground spaces}

The proof for the uniqueness of the ground state of parent Hamiltonians breaks down once the correspondence between physical and virtual system captured by the concept of injectivity is lost -- that is, if there are degrees of freedom in the virtual space which have no physical correspondence.
For instance, this happens for MPS which have more than one block ($r>1)$ in their  canonical form
Eq.~\eqref{eq:II_CF1}: In that case, one can easily see that any state corresponding to a single block 
$A_k^i$ is a ground state, as they are all supported on $\mathcal G_L$, Eq.~\eqref{eq:4:GL-for-parent}. 
In this scenario, we can define a generalization of injective tensors, where 
the physical-virtual correspondence holds blockwise (that is, for block-diagonal boundary conditions $X$ in \eqref{eq:4:GL-for-parent}).
\begin{defn}
A tensor $A$ is in \emph{block injective canonical form} if it is in canonical form, and for each element $X\in \bigoplus_{j=1}^g {\cal M}_{D_j\times D_j}$ there exists a vector $c(X)$ such that
$X=\sum_i c_i(X) \tilde A^i$, where $\tilde A^i:=\bigoplus_{i=1}^g A^i_j$, and $A_j$ are a basis of normal tensors 
(cf.\ Def.~\ref{def:4:BNT}) of $A$.
\end{defn}
\noindent
Using the quantum version of Wielandt's theorem \cite{sanz:wielandt}, it was shown by \citet{perez-garcia:mps-reps} and \citet{cirac:mpdo-rgfp}
that after blocking at most $L_0\le 3D^5$ spins, any tensor $A$ in canonical form acquires block injective canonical form.  One can then  prove
a generalization of Theorem~\ref{thm:4:unique-gs-2L0}.
\begin{thm}[\citet{fannes:FCS,perez-garcia:mps-reps}]
For any $N\ge 2L_0$, the ground space of any parent Hamiltonian is exactly the vector space generated by a basis of normal tensors  of the initial tensor $A$. 
\end{thm}
Following the same steps as following Eq.~\eqref{eq:4:initiate-intersection-proof} (inverting and re-growing tensors, where the inverse projects on the space of block-diagonal matrices, and restricting to boundary conditions $L$ and $R$ which are themselves block-diagonal), we can strengthen this result in analogy to Theorem~\ref{thm:4:unique-gs-L0_plus_1}:
\begin{thm}
For any $N\ge L_0+1$, the ground space of any parent Hamiltonian is exactly the vector space generated by a basis of normal tensors  of the initial tensor~$A$. 
\end{thm}

The reason for the degenerate ground space can be understood from the fact that such MPS have decoupled blocks in the virtual space. Alternatively, this can be explained from a symmetry $[A^i,U_g]=0$ of the MPS tensor $A^i$ with $U_g$ a unitary representation of an abelian group (where the blocks are supported on the different irreducible representations): In that case, $[A^i,U_g]=0$ implies that projectors $P_\alpha=\sum \overline{\chi_k(g)} U_g$ onto irreps (blocks) $k$ commute with $A$, and thus cannot be detected by the parent Hamiltonian; placing them on a link thus selects different ground states. These two perspectives suggest two different generalizations to two dimensions: First, we can choose a 2D PEPS tensor with a ``direct sum'' block structure over \emph{all} virtual indices as in  \eqref{eq:II_CF1}, such as in the PEPS for the GHZ state (Sec.~\ref{sec:examples:2D:GHZ}).
Such a PEPS will have a GHZ-type structure; in particular, if under blocking injectivity of all block is reached, the parent Hamiltonian with terms acting on two blocks will have a ground space spanned by the individual blocks. 

The second generalization to PEPS is based on generalizing the commutation relation $U_gA^iU_g^\dagger=A^i$ to tensors which are invariant under the action of some symmetry on all the indices simultaneously.  This in particular encompasses the $G$-injective PEPS and the MPO-injective PEPS introduced in Sec.~\ref{sec:3:2D}. In that case, the pulling though condition is exacly what allows to create projections onto different sectors which can be commuted (i.e.\ moved) through the tensor network and are thus invisible to the Hamiltonian.  In the case of $G$-injective PEPS, the condition for a controlled ground space is precisely that the blocked PEPS tensor in injective on the subspace which is invariant under the symmetry action (``$G$-injective'').  We refer the reader to \onlinecite{schuch:peps-sym} for the formal result, and to \onlinecite{buerschaper:twisted-injectivity}  for the generalization to twisted $G$-injective PEPS and to 
\onlinecite{sahinoglu:mpo-injectivity,bultinck:mpo-anyons} for the generalization to 
MPO-injective PEPS, respectively. Let us also note that tricks similar to the ``re-growing'' used in 1D to construct parent Hamiltonians on $L_0+1$ sites can also be used in the case of topologically ordered PEPS in 2D to obtain smaller Hamiltonians, this has e.g.\ been carried out by \citet[][Appendix D]{schuch:rvb-kagome} for the kagome RVB model to obtain a two-star Hamiltonian (which can be broken down to a one-star Hamiltonian by direct inspection \cite{zhou:rvb-parent-onestar}, similar to the 2-body Hamiltonian for the AKLT model).

\paragraph{Converse: MPS ground states for frustration free Hamiltonians}

As we have seen, every MPS is the ground state of a frustration free Hamiltonian. Remarkably, under certain conditions the converse holds as well: Every frustration free Hamiltonian has an MPS ground state.  

The following result is due to \textcite{matsui:gapped-ham-vs-mps} 
(generalizations of this connection have been shown recently by 
\textcite{ogata:gapped-ham-vs-mps-I,ogata:gapped-ham-vs-mps-II,ogata:gapped-ham-vs-mps-III}).

\begin{thm}
Let $h\ge0$ be a Hamiltonian acting on $r+1$ sites, and let
$H_{[M,N]}=\sum_{i=M}^{N-r} h_i$, where $h_i$ is $h$ acting on sites $i,\dots,i+r$ (i.e., the translational invariant open boundary condition Hamiltonian with interaction $h$).

If there exist $C>0$ so that the dimension of the kernel of $H_{[M,N]}$ is $\le C$ for all $M,N$, then any translational invariant frustration free ground state $\ket\phi$ in the thermodynamic limit can be described by an MPS with a normal tensor.
\end{thm}
We refer the reader to the original works for the precise mathematical formulation of the result in the thermodynamic limit.

\subsubsection{Gaps\label{sec:4:hams-gaps}}

Having defined parent Hamiltonians and given conditions which control their ground space structure, let us now turn towards the question when these Hamiltonians are gapped.  As we have seen, parent Hamitonians of MPS are frustration free, i.e., the ground state minimizes each interaction term individually.  There are two techniques to make statements about gaps of frustration free Hamiltonians: The martingale method and Knabe-type bounds. Both of them relate the gap in the thermodynamic limit to the gap of a finite-size problem, which can subsequently be solved numerically or analytically.  Most importantly, it turns out that the martingale method allows to prove the existence of a gap, and to provide explicit lower bounds on it, for all MPS parent Hamiltonians.

Recently, these methods have been used to prove the gap of a class of decorated AKLT models \cite{abdul2019class}, as well as to numerically show the gap of a range of models (among others, the honeycomb AKLT model) by numerically checking the corresponding finite-size problems \cite{lemm2019existence,pomata2019demonstrating}.

\paragraph{The martingale method}

The martingale method \cite{fannes:FCS,nachtergaele:degen-mps,kastoryano:martingale} relates the minimum non-zero angle between the ground spaces of overlapping regions with the gap: The system is gapped in the thermodynamic limit if and only if by blocking, the overlaps of vectors in overlapping ground spaces become sufficiently large (which intuitively allows to detect excitations locally).

More precisely, consider a frustration-free Hamiltonian $H=\sum h_i$ with w.l.o.g.\ projectors $h_i$. (If the $h_i$ are not projectors, they are still lower and upper bounded by projectors up to a constant.) Having a gap $\gamma$ is equivalent to $H^2 \ge \gamma H$, which (using $h_i^2=h_i$) is equivalent to
\begin{equation}
\label{eq:4:martingale-1}
\sum h_i + \sum\nolimits' h_i h_j + \sum\nolimits'' h_ih_j \ge \gamma \sum h_i\ ,
\end{equation}
where $\sum'$ and $\sum''$ denote sums over overlapping and non-overlapping $h_i$, respectively.  $\sum''h_ih_j\ge0$, and thus, \eqref{eq:4:martingale-1} is satisfied as long as 
\begin{equation}
\label{eq:4:martingale-2}
h_ih_j+h_jh_i \ge -c_{ij} (1-\gamma)(h_i+h_j)
\end{equation}
for all overlapping pairs $(i,j)$, where $c_{ij}$ has to be chosen to add up to $1$ (e.g., if each $h_i$ overlaps with three others, $c_{ij}=1/3$).  Thus, finding a blocking for which \eqref{eq:4:martingale-2} is thus sufficient to prove a gap. Note that \eqref{eq:4:martingale-2} effectively poses a lower bound on the smallest non-zero angle between the ground spaces of $h_i$ and $h_j$.

Remarkably, the martingale condition \eqref{eq:4:martingale-2} is also sufficient: As shown by \textcite{kastoryano:martingale}, whenever a frustration free Hamiltonian is gapped, $\delta(\ell):=\|h_ih_j-P\|$ (with $P$ the projector onto the kernel of $h_i+h_j$) goes to zero exponentially with universal constants in the size $\ell$ of the overlap region, which in particular implies validity of \eqref{eq:4:martingale-2} (as both quantities only depend on the principal angles between $\ker h_i$ and $\ker h_j$).

Using the martingale method, one can prove that all MPS parent Hamiltonians (both for normal and block-injective MPS) have a gap \cite{fannes:FCS,nachtergaele:degen-mps}:
\begin{thm}
All MPS parent Hamiltonians are gapped; this is, there exists a $\gamma>0$ such that for the parent Hamiltonian $H_N$ on $N$ sites, the smallest non-zero eigenvalue $\lambda(H_N)\ge \gamma$ uniformly in $N$.
\end{thm}
For the dependence of $\gamma$ on the MPS tensor, we refer the reader to \citet*{fannes:FCS} and \textcite{nachtergaele:degen-mps}.

For two- and higher-dimensional systems, no comparably strong result is known. In particular, injectivity does not imply a gap, since examples of injective PEPS are known which exhibit power law correlations (such as the ``Ising PEPS'' \cite{verstraete:comp-power-of-peps} on the honeycomb lattice at the critical point, see Appendix.~\ref{app:examples}) and thus cannot be ground states of gapped Hamiltonians \cite{hastings:gap-and-expdecay,nachtergaele:exp-clustering}.

One can, however, derive a lower bound on the gap of the parent Hamiltonian for PEPS whose tensor $A$ is sufficiently close to a PEPS $B$ with parent Hamiltonian $h$ which satisfies the martingale condition \eqref{eq:4:martingale-2} (such as a commuting Hamiltonian), in the sense that $A^i=\sum\Lambda_{ij} B^j$ with $\Lambda-\openone$ small. This is based on the fact that following the logic of the previous subsection around Eq.~\eqref{eq:4:deformed-parent-1}, $h'=({\Lambda^{-1}}^\dagger)^{\otimes k}h(\Lambda^{-1})^{\otimes k}$ is a parent Hamiltonian for $A$ (with $k$ the locality of the Hamiltonian), and the bound $\gamma$ on the gap in the martingale condition changes smoothly with $\Lambda$. Alternatively, this can be seen from the fact that \eqref{eq:4:martingale-2} lower bounds the angle between the ground spaces $\mathcal G_{i,j}$ of $h_i$ and $h_j$, which changes smoothly under deformations $\Lambda^{\otimes k}\mathcal G_{i,j}$. This has been carried out explicitly by \textcite[][Appendix E]{schuch:mps-phases} for a commuting Hamiltonian $h$ acting on $2\times 2$ blocks, and it has been found that the gap is stable as long as the ratio of the smallest and largest singular value of $\Lambda$ is above $\approx 0.967$.

\paragraph{The Knabe bound} The Knabe bound relates the existence of a gap of a translational invariant and frustration free Hamiltonian in the thermodynamic limit with the scaling of the gap of the same  Hamiltonian on a finite chain.  In particular, \textcite{knabe:knabe-bound} showed that if the gap of a open boundary 1D chain with a nearest neighbor projector Hamiltonian is larger than $1/(n-1)$  for some $n>2$, then the system with periodic boundaries is gapped in the thermodynamic limit; this result was later improved  to $6/n(n+1)$ by \textcite{gosset:knabe-bound}, and generalized to two dimensions. The method has also been extended to frustration-free Hamiltonians with open boundary conditions (stating that for gapless systems, the gap must close at least as $n^{-3/2}$, showing the impossibility of chiral edge modes with frustration free Hamiltonians whose gap should scale as $1/n$) by \citet{mozgunov:knabe-bound-chiral}.

\subsubsection{Stability}

A key question in the definition of quantum phases is their stability against arbitrary small perturbations (or, when considering SPT phases, against arbitrary symmetry-preserving perturbations). Here, stability can refer to different properties, such as a smooth dependence of various physical properties on the perturbation.  The most commonly considered property is the stability of the spectral gap, as it implies stability of local properties through quasi-adiabatic continuation \cite{hastings:quasi-adiabatic}. It is easy to see that parent Hamltonians of MPS and PEPS do not necessarily have such a stability property: The parent Hamiltonian of a GHZ state (which is a PEPS, Section~\ref{app:examples}) is a ferromagnetic Ising Hamiltonian, whose two-fold degenerate ground space is susceptible to small perturbations. 

In the following, we will discuss conditions under which the ground space of a MPS/PEPS parent Hamiltonian can be shown to be stable under perturbations.

\paragraph{The LTQO condition}

For frustration free Hamiltonians, stability of the spectral gap is implied by a conditions known as LTQO (local topological quantum order) condition \cite{bravyi:tqo-long,bravyi:local-tqo-simple,michalakis:local-tqo-ffree}.
Roughly speaking, it states that the effect of boundary conditions is exponentially suppressed in the bulk (as a function of the distance of the boundary): 

\begin{defn}
Consider a translational invariant frustration free Hamiltonian on on a 2D square lattice.  We say that a \emph{region $A$ satisfies LTQO}, 
if there is a superpolynomially decaying function $f_A(m)$ (i.e., 
$\lim_{m\rightarrow\infty}m^kf_A(m) = 0$ for all $k>0$), such that for
for any observable $O_a$ supported on $A$, and all bounded regions 
$B$ containing $A$, 
it holds that for any pair of normalized ground states $\ket{\Psi}$, $\ket{\Psi'}$ of the Hamiltonian restricted to region $B$, 
\begin{equation}
\label{eq:4:ltqo}
 \Big|\langle \Psi | O_a |\Psi\rangle
 -\langle \Psi'|O_a|\Psi'\rangle
\Big| \le \|O_a\| f_A(m)\ ,
 \end{equation}
where $m$ is the distance between $A$ and $\partial B$ (the boundary of $B$).

We say that a particular \emph{observable $O_a$ satisfies LTQO} if it verifies (\ref{eq:4:ltqo}). 

We finally say that a \emph{system satisfies LTQO} if all its regions $A$ satisfy it and the function $f$ in (\ref{eq:4:ltqo}) is independent of $A$.
\end{defn}

LTQO implies stability of the gap under local perturbations, under some additional local gap conditions. 
\begin{thm}[\textcite{michalakis:local-tqo-ffree}]
\label{thm:Spiros-stability}
Let $H_N=\sum h_i$ be a local Hamiltonian which satisfies LTQO, and let $V=\sum_{k} v_k$, with $v_k$ a bounded local term centered at site $k$. Moreover, assume there is a $\gamma>0$ such that $H_N$ has a gap $\Delta_N\ge \gamma$ with periodic boundaries, and let the spectral gap of $H_N$ restricted to open boundaries decay at most polynomially with the system size. Then, there exist $N_0$ and $\epsilon_0>0$ such that $H_N+\epsilon V$ has a gap at least $\gamma/2$ for any $N\ge N_0$ and $\epsilon\le \epsilon_0$.
\end{thm}
We refer the reader for a more precise formulation of the theorem, including several generalizations, to \textcite{michalakis:local-tqo-ffree,nachtergaele2020quasi}.

In the context of PEPS, the LTQO conditions has two additional advantages: First, it can be checked numerically for specific regions (and possibly observables), by relating it to an eigenvalue problem; and second, it allows for direct conclusions about the stability of physical observables under the class of natural PEPS perturbations $A^i\to \sum\Lambda(\epsilon)_{ij}A^j$, where $\Lambda(\epsilon)\to1$ smoothly (as $\epsilon\to0$): In that case, the derivative of any observable $O_a(\epsilon)$ changes smoothly in around $\epsilon=0$ as well \cite{cirac:itb}.

\paragraph{Stability in one dimension} In one dimension, one can prove for one-dimensional MPS with normal tensors that the LTQO condition always holds \cite{cirac:itb}.  This implies that
\begin{thm}
For MPS with normal tensors, the gap of the parent Hamiltonians is stable under perturbations $V_N=\sum_k v_k$, with $v_k$ a bounded local perturbation centered around $k$. This is, there exists an $\epsilon_0>0$ and $\gamma>0$ such that $H_N+\epsilon V_N$ has a unique ground state with a gap $\Delta_N>\gamma$ for all $\epsilon<\epsilon_0$ and all $N$.
\end{thm}

An alternative proof of this stability, which does not build on the LTQO condition, has been given by \textcite{szehr:gap-stability-1D}.

\paragraph{Stability in two dimensions}

In two dimensions, the LTQO condition is generally hard to prove.  Specific cases in which it holds are PEPS with commuting parent Hamiltonians, such as isometric PEPS, $G$-isometric PEPS \cite{schuch:peps-sym}, or MPO-isometric PEPS \cite{sahinoglu:mpo-injectivity}, which are therefore robust against local perturbations. 

\paragraph{Perturbations of the tensor}

An alternative way to perturb MPS and PEPS is to perturb the tensor, rather than the Hamiltonian.  There are two types of these perturbations:

\emph{(i) Physical perturbations} are perturbations which can be understood as acting only on the physical index, $A^i\to \sum\Lambda_{ij} A^j$. As discussed, these correspond to perturbations of the parent Hamiltonian of the form 
$h\to({\Lambda^{-1}}^\dagger)^{\otimes k}h(\Lambda^{-1})^{\otimes k}$. They are thus ``physical'' in the sense that they correspond to a physical perturbation of the Hamiltonian. As discussed in the preceding subsections, immediately implies that for normal MPS in 1D, and in the presence of LTQO in 2D, these perturbations only give rise to smooth changes in the properties of the system, and do not close the gap; alternatively, this also follows from the stability of the martingale condition.  Note that for normal MPS and PEPS, any perturbation of the tensor is a physical perturbation on injective blocks -- e.g., one can first invert the injective tensor by acting on the physical block, and then put the perturbed tensor instead. 

\emph{(ii) Unphysical perturbations} are perturbations of the tensor which cannot be understood as a physical perturbation, i.e.\ one which only acts on the physical index. Clearly, such perturbations only exist for non-normal tensors. It has been shown that unlike physical perturbations, unphysical perturbations of the tensor can have a drastic effect: Perturbing a $G$-injective PEPS immediately breaks topological order \cite{chen:topo-symmetry-conditions} by condensing the anyons (cf.~Section~\ref{sec:3}), and the same is true for MPO-injective PEPS \cite[][though MPO-injective PEPS are stable against certain unphysical perturbations]{shukla:tns-stringnet-pert-condensation}. This immediately implies that unphysical perturbations cannot be related to physical perturbations of the parent Hamitonian, since e.g.\ the Toric Code is stable against any perturbation of the Hamiltonian.  Within the framework of parent Hamiltonians, these perturbations are thus generally unphysical and not of direct interest.  It is however possible to introduce other types of Hamiltonians, such as the \emph{uncle Hamiltonians} discussed below, which are stable under general perturbations. At the same time, understanding which perturbations are unphysical is relevant for the numerical simulation of topological phases, since one needs to protect against breaking of these symmetries (i.e., enforce $G$-symmetry or MPO-symmetry) in order to obtain systems which exhibit topological order.

\subsubsection{Alternative Hamiltonians}

There are ways to obtain Hamiltonians with properties different from parent Hamiltonians. This can be achieved in at least two ways: Either by considering tensors which are not in canonical form, or by considering an alternative construction for the Hamiltonian.

\paragraph{Product Vacua with Boundary States (PVBS)}
PVBS \cite{bachmann:pvbs-prb,bachmann:pvbs-cmp} are models which are constructed from MPS which are not in their canonical form, such as 
$$
A^0=(1+\lambda) |0)(0|+|1)(1|\,,\ A^1=|0)(1|\ ,
$$
where $\lambda\in[-1,1]$.
The parent Hamitonian of this MPS will have the MPS on an open boundary condition chain as its ground space.  This model will support two ground states: Either the all-$0$ state, or a state with a single $1$ state, which ``binds'' to the left ($\lambda<0$) or right ($\lambda>1$) edge -- the edge states. On the other hand, the PBC ground space only supports the all-$0$ state -- the product vacuum. PVBS models demonstrate that the classification of phases is different on a system with boundaries, or alternatively, that one has to consider whether a closing gap affects the bulk behavior (which in the above model is smooth even when $\lambda$ changes sign).

Note that these models have been generalized to higher dimensions outside the framework of tensor networks \cite{bachmann:pvbs-higherdim}.

\paragraph{Uncle Hamiltonians}
Uncle Hamiltonians \cite{fernandez:1d-uncle,fernandez-gonzalez:uncle-long} are defined to overcome the non-continuity of the parent Hamiltonian under physical perturbations. They are defined by first applying a 
(potentially unphysical) perturbation to the tensor, constructing the parent Hamiltonian, and subsequently taking the limit of the perturbation to zero.  The obtained Hamiltonians are termed \emph{uncle Hamiltonians} and are by construction continuous under the perturbation considered.  On the other hand, depending on the bond dimension of the MPS/PEPS they might not be unique, but depend on the path of perturbations considered.  Uncle Hamiltonians have very different properties, in particular, they are generally gapless for non-injective MPS/PEPS, such as for the 1D Ising or the 2D Toric Code model (where momentum eigenstates of domain walls or anyons, respectively, through which the perturbation destroys the conventional or topological order, have low energy).

\section{Outlook\label{sec:5-outlook}}

In this review we have covered some of the basic concepts in the field of tensor networks and many-body quantum systems, paying special attention to MPS and PEPS on regular lattices. While we have revised in certain depth many results that have been obtained until recently, we have left out whole areas of research on tensor networks that are rapidly developing and that could be the subject of one or several review by themselves. In this outlook we briefly list some of those areas, as well as some open research directions.

Let us start out with the numerical algorithms built on tensor networks to describe different aspects of many-body quantum systems. While the present paper exclusively deals with analytical results, the fact that tensor network sates efficiently approximate many-body systems immediately provides a powerful playground for addressing complex problems with that technique. In one spatial dimension, the success of the density matrix renormalization group (DMRG) \cite{white:DMRG} method in addressing the physical properties of one-dimensional spin chains at zero temperature can be traced back to the fact that it can be viewed as a variational method over the manifold of MPS. One can naturally extend this algorithm to higher spatial dimensions through PEPS, although the scaling of the computational complexity with the bond dimension is not so friendly as in 1 dimension, and one has to use approximate techniques in order to compute expectation values of physical observables \cite{verstraete:2D-dmrg}. Thus, arguably the most important subjects of research in tensor networks is the development of powerful algorithms in more than one spatial dimensions. One can also extend those methods to finite temperatures by using MPDOs or PEPOs \cite{czarnik2015variational}, and to time-dependent problems \cite{czarnik2019time}. The latter can be carried out using either a variational method or a Trotterization of the evolution operator followed by truncations of the the time evolved states after each time step. Tensor networks have also been employed to express many-body operators, like Hamiltonians, to compute elementary excitations, spectral functions, densitiy of states, etc. \cite{pirvu2010matrix,mcculloch2007density}. Again, the extension of those methods to higher dimensions remains as one of the main challenges. Other tensor network states, like TTN or MERA have also played an important role in certain many-body problems and are particularly appropriate to describe critical systems \cite{evenbly2014algorithms,silvi2010homogeneous}. Methods based on MPS and PEPS have also been recently developed that allow one to compute physical properties of critical systems based on the scaling as a function of the bond dimension \cite{vanhecke2019scaling,tagliacozzo2008scaling,corboz2018finite,pirvu2012matrix,rader2018finite,stojevic2015conformal}. In a parallel effort, mathematicians have also investigated different concepts to apply tensor trains (which are analogous to MPS) to diverse problems \cite{oseledets2011tensor,hackbusch2012tensor,grasedyck2010hierarchical}. Here, the idea is also to compress high-rank tensors in terms of smaller ones, thus saving time and memory in computations.

Another very active area of research is continuous tensor networks. We have reviewed here the theory underlying the one-dimensional version, cMPS. Constructing algorithms that integrate the related Quantum-Gross-Pitaesvskii equation \cite{haegeman2017quantum} is very challenging, although considerable progress in this direction has very recently been made by \citet{tuybens2020variational}. Many efforts are also devoted nowadays to construct the higher dimensional versions \cite{tilloy2019continuous}, or the continuous MERA \cite{haegeman2013entanglement,zou2019magic,cotler2019entanglement,fernandez2019entanglement,fernandez:cmera-qft}, for computations in quantum field theories. Here, the main challenge is to make practical algorithms to deal with quantum field theories. Another relatively unexplored direction is the use of tensor networks with infinite bond dimension, like iMPS, in order to describe critical or chiral topological systems \cite{cirac2010infinite,nielsen2013local,tu2014lattice,nielsen2012laughlin}. 

As for application of the computational techniques, tensor networks have been used in problems in atomic, condensed matter and, more recently, high-energy physics. In particular the fact that symmetries (both global and local) can be easily incorporated into the tensors appears as a very attractive feature to investigate symmetry protected phases, topological models, and lattice gauge theories numerically. Again, the main challenge here is to extend current methods to higher spatial dimensions. MPS and TTN have also been applied to problems in quantum chemistry, yielding very promising results, although there are still open questions about the suitability of different tensor networks for different chemical structures \cite{white1999ab,chan2011density,szalay2015tensor}. More recently, tensor networks have been used to construct toy models of holographic principles in hyperbolic geometries \cite{swingle:mera-ads-cft,pastawski:holography,hayden:holographic-duality-from-tn}. This was triggered by the observation that MERA explicitly leads to the Ryu-Takayanagi formula for the entanglement entropy of critical states in 1+1 dimensions, where the renormalization direction can be interpreted as the radial coordinate in an AdS bulk. The language of tensor networks seems to be appropriate to construct and analyze simple models displaying some of the expected features of the AdS/CFT correspondence. Furthermore, it has also been used to describe some of the physics expected in black and worm holes.

Tensor networks are also being widely used in machine learning \cite{huggins2019towards,stoudenmire2016supervised,carleo2019machine,glasser2019expressive,glasser2018supervised}. In fact, some of the most traditional methods in that field are very closely related to those networks. There is also an intimate connection between tensor networks and so-called neural network states (and string-bond states, entangled plaquette states, etc), which in their simplest incarnation are just MPS. However, those states can be extended into other set of states which have the property that physical observables can be computed using Monte-Carlo methods, and thus they can be employed to study the ground state of many-body systems with variational Monte Carlo techniques. All those states are also intimately related to graph models in the field of machine learning. This connection is being successfully exploited in both directions: on the one hand, the techniques of deep neural networks can be applied to construct powerful computational methods for many-body quantum systems; on the other, the theory of tensor networks and its connection with entanglement can help to devise better methods in machine learning.

Tensor network techniques have also been proposed and used in quantum optics experiments. For instance, quantum tomography can become much more efficient if the states one deals with can be approximated by MPS or MPDO, as with fewer measurements one can fully characterize the many-body state \cite{cramer2010efficient}. Furthermore, in many physical systems MPS appear in a very natural way. For instance, in sequential generation, where a physical system produces or interacts with other subsystems sequentially \cite{schoen:hen-and-egg,osborne2010holographic}. This occurs, for instance, when atoms cross a cavity where they interact with one or few optical modes or, when an emitter generates photons one after each other. 

Tensor networks appear naturally in the field of quantum computing in different incarnations. First, measurement based quantum computing can be easily explained in terms of a simple PEPS, the cluster state, and teleportation-based gates acting on the auxiliary indices of the tensor whenever one performs a measurement \cite{verstraete:mbc-peps}. Quantum circuits have a natural expression in terms of tensor networks, so that the analysis of different quantum algorithms, and even the effects of the errors can sometimes be easily traced \cite{nielsen-chuang}. Additionally, quantum error correcting codes have typically simple characterizations as tensor networks \cite{terhal2015quantum}. This is the case of surface codes, for instance, which are the basis of physical implementations where gates occur locally. Tensor network techniques also seem to be essential to analyse more sophisticated quantum error correcting codes bases on e.g. string nets.  

Finally, there are very intriguing connections between tensor networks and some areas in Mathematics. For instance, MPOs can be used to construct representations of fusion categories, weak Hopf algebras and subfactors \cite{lootens2020matrix,kawahigashi2020remark,molnar:WHA}, which in turn are related to topological field theories, conformal field theories and integrability through the Yang-Baxter equation. 
\appendix

\section{Examples%
\label{app:examples}}

This appendix collects the MPS and PEPS descriptions for a variety of widely used tensor network states.

\subsection{One dimension: MPS}
\TOCstop

We start by giving a range of examples of one-dimensional MPS.

\subsubsection{Product states} Any product state $\ket{\psi}=\ket{\phi^1}\otimes\ket{\phi^2}\otimes\cdots\otimes\ket{\phi^N}$ is a trivial MPS with $D=1$. With the convention that $\ket{\phi^s}=\sum_i a^{i,[s]}\ket{i}$, we have that
\begin{equation}
\ket\psi = \sum_{i_1,\dots,i_N} a^{i_1,[1]}\cdots a^{i_N,[N]}\ket{i_1,\dots,i_N}\ .
\end{equation}

\subsubsection{The GHZ state}

The GHZ state on  a $d$-level system, 
\[
\ket{\mathrm{GHZ}}=\sum_{i=0}^{d-1}\ket{i,i,\dots,i}\ ,
\]
is an MPS with $A^{i}_{\alpha\beta}=\delta_{i=\alpha=\beta}$.

\subsubsection{The W state} The $W$ state 
\[
\ket{W}=\ket{100\dots}+\ket{010\dots}+\dots+\ket{0\dots001}
\]
is an MPS with open boundary conditions and $D=2$, with 
\begin{equation}
    A^{0}=\begin{pmatrix}1&0\\0&1\end{pmatrix}\ ,\quad 
    A^{1}=\begin{pmatrix}0&1\\0&0\end{pmatrix}\ ,
\end{equation}
and left and right boundary conditions $(l|=(0|$ and $|r)=|1)$, respectively, i.e.
\begin{equation}
    \ket W = \sum (l|A^{i_1}A^{i_2}\cdots A^{i_N}|r)\,
                 \ket{i_1,\dots,i_N} \ .
\end{equation}
Note that this is not a translationally invariant representation of the MPS due to the non-periodic boundary condition. This opens the question of which is the optimal bond dimension to represent the W state as a translationally invariant MPS. Remarkably, in this case the bond dimension $D$ must scale polynomially with the system size $N$. In particular, combining results of \citet{perez-garcia:mps-reps} and \citet{michalek:wielandt-Dsquare}, one gets that $D$ must fulfill a bound of the form $D^3\log D =\Omega(N)$ which implies in particular that, for each $\delta>0$, $D=\Omega(N^{\frac{1}{3+\delta}})$.

\subsubsection{The cluster state} The 1D cluster state~\cite{raussendorf:cluster-short} is an MPS with
\[
A^0=|0)(+|\ ,\quad A^1=|1)(-|
\]
\cite{verstraete:mbc-peps}.
This can be derived, e.g., by using the fact that the cluster state can be constructed by acting with a controlled-$Z$ between nearest neighbors, starting from a $\ket{+}^{\otimes N}$ state.

\subsubsection{The AKLT state} The 1D AKLT state~\cite*{affleck:aklt-prl} is  constructed by taking spin-$\tfrac12$ singlets as bonds and projecting the two spin-$\tfrac12$ at each site on the joint spin-$1$ subspace. The resulting tensor is (labelling the physical states as $S_z=0,\pm1$)
\[
A^{+1}=\begin{pmatrix}1&0\\0&0\end{pmatrix}Y\,,
\
A^{0}=\frac{1}{\sqrt{2}}\begin{pmatrix}0&1\\1&0\end{pmatrix}Y\,,
\ 
A^{-1}=\begin{pmatrix}0&0\\0&1\end{pmatrix}Y\,,
\]
where 
\begin{equation}
\label{eq:app:singlet-Y}
Y=\left(\begin{matrix}0&-1\\1&0\end{matrix}\right)
\end{equation}
encodes the singlet.
When expressed in the basis $\ket{+}=i(\ket{-1}+\ket{+1})/\sqrt{2}$,
$\ket{-}=(\ket{-1}-\ket{+1})/\sqrt{2}$,
and $\ket0$, this becomes
\[
A^{-} = \tfrac{1}{\sqrt{2}}\sigma_x\,,\ 
A^{+} = \tfrac{1}{\sqrt{2}}\sigma_y\,,\ 
A^{0} = \tfrac{1}{\sqrt{2}}\sigma_z\,.
\]

\subsubsection{The Majumdar-Ghosh model} The Majumdar-Ghosh model 
is the 1D version of the RVB state, and appears as the ground state of the spin-$\tfrac12$
Hamiltonian $H=\sum \bm S_i\cdot \bm S_{i+1} + \tfrac12 
\sum \bm S_i\cdot \bm S_{i+2}$
\cite{majumdar:majumdar-ghosh-model}. Its ground state is a superposition of singlet pairs $(1,2),(3,4),\dots$ and $(2,3),(4,5),\dots,(N,1)$, and can be written as an MPS with
\[
A = \Big[\ket0\big[(02|+(20|\big] + \ket1\big[(12|+(21|\big]\Big]\otimes Y
\]
\cite{verstraete:comp-power-of-peps}, with $Y$ as in Eq.~\eqref{eq:app:singlet-Y}.

\TOCstart
\subsection{Two dimensions: PEPS}
\TOCstop

Next, we give a range of examples for two-dimensional PEPS, where we follow the convention to order the virtual indices top--right--down--left, as introduced in Sec.~\ref{MPSandPEPS}.

\subsubsection{The GHZ state\label{sec:examples:2D:GHZ}} Just as in 1D, the 2D GHZ state 
\[
\ket{\mathrm{GHZ}}=\sum_{i=0}^{d-1}\ket{i,i,\dots,i}
\]
can be written as a PEPS with $D=d$ and $A^i_{\alpha\beta\gamma\delta}=\delta_{i=\alpha=\beta=\gamma=\delta}$.

\subsubsection{The cluster state} The 2D cluster state \cite{raussendorf:cluster-short} on the square lattice can be written as a PEPS with 
\[
A=\ket0(00++|+\ket{1}(11--|\ ,
\]
where $(\pm|=\big[(0|\pm(1|\big]/\sqrt{2}$.
This can again be understood by rewriting the circuit preparing the cluster state -- controlled-Z's between nearest neighbors acting on the $\ket{+}^{\otimes N}$ state -- as a tensor network \cite{verstraete:mbc-peps}. Indeed, this tensor network description straightforwardly generalized to arbitrary graphs.

\subsubsection{The AKLT model}
The 2D AKLT model \cite*{affleck:aklt-cmp} is obtained by placing singlets on the links of the lattice (commonly honeycomb or square) and projecting onto the symmetric subspace. This can be directly translated into a PEPS, by absorbing the singlets $Y=\left(\begin{smallmatrix}0&-1\\1&0\end{smallmatrix}\right)$ into the projectors $\Pi_\mathrm{sym}$ onto the symmetric space. For the square lattice, this yields tensors
\[
A= \Pi_\mathrm{sym}(\openone\otimes\openone\otimes Y\otimes Y)\ .
\]

\subsubsection{The RVB state} The (nearest neighbor) RVB state on a 2D lattice is the superposition of all ways of covering the lattice with nearest neighbor singlets. The corresponding $D=3$ PEPS tensor \cite{verstraete:comp-power-of-peps} is given by combining the projector
\[
P = \ket0\big[(0222|+(2022|+\dots\big] + 
        \ket1\big[(1222|+(2122|+\dots\big]  
\] (illustrated here for coordination number $4$) with $\pm Y$ tensors for each link (with the sign corresponding to the orientation of the singlet); for the square lattice with a translational invariant orientation of singlets, the tensor would e.g.\ be
\[
A=P(\openone\otimes\openone\otimes Y\otimes Y)\ .
\]

\subsubsection{The Toric Code and quantum double models}

The quantum double model for a finite group $G$~\cite{kitaev:toriccode} on an oriented square lattice has spins with basis $\{\ket{g}\}_{g\in G}$ assigned to every edge, and is the equal weight superposition of all basis configurations which satisfy a Gauss' law across a vertex, $g_1g_2g_3^{-1}g_4^{-1}=0$, where the inverses relate to the orientation of the edges.  A PEPS representation can be obtained by blocking every other plaquette (containing four edges) into a tensor (aligned diagonally w.r.t.\ the lattice), and using the virtual indices (which sit at the vertices of the original lattice) to enforce the Gauss' law, i.e.,
the non-zero configurations are
\begin{equation}
\label{eq:app:tcode-rep-primal}
\raisebox{-.5cm}{\includegraphics{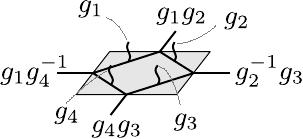}}
\end{equation}
(where the lines inside the tensor indicate the original lattice). 

Alternatively, a dual PEPS representation can be obtained by assigning dual ``color'' variables $g\in G$ to the plaquettes and defining the physical spins as the difference of adjacent plaquette colors~\cite{schuch:peps-sym}. The equal weight superposition of all plaquette colors then corresponds to the equal weight superposition of all Gauss' law configurations. The corresponding tensor is thus
\begin{equation}
\raisebox{-.5cm}{\includegraphics{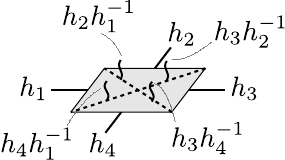}}
\label{eq:app:tcode-rep-dual}
\end{equation}
Note that this representation is $G$-injective (in fact, $G$-isometric) with respect to the regular representation, as shifting all plaquette colors does not affect the physical state. On the other hand, in the representation \eqref{eq:app:tcode-rep-primal}, the four virtual indices in the group basis fuse to the identity and thus, they possess a symmetry under the action of any irreducible representation.

From the point of view of bimodule categories  \cite{lootens2020matrix}, the case \eqref{eq:app:tcode-rep-primal} corresponds to $\mathcal{D}=G, \mathcal{M}=\mathrm{Vec}, \mathcal{C}=\mathrm{Rep}_G$, hence the MPO symmetries are labelled by the irreps. The case \eqref{eq:app:tcode-rep-dual} corresponds to $\mathcal{C}=\mathcal{M}=\mathcal{D}=G$ up to a blocking, and the MPOs are hence labelled by the group elements.

\subsubsection{String-net models}

The string net picture provides the most natural description of topological phases of matter in terms of MPO-symmetric tensors. As discussed in Sec. \ref{CSS} and \ref{sec:3:2D}, the PEPS description involves a $(\mathcal{C},\mathcal{D})$-bimodule category $\mathcal{M}$ with labels $\{a,b,c,\ldots\}\in I_\mathcal{C}$, $\{A,B,C,\ldots\}\in I_\mathcal{M}$ and $\{\alpha,\beta,\gamma\ldots\}\in I_\mathcal{D}$. As a special case, the categories might all be chosen equal to each other. For the case of a bipartite hexagonal lattice and a gauge in which all $F$-symbolds are unitary, one can choose the A-type PEPS tensors of the bipartite lattice as

\[
\left(\frac{d_\alpha d_\beta}{d_\gamma
d_C^2}\right)^{1/4}\left(^3\!F^{A\alpha\beta}_{B}\right)^{\gamma,km}_{C,jn}
= \parbox[c]{0.07\textwidth}{ \includegraphics[scale=0.75]{fig3_peps.pdf} }
\]
and the PEPS tensors on the B-type sublattice obtained by reflecting the above tensor around the x-axis and reversing all arrows, as the complex conjugate. The $d_i$ are the quantum dimensions of the different categorical objects. Additionally, an extra factor $d_A$ has to be introduced for every closed loop of virtual labels $\mathcal{M}$.

The quantum double description for a group $G$ as discussed in the previous section is a special case of this string net representations. The two options discussed above correspond to $\mathcal{D}=\mathcal{M}=\mathcal{C}=\mathrm{Vec}_{G}$ and to $\mathcal{D}=\mathrm{Vec}_{G},\,\mathcal{M}=\mathrm{Vec},\,\mathcal{C}=\mathrm{Rep}(G)$. Additionally, one can define two more Morita equivalent PEPS with physical labels in $\mathrm{Rep}(G)$ as $\mathcal{D}=\mathcal{M}=\mathcal{C}=\mathrm{Rep}(G)$ or as $\mathcal{D}=\mathrm{Rep}(G),\,\mathcal{M}=\mathrm{Vec},\,\mathcal{C}=\mathrm{Vec}_G$. Here, $\mathrm{Vec}_G$ is the category with the group elements as labels, and $\mathrm{Vec}$ is the trivial category consisting out of only 1 element.

\subsubsection{PEPS from classical models}

To each classical model $H$ and a finite inverse temperature $\beta$ there is associated PEPS that reproduces the 
the expectation value of any diagonal observable (in particular, the classical correlation functions)
present in the Gibbs state $\frac{e^{-\beta H}}{Z}$. For simplicity, let us restrict to the case of nearest neighbor interactions on a square lattice $H(\sigma_1,\ldots, \sigma_N)=\sum_{(i,j)} h_{i,j}$. Define the matrix 
$M=\sum_{i,j}e^{-\frac{\beta}{2}h(i,j)} |i)(j|$. Then, the corresponding  PEPS is given by the tensor \cite{verstraete:comp-power-of-peps}
$$
A=\Big[\sum_{i} \ket{i} (i\,i\,i\,i|\Big]( \Id \otimes \Id\otimes M\otimes M)\ .
$$
It is clear from this that expectation values of the classical Gibbs state correspond to expectation values of the associated PEPS for diagonal observables. In particular, for the critical temperature $\beta_c$ the associated PEPS has power law decaying correlations and hence its parent Hamiltonian must be gapless~\cite{hastings:gap-and-expdecay,nachtergaele:exp-clustering}.

\subsubsection{The CZX model}

The CZX model \cite{chen:2d-spt-phases-peps-ghz} is a product state of GHZ states of qubits ($d=2$) across plaquettes of a square lattice, 
placed on a torus of size $2N\times 2M$:
$$\bigotimes_{i,j=1}^{N,M} \ket{GHZ_{ij}}\ ,$$
with $\ket{GHZ_{ij}}$ the GHZ state on sites $(2i,2j), (2i+1,2j),(2i+1,2j+1), (2i, 2j+1)$.
The CZX is just the state resulting from considering blocked sites formed by $(2i-1,2j-1), (2i, 2j-1), (2i,2j), (2i-1,2j)$,
\begin{equation*}
\label{fig:CZX}
\includegraphics[scale=1]{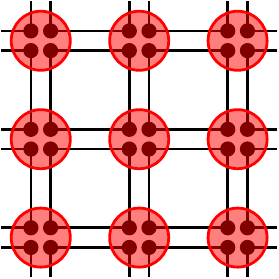}
\end{equation*}
Using the description of the GHZ given above, the PEPS tensor, with bond dimension $D=d^2=4$ is then given by
$$\sum_{i,j,k,l=0}^1\ket{ijkl}((i,j),(j,k),(k,l),(l,i)|\, $$

As explained in Section~\ref{sec:3:2D-SPT}, the CZX model belongs to the non-trivial SPT sector of a global on-site $\mathbb{Z}_2$ symmetry.

\TOCstart
\subsection{Fermionic MPS and PEPS}
\TOCstop

\subsubsection{The Kitaev chain}

The Kitaev chain, or Majorana chain~\cite{kitaev:majorana-chain} consists of a chain of spinless fermions. Each fermion consists of two Majorana fermions, which can be paired up either within a site or across adjacent sites, which can be changed by tuning the Hamiltonian. Here, we are interested in the limit where the Majorana modes pair up solely across sites, as this corresponds to a non-trivial (topological) phase.

To describe the corresponding ground state as a fermionic MPS, we start from $2N$ Majorana modes $c_j$, $\{c_j,c_k\}=2\delta_{ij}$. 
Denote by $\ket\Omega$ the vacuum, defined via $c_{2n}c_{2n+1}\ket{\Omega}=0$, $n=1,\dots,N$. Then, the non-trivial fixed point of the Kitaev chain is the state of the $N$ complex (Dirac) fermions $a_n=(c_{2n-1}+ic_{2n})/2$ in the state $\ket\Psi=\ket\Omega$. The corresponding description in terms of graded tensor networks is given in Eq.~\eqref{Majchain} and has tensors \cite{bultinck:fermionic-mps-phases} $A^0=\left(\begin{smallmatrix}1&0\\0&1\end{smallmatrix}\right)$ and 
$A^1=\left(\begin{smallmatrix}0&1\\-1&0\end{smallmatrix}\right)$, with a twist $Y=A^1$ at the boundary.

\subsubsection{Free fermionic and chiral PEPS}

Free (or non-interacting) fermions are fermionic systems governed by a Hamiltonian which is quadratic in the fermionic creation and annihilation operators $a^\dagger_x$ and $a_x$, where $x$ denotes the lattice position (and possibly other degrees of freedom such as spin). Ground and thermal states $\rho$ of such Hamiltonians (``Gaussian states'') are fully characterized by their second moments $\gamma_{xy} = \tfrac{\mathrm{i}}{2}\mathrm{tr}\big(\rho [c_x,c_y]\big)$ due to Wick's theorem, where we use a  Majorana representation $c_{2x-1}=a_x+a_x^\dagger$, $c_{2x}=-\mathrm{i}(a_x-a_x^\dagger)$. A special case are PEPS constructed using Gaussian states as bonds and Gaussian maps as PEPS tensors (that is, maps which map Gaussian states to Gaussian states). Due to their compact representation and the possibility to exactly solve for the ground state of e.g.\ translational invariant quadratic Hamiltonians, those Gaussian fermionic PEPS form an important testbed for the investigation of PEPS and their ability to describe certain types of systems. 

A translational invariant Gaussian fermionic PEPS in $D$ spatial dimensions with $n$ physical Majorana modes per site and $m$ Majorana modes per bond is specified by a $(n+2Dm)\times (n+2Dm)$ real antisymmetric matrix with a block structure
\[
\Gamma = \begin{pmatrix}X & Y \\ -Y^T & Z \end{pmatrix}
\]
(where $X$ is $n\times n$ and $Z$ is $2Dm\times 2Dm$) which satisfies $\Gamma^2=-\openone$~\cite{kraus:fPEPS}.  It describes a Gaussian state with non-zero correlations
\begin{align*}
\hat\gamma(k) &= X+Y(Z+\omega(k))^{-1}Y^T\ ,
\\
    \omega(k) &= \bigoplus_{\alpha=1}^D \begin{pmatrix} 0 & e^{ik_\alpha}\openone_m \\
                    e^{-ik_\alpha}\openone_m & 0 \end{pmatrix}
\end{align*}
in momentum space, $\hat\gamma(\vec k) =
\tfrac{\mathrm i}{2}\mathrm{tr}\big(\rho[\hat c_k,\hat c_{-k}]\big)$, $\hat c_k=\sum e^{i\,k\cdot x}c_x/\sqrt{N}$.
 Importantly, since the entries of the inverse of a matrix $M=Z+\omega(k)$ are the quotient of the determinant of minors of $M$ and of $\det(M)$, any Gaussian fermionic PEPS has the special property that $\hat\gamma(k)$ is the ratio of polynomials of degree at most $2Dm$ in $e^{\pm i k_\alpha}$~\cite{schuch:GMPS}.

One key example is a Gaussian fermionic PEPS which describes a topological superconductor in $D=2$ spatial dimensions, that is, a system with chiral order~\cite{wahl:chiral-fPEPS,wahl:chiral-boundaries}, for which 
$n=2$, $m=1$ (i.e.\ each bond only consists of a single Majorana mode), and
\begin{align*}
X &= \left(\begin{array}{cc}
0&1-2\lambda\\
-1+2\lambda&0
\end{array}\right), \notag \\
Y &=  \sqrt{\lambda - \lambda^2}\left(\begin{array}{cccc}
1&-1 &0&-\sqrt{2}\\
-1&-1&-\sqrt{2}&0
\end{array}\right), \notag \\
Z &= \left(\begin{array}{cccc}
0&     1-\lambda     &-\frac{\lambda}{\sqrt 2}&-\frac{\lambda}{\sqrt{2}}\\
-1+\lambda&0&\frac{\lambda}{\sqrt 2}&-\frac{\lambda}{\sqrt{2}}\\
\frac{\lambda}{\sqrt 2}&-\frac{\lambda}{\sqrt 2}&0&1-\lambda\\
\frac{\lambda}{\sqrt 2}&\frac{\lambda}{\sqrt 2}&-1+\lambda&0
\end{array}\right) \ .
\end{align*}
where $0<\lambda<1$.

\TOCstart
\subsection{MPOs and MPUs\label{sec:app-examples:mpo-mpu}}
\TOCstop

\subsubsection{The CZX MPU}

As explained in Section \ref{sec:3:2D-SPT}, the reason behind the fact that the CZX model is a non-trivial SPT phase is the existence of a non trivial MPU symmetry in the associated PEPS determined by a nontrivial 3-cocycle. This MPU is given by the tensor
$$\ket{0}\bra{1} \otimes |0)(+|+ \ket{1}\bra{0}\otimes |1)(-|\ ,$$
and can be understood as a product of overlapping controlled-$Z$'s $\mathrm{CZ}=\mathrm{diag}(1,1,1,-1)$ between all adjacent sites (as they commute, the ordering does not matter), followed by a Pauli $X$ on all sites (thus the name)~\cite{chen:2d-spt-phases-peps-ghz}.

This MPO, that we denote $O(A)$, is injective, as can easily be seen by considering the algebra generated by the matrices 
$$
A^{01}= \begin{pmatrix}1&1 \\ 0&0 \end{pmatrix} \text{ and }  A^{10}=\begin{pmatrix}0&0\\1&-1\end{pmatrix}\;.
$$

Let us now square this MPO, thereby getting an MPO $O(B)$ with bond dimension 4, given by the matrices:
\[B^{00}=\begin{pmatrix}1 & 1 \\0 & 0\end{pmatrix}\otimes \begin{pmatrix}0 & 0\\1 & -1 \end{pmatrix}=\begin{pmatrix} 0 & 0 & 0 & 0\\1 & -1 & 1 & -1\\0 & 0 & 0 & 0\\0 & 0 & 0 & 0\end{pmatrix} \]
\[B^{11}=\begin{pmatrix}0 & 0\\1 & -1 \end{pmatrix}\otimes\begin{pmatrix}1 & 1\\0 & 0\end{pmatrix}=\begin{pmatrix}0 & 0 & 0 & 0\\0 & 0 & 0 & 0\\1 & 1 & -1 & -1\\0 & 0 & 0 & 0 \end{pmatrix}\]
This MPO is clearly not injective, and it has an invariant subspace given by the projector
\[P=\begin{pmatrix}0 & 0 & 0 & 0\\0 & 1 & 0 & 0\\0 & 0 & 1 & 0\\0 & 0 & 0 & 0\end{pmatrix}\]
By the canonical form construction (Section \ref{sec_4}), we can therefore as well work with the block
\[B^{00}=\begin{pmatrix}-1 & 1\\0 & 0\end{pmatrix},\hspace{1cm} B^{11}=\begin{pmatrix}0 & 0\\1 & -1\end{pmatrix}\]
But this MPO again has an invariant subspace given by the projector $$Q=1/2\begin{pmatrix}1 & -1\\-1 & 1\end{pmatrix}.$$ Applying once more the canonical form construction, we arrive at the following canonical form for the MPO
 $B^{ij}= (-1)\delta_{ij}$, which globally means ${O}(A)^2=(-1)^N I$. The CZX MPO hence provides a very nice example of how the canonical form construction works.

\subsubsection{The shift MPU}

The shift is the paradigmatic example of an MPU that cannot be approximated by a short range time evolution, see Section \ref{sec:2:MPU}~\cite{cirac2017matrix}.
Its tensor is given by 
\[\sum_{i,j} \ket{i}\bra{j}\otimes |i)(j|\ .\]

\subsubsection{The MPO for the Fibonacci model}

 String nets in the PEPS picture are described by a $(\mathcal{C},\mathcal{D})$-bimodule category $\mathcal{M}$ with labels $\{a,b,c,\ldots\}\in I_\mathcal{C}$, $\{A,B,C,\ldots\}\in I_\mathcal{M}$ and $\{\alpha,\beta,\gamma\ldots\}\in I_\mathcal{D}$.
 
The MPOs are given by

\[ \parbox[c]{0.07\textwidth}{ \includegraphics[scale=0.75]{fig3_mpo.pdf} } \hspace{.4cm}=\frac{1}{\sqrt{d_Ad_D}}\left(^2\!F^{aC\alpha}_B\right)^{D,nk}_{A,jm} \]
with $^2\!F$ a solution of the 6 coupled pentagon equations and $d_i$ the quantum dimensions of the categorical objects. For the Fibonacci model, we can take $\mathcal{C}=\mathcal{M}=\mathcal{D}$ and hence $^1\!F=^2\!F=^3\!F=^4\!F=^5\!F=F$. The categorical objects are $I_{\mathcal{C}}=\{1,\tau\}$ with quantum dimensions $d_1=1$, $d_\tau=(1+\sqrt{5})/2$ and the fusion rules are given by 
\[N^1_{11}=N^\tau_{\tau 1}=N^\tau_{1 \tau }=N^1_{\tau\tau}=N^\tau_{\tau\tau}=1.\]
The elements of $\left(F^{abc}_d\right)^f_e$ are zero unless $N_{ab}^c>0,N_{cd}^e>0,N_{ad}^f>0,N_{bc^f}>0$, and the only allowed elements that can not be chosen equal to $1$ by an appropriate gauge choice are 

\[\left(F^{\tau\tau\tau}_\tau\right)^a_b=\mat{cc}{\frac{1}{d_\tau} & \frac{1}{\sqrt{d_\tau}}\\ \frac{1}{\sqrt{d_\tau}} & -\frac{1}{d_\tau}}_{ab}\]

\TOCstart
\begin{acknowledgments}
This review would not have been possible without the numerous colleagues  -- too many to list in person -- with whom we have both 
collaborated on and extensively discussed about the various aspects of tensor networks, ranging all the way from their mathematical structure to their physical use and numerical utility.  We are deeply grateful to each and every one of them. 
We also want to thank all people who kept encouraging us both to start and to finish this review. 
Finally, we are very grateful to  Jos\'e Garre Rubio for creating most figures in the manuscript.

\vspace{0.5cm}

This work has received support from the European Research Council (ERC) under the European Union’s Horizon 2020 program (grant agreements No.\  636201 (WASCOSYS),  647905 (QUTE),  648913 (GAPS),   742102 (QENOCOBA),  and 863476 (SEQUAM)), from the DFG (German Research Foundation) under Germany’s Excellence Strategy (EXC2111-390814868), and through the Severo Ochoa project SEV-2015-0554 (MINECO, Spain).
\end{acknowledgments}

\end{document}